\documentclass[a4paper, 10pt]{article}

\usepackage{jheppub}

\usepackage{amsthm}

\usepackage[numbers,sort&compress]{natbib}
\usepackage{dsfont}
\usepackage{slashed}
\usepackage{color}
\usepackage{empheq}

\usepackage{adjustbox}

\usepackage{bibgerm}

\usepackage{soul}

\usepackage[english]{babel}

\usepackage[utf8x]{inputenc}

\usepackage{graphicx,epsfig}
\usepackage{pstricks}

\usepackage{tikz}

\usepackage{bashful}

\usepackage{subfigure}

\usepackage{setspace}

\usepackage{booktabs}

\usepackage{float}

\usepackage{nextpage}

\usepackage{pdfpages}

\usepackage{hypcap}

\usepackage[small,bf]{caption}

\usepackage{longtable}

\usepackage{fancyhdr}

\usepackage[makeroom]{cancel}

\usepackage{microtype}

\usepackage{feynmp-auto-mod}

\usepackage{xparse}
\usepackage{etoolbox}

\usepackage[subfigure]{tocloft}

\usepackage{listings}

\usepackage{enumitem}

\usepackage{simpler-wick}

\newcommand{\p}{\partial}
\newcommand{\leftp}{{\overleftarrow \p\!}}

\newcommand{\<}{\langle}
\renewcommand{\>}{\rangle}

\renewcommand{\O}{\mathcal{O}}

\newcommand{\tr}{\mathrm{Tr}}

\newcommand{\A}{\mathcal{A}}
\newcommand{\B}{\mathcal{B}}
\newcommand{\G}{\mathcal{G}}
\newcommand{\D}{\mathcal{D}}
\newcommand{\E}{\mathcal{E}}
\newcommand{\R}{\mathcal{R}}
\renewcommand{\L}{\mathcal{L}}
\newcommand{\nn}{\nonumber\\}

\newcommand{\dD}{d^D\!}
\newcommand{\q}{\mathsf{q}}
\newcommand{\lt}{l_t}

\newcommand{\msbar}{$\overline{\text{MS}}$}
\newcommand{\hc}{\mathrm{h.c.}}

\NewDocumentCommand{\g}{ O{} }{
	\ifblank{#1}{\bar g}{\bar g_{#1}}
}

\newcommand{\op}[3]{\O^{#2,#3}_{#1}}

\NewDocumentCommand{\Op}{ m m O{} o }{
	\O^{\ifblank{#3}{}{#3,}#2 }_{\IfNoValueTF{#4}{#1}{\substack{#1\\#4}}}
}
\NewDocumentCommand{\EOp}{ m m O{} o }{
	\E^{\ifblank{#3}{}{#3,}#2 }_{\IfNoValueTF{#4}{#1}{\substack{#1\\#4}}}
}
\NewDocumentCommand{\lwc}{ m m O{} o }{
	L^{\ifblank{#3}{}{#3,}#2 }_{\IfNoValueTF{#4}{#1}{\substack{#1\\#4}}}
}

\newcommand{\mytag}{\\[-\baselineskip] \stepcounter{equation}\tag{\theequation}}

\definecolor{darkgreen}{rgb}{0,0.5,0}
\definecolor{darkblue}{rgb}{0,0,0.5}
\definecolor{darkred}{rgb}{0.5,0,0}
\definecolor{beige}{rgb}{0.7,0.4,0.3}

\makeatletter
  \def\my@tag@font{\normalsize}
  \def\maketag@@@#1{\hbox{\m@th\normalfont\my@tag@font#1}}
  \let\amsmath@eqref\eqref
  \renewcommand\eqref[1]{{\let\my@tag@font\relax\amsmath@eqref{#1}}}
\makeatother

\newenvironment{myfmf}[1]
{\begin{fmffile}{#1}
\fmfcmd{%
  style_def momins expr p =
  drawarrow p;
  enddef;}
\fmfcmd{%
  style_def marrowc expr p =
  drawarrow subpath (1/4, 3/4) of p withpen pencircle scaled 0.4;
  enddef;}
\fmfcmd{%
  style_def marrowd expr p =
  drawarrow subpath (1/4, 3/4) of p shifted 10 down withpen pencircle scaled 0.4;
  enddef;}
\fmfcmd{%
  style_def marrowu expr p =
  drawarrow subpath (1/4, 3/4) of p shifted 10 up withpen pencircle scaled 0.4;
  enddef;}
\fmfcmd{%
  style_def marrowl expr p =
  drawarrow subpath (1/4, 3/4) of p shifted 10 left withpen pencircle scaled 0.4;
  enddef;}
\fmfcmd{%
  style_def mlarrowd expr p =
  drawarrow subpath (1/8, 7/8) of p shifted 10 down withpen pencircle scaled 0.4;
  enddef;}
\fmfcmd{%
  style_def mlarrowu expr p =
  drawarrow subpath (1/8, 7/8) of p shifted 10 up withpen pencircle scaled 0.4;
  enddef;}
\fmfcmd{%
  style_def mlarrowl expr p =
  drawarrow subpath (1/8, 7/8) of p shifted 10 left withpen pencircle scaled 0.4;
  enddef;}
\fmfcmd{%
  style_def marrowr expr p =
  drawarrow subpath (1/4, 3/4) of p shifted 10 right withpen pencircle scaled 0.4;
  enddef;}
\fmfcmd{%
  style_def darrowd expr p =
  drawarrow subpath (1/4, 3/4) of p shifted 10 down dashed evenly withpen pencircle scaled 0.4;
  enddef;}
\fmfcmd{%
  style_def darrowu expr p =
  drawarrow subpath (1/4, 3/4) of p shifted 10 up dashed evenly withpen pencircle scaled 0.4;
  enddef;}
\fmfcmd{%
  style_def darrowl expr p =
  drawarrow subpath (1/4, 3/4) of p shifted 10 left dashed evenly withpen pencircle scaled 0.4;
  enddef;}
\fmfcmd{%
  style_def darrowr expr p =
  drawarrow subpath (1/4, 3/4) of p shifted 10 right dashed evenly withpen pencircle scaled 0.4;
  enddef;}
\fmfcmd{
    path quadrant, q[], otimes;
    quadrant = (0, 0) -- (0.5, 0) & quartercircle & (0, 0.5) -- (0, 0);
    for i=1 upto 4: q[i] = quadrant rotated (45 + 90*i); endfor
    otimes = q[1] & q[2] & q[3] & q[4] -- cycle;
}
\fmfwizard
}
{
\end{fmffile}
}

\makeatletter
\renewcommand\paragraph{\@startsection{paragraph}{4}{\z@}%
  {-3.25ex\@plus -1ex \@minus -.2ex}%
  {1.5ex \@plus .2ex}%
  {\normalfont\normalsize\bfseries}}
\makeatother

\cftsetindents{section}{0em}{2em}
\cftsetindents{subsection}{2em}{3em}
\cftsetindents{subsubsection}{5em}{4em}

\allowdisplaybreaks[1]

\preprint{
\mbox{}\hfill{} ZU-TH 02/26
}

\title{\boldmath One-loop matching of the LEFT to the QCD~gradient~flow}

\author[]{\`Oscar L. Crosas,}

\author[]{Peter Stoffer}

\emailAdd{oscar.laracrosas@physik.uzh.ch}
\emailAdd{stoffer@physik.uzh.ch}

\affiliation[]{Physik-Institut, Universit\"at Z\"urich, Winterthurerstrasse 190, 8057 Z\"urich, Switzerland}
\affiliation[]{PSI Center for Neutron and Muon Sciences, 5232 Villigen PSI, Switzerland}

\abstract{We present the complete one-loop matching of the baryon- and lepton-number-conserving low-energy effective field theory (LEFT) to the QCD gradient flow. Using Euclidean conventions and the background-field formulation of the gradient flow, we derive the short-flow-time expansion for the full LEFT operator basis up to mass dimension six. The matching is performed in dimensional regularization in the algebraically consistent ’t Hooft--Veltman scheme, including a systematic treatment of evanescent operators and the finite counterterms required to restore chiral symmetry in the spurion sense. Keeping fully generic flavor structures, we verify the cancellation of spurious chiral-symmetry-violating terms with the known finite symmetry-restoring counterterms. This demonstrates that the gradient flow as a gauge-invariant ultraviolet regulator enables an efficient extraction of both divergent and finite counterterms in addition to the matching contributions. We provide the matching coefficients both before and after field redefinitions that remove redundant operators, as well as power-divergent mixings into lower-dimensional operators. Our results establish a consistent perturbative link between continuum LEFT calculations and gradient-flow-based lattice-QCD matrix elements, enabling precision low-energy phenomenology beyond leading-logarithmic accuracy.
}

\numberwithin{equation}{section}

\begin{document}

	\maketitle

% ============================================================================

	\begin{myfmf}{diags/diags}

	% !TEX root = ../Paper-LEFT-GF.tex

\section{Introduction}

Effective field theories (EFTs) for physics beyond the Standard Model (SM) have matured significantly over the past years and have become an accurate and efficient tool for indirect low-energy searches. The description of the effects of physics beyond the SM in low-energy observables is affected by widely separated scales: the assumed very high scale of new physics and the typical energy scale of low-energy precision experiments. EFTs exploit this large ratio of scales: on the one hand, they enable a resummation of large logarithms of the ratio of scales, on the other hand they also allow us to split the calculation into several simpler parts depending on fewer scales. Above the electroweak scale but well below the scale of new physics, one usually employs the SMEFT~\cite{Buchmuller:1985jz,Grzadkowski:2010es,Jenkins:2013zja,Jenkins:2013wua,Alonso:2013hga}, which at the electroweak scale can be matched to the low-energy EFT (LEFT) by integrating out the heavy SM particles~\cite{Jenkins:2017jig,Jenkins:2017dyc,Dekens:2019ept}. In recent years, much progress has been achieved towards the extension of these EFTs to complete next-to-leading-logarithmic (NLL) accuracy, with the calculation of two-loop renormalization-group (RG) equations~\cite{Buras:1989xd,Buras:1991jm,Buras:1992tc,Ciuchini:1993vr,Ciuchini:1993fk,Buchalla:1995vs,Chetyrkin:1997gb,Buras:2000if,Gorbahn:2016uoy,Panico:2018hal,deVries:2019nsu,Bern:2020ikv,Aebischer:2022anv,Fuentes-Martin:2022vvu,Aebischer:2023djt,Jenkins:2023rtg,Jenkins:2023bls,Naterop:2023dek,DiNoi:2023ygk,Fuentes-Martin:2023ljp,Morell:2024aml,Manohar:2024xbh,DiNoi:2024ajj,Born:2024mgz,Naterop:2024cfx,Fuentes-Martin:2024agf,Aebischer:2025hsx,Duhr:2025zqw,Haisch:2025lvd,Zhang:2025ywe,Naterop:2025lzc,Naterop:2025cwg,Haisch:2025vqj,Duhr:2025yor,Banik:2025wpi}, the one-loop matching between the two EFTs~\cite{Dekens:2019ept}, and the automation of the one-loop matching to UV models~\cite{Carmona:2021xtq,Fuentes-Martin:2022jrf,Fuentes-Martin:2023ljp,Guedes:2023azv,Aebischer:2023nnv,Thomsen:2024abg,Guedes:2024vuf}. The LEFT (with varying number of active degrees of freedom) is valid down to the hadronic scale, where QCD becomes strongly coupled. The description of the low-energy effects of the strong interaction requires non-perturbative methods, such as lattice QCD. The EFT framework allows one to tackle this problem independently of the UV models by computing hadronic matrix elements of higher-dimension effective operators in QCD (and possibly QED). Since the EFT framework is commonly formulated in dimensional regularization (DR), the use of lattice-QCD input requires another matching calculation between the continuum DR scheme and a scheme amenable to lattice computations. An increasingly popular scheme is given by the gradient flow~\cite{Luscher:2010iy,Luscher:2011bx,Luscher:2013cpa}. It offers several advantages, such as manifest gauge invariance or the disentanglement of power divergences from the continuum limit, which can be taken for fixed flow times. The matching from a DR continuum scheme to the gradient flow can be performed at a scale where perturbation theory is still expected to work well, provided that sufficiently short flow times can be reached on the lattice. A particularly interesting application are lattice-QCD matrix elements for the neutron electric dipole moment (nEDM)~\cite{Shintani:2005xg,Berruto:2005hg,Shindler:2014oha,Guo:2015tla,Shindler:2015aqa,Alexandrou:2015spa,Shintani:2015vsx,FlavourLatticeAveragingGroupFLAG:2021npn,Dragos:2019oxn,Bhattacharya:2021lol,Bhattacharya:2023qwf,Bhattacharya:2025aht}, which are notoriously difficult due to the presence of power divergences and a large operator basis in particular when using more traditional momentum-subtraction schemes~\cite{Bhattacharya:2015rsa,Cirigliano:2020msr}. The one-loop matching to the gradient flow has already been calculated for all operators contributing to the nEDM up to mass dimension six~\cite{Mereghetti:2021nkt,Buhler:2023gsg,Crosas:2023anw}. For other observables, the matching to the gradient flow has been performed to higher perturbative orders~\cite{Harlander:2016vzb,Harlander:2018zpi,Artz:2019bpr,Harlander:2020duo,Harlander:2022tgk,Borgulat:2023xml,Borgulat:2025gys,Harlander:2025qsa,Georg:2026ozz}.

In the present work, we make a step towards the one-loop matching of the complete LEFT to the gradient flow. We disregard QED corrections, which for many phenomenological applications are negligible, and we do not consider baryon- and lepton-number-violating operators, which will be treated in an upcoming work~\cite{Crosas:2026}. Compared to previous work, we aim at a unified treatment of the whole theory: we include the full flavor structure, including both flavor-conserving and flavor-changing operators, thus covering all conceivable low-energy processes. We perform the matching to the LEFT in the algebraically consistent 't~Hooft--Veltman (HV) scheme~\cite{Breitenlohner:1977hr,tHooft:1972tcz} as defined in Ref.~\cite{Naterop:2023dek}. This scheme avoids any ill-defined expressions related to the intrinsically four-dimensional quantities $\gamma_5$ or the Levi-Civita symbol. The employed scheme also includes finite counterterms that on the one hand separate the evanescent sector from the physical one, and on the other hand restore global chiral symmetry in the spurion sense. We show that the matching to the gradient flow enables an efficient calculation of both the divergent as well as the finite symmetry-restoring counterterms. We avoid gauge-variant off-shell operators by employing the background-field formulation of the gradient flow~\cite{Suzuki:2015bqa}. Thus, the gradient flow is not only of interest for lattice-QCD applications, but also provides access to the renormalization of the theory.

The article is structured as follows. In Sect.~\ref{sec:LEFT}, we define our conventions for the LEFT in Euclidean space. In Sect.~\ref{sec:Scheme}, we review the continuum HV scheme of Ref.~\cite{Naterop:2023dek}. We discuss the gradient flow and its background-field formulation in Sect.~\ref{sec:GradientFlow}.
In Sect.~\ref{sec:Computation}, we explain some aspects of the calculation and we discuss the results in Sect.~\ref{sec:Results}, before concluding in Sect.~\ref{sec:Conclusions}. The appendices contain several technical details and the explicit results: in App.~\ref{sec:Conventions}, we provide the definition of our conventions and we give the relation between our present Euclidean formulation of the LEFT and the operator coefficients of Refs.~\cite{Jenkins:2017jig,Naterop:2023dek} in App.~\ref{sec:LEFTBasis}. In App.~\ref{sec:FeynmanRules}, we list the Feynman rules for the gradient flow in the background-field formulation. In App.~\ref{sec:MatchingEquations}, we provide the full set of matching equations, both before and after field redefinitions that remove redundant operators.

	% !TEX root = ../Paper-LEFT-GF.tex

\section{LEFT}

\label{sec:LEFT}

\subsection{LEFT in Euclidean space-time}

As the gradient flow provides a Gaussian damping only in Euclidean space, in the present work we employ Euclidean conventions, which is also more convenient for making contact with lattice-QCD literature. In particular, we choose slightly different conventions than Refs.~\cite{Jenkins:2017jig,Naterop:2023dek} in Minkowski space---the translation between the conventions is provided in App.~\ref{sec:Conventions} and App.~\ref{sec:LEFTBasis}.

The LEFT is a $SU(3)_c \times U(1)_\mathrm{em}$ gauge theory and its leading order is given by the QCD and QED Lagrangian
\begin{align}
	\label{eq:QCDQEDLagrangian}
	\L_\mathrm{QCD+QED} &= \frac{1}{4g^2} G^A_{\mu\nu} G^A_{\mu\nu} + \frac{1}{4e^2} F_{\mu\nu} F_{\mu\nu} + i \frac{\theta_\mathrm{QCD}}{32\pi^2} G^A_{\mu\nu} \widetilde G^A_{\mu\nu} + i \frac{\theta_\mathrm{QED}}{32\pi^2} F_{\mu\nu} \widetilde F_{\mu\nu} \nn
		&\quad + \sum_{\psi=u,d,e} \bar\psi ( \slashed D + M_\psi P_L + M_\psi^\dagger P_R ) \psi + \bar\nu_L \slashed\p \nu_L + \Lambda \, ,
\end{align}
including a cosmological constant $\Lambda$. Here, $g$ and $e$ denote the bare QCD and QED gauge couplings, respectively, and the dual field-strength tensors are defined as $\widetilde G_{\mu\nu}^A = \frac{1}{2} \epsilon_{\mu\nu\alpha\beta} G^A_{\alpha\beta}$, $\widetilde F_{\mu\nu} = \frac{1}{2} \epsilon_{\mu\nu\alpha\beta} F_{\alpha\beta}$ with $\epsilon_{1234} = +1$. The covariant derivative is defined as
\begin{equation}
	D_\mu = \p_\mu + G_\mu - i Q A_\mu \, , \quad G_\mu = T^A G_\mu^A \, ,
\end{equation}
with anti-Hermitian $SU(3)_c$ generators $T^A$, see App.~\ref{sec:Conventions}, and the $U(1)_\mathrm{em}$ charge operator $Q$ with eigenvalues $\q_\psi$.

Beyond leading order in the LEFT power counting, the QCD and QED Lagrangian is supplemented by an infinite tower of higher-dimension effective operators,
\begin{equation}
	\label{eq:LEFTLagrangian}
	\L_\mathrm{LEFT} = \L_\mathrm{QCD+QED} + \sum_i L_i \O_i + \sum_i L_i^\text{red} \O_i^\text{red} + \sum_i K_i \E_i \, .
\end{equation}
In the present work, we disregard operators that violate baryon or lepton number. The operators $\O_i$ form an on-shell basis, whereas the redundant operators $\O_i^\text{red}$ appear in off-shell calculations and can be removed by field redefinitions. In App.~\ref{sec:LEFTEuclideanBasis}, we reproduce the operators $\O_i$ and $\O_i^\text{red}$ up to mass dimension six~\cite{Jenkins:2017jig}. The operators $\E_i$ are evanescent and hence vanish in $D=4$ space-time dimensions, but they are generated in DR. We use the definition of evanescent operators in the HV scheme of Ref.~\cite{Naterop:2023dek}.

\subsection{Background-field method}

In the background-field method (BFM)~\cite{Abbott:1980hw,Abbott:1983zw}, we split all fields $F$ into the sum of a classical background field $\hat F$ and a quantum fluctuation $\mathcal{F}$, explicitly
\begin{equation}
	G_\mu = \hat G_\mu + \G_\mu \, , \quad
	A_\mu = \hat A_\mu + \A_\mu \, , \quad
	q = \hat q + \tilde q \, , \quad
	l = \hat l + \tilde l \, ,
\end{equation}
where we denote quarks and leptons collectively by $q$ and $l$, and where the quantum fields $\G_\mu$, $\A_\mu$, $\tilde q$, $\tilde l$ are the integration variables in the path integral. For ordinary Euclidean QCD and QED, we use the quantum gauge-fixing term
\begin{equation}
	\L_\mathrm{QGF}^\mathrm{QCD+QED} = \frac{1}{2g_0^2 \xi_g} (\G^A)^2 + \frac{1}{2e_0^2\xi_\gamma} (\p_\mu \A_\mu)^2  \, , \quad \G^A = (\hat D_\mu \G_\mu)^A = \p_\mu \G_\mu^A + f^{ABC} \hat G^B_\mu \G^C_\mu \, ,
\end{equation}
where the background-covariant derivative in the adjoint representation is
\begin{equation}
	\hat D_\mu (\cdot) = \p_\mu (\cdot) + [ \hat G_\mu, \,\cdot\, ] \, .
\end{equation}
The quantum gauge-fixing term preserves background-gauge invariance. Since we will match the off-shell 1PI effective action, there is no need to fix the gauge of the background fields.

The ghost Lagrangian is given by
\begin{equation}
	\L_\mathrm{gh} = (\hat D_\mu \bar\eta^A) (D_\mu \eta^A) \, .
\end{equation}

Although we include QED in the discussion, in the present work we will neglect $\O(\alpha_\mathrm{QED})$ corrections, i.e., we will not include dynamical photons in the matching calculation.

	% !TEX root = ../Paper-LEFT-GF.tex

\section{Scheme definition}
\label{sec:Scheme}

In this work, we perform the one-loop matching of the LEFT to flowed operators using dimensional
regularization in the HV scheme. While the HV scheme provides a
mathematically consistent treatment of $\gamma_5$ and the Levi-Civita tensor in DR, it leads to spurious violations of chiral symmetry when combined with minimal
subtraction (MS). Following Ref.~\cite{Naterop:2023dek}, we therefore define a renormalization scheme
that deviates from pure MS by including appropriate finite counterterms that
restore chiral symmetry.

Throughout this paper, we work in $D = 4 - 2\varepsilon$ dimensions and we adopt Euclidean conventions,
as introduced in Sect.~\ref{sec:LEFT}. The continuum scheme definition consists of three elements: the implementation of
DR in the HV scheme, the treatment of evanescent structures, and the
inclusion of finite counterterms restoring chiral symmetry.

\subsection{Dimensional regularization in the HV scheme}

In the HV scheme, the Levi-Civita symbol $\epsilon_{\mu\nu\rho\sigma}$ and $\gamma_5$ are treated as
intrinsically four-dimensional objects. The $D$-dimensional metric is decomposed as
\begin{equation}
	g_{\mu\nu} = g_{\bar\mu\bar\nu} + g_{\hat\mu\hat\nu} \, ,
\end{equation}
where $g_{\bar\mu\bar\nu}$ projects onto a four-dimensional subspace and $g_{\hat\mu\hat\nu}$ onto the
remaining $-2\varepsilon$ dimensions, satisfying
\begin{equation}
	g_{\bar\mu\bar\mu} = 4 \, , \qquad g_{\hat\mu\hat\mu} = -2\varepsilon \, , \qquad
	g_{\bar\mu\bar\rho} g_{\hat\rho\hat\nu} = 0 \, .
\end{equation}
Correspondingly, Dirac matrices are decomposed as
\begin{equation}
	\gamma_\mu = \gamma_{\bar\mu} + \gamma_{\hat\mu} \, ,
\end{equation}
where $\gamma_5$ anticommutes with $\gamma_{\bar\mu}$ but commutes with $\gamma_{\hat\mu}$.

The renormalizable part of the LEFT Lagrangian is promoted to $D$ dimensions in a
background-gauge-invariant manner. In contrast, the basis of higher-dimension physical operators is defined
strictly in four space-time dimensions, with all Lorentz indices in effective operators contracted using
$g_{\bar\mu\bar\nu}$ and $\gamma_{\bar\mu}$. In particular, tensor bilinears are defined using
\begin{equation}
	\sigma_{\mu\nu} \;\to\; \sigma_{\bar\mu\bar\nu}
	:= \frac{i}{2} [\gamma_{\bar\mu},\gamma_{\bar\nu}] \, .
\end{equation}
This convention is consistent with the operator basis
used in Ref.~\cite{Naterop:2023dek}.

\subsection{Evanescent structures}

Loop calculations in DR generate intermediate Dirac and Lorentz structures that vanish
in four space-time dimensions. The presence of such evanescent operators $\E_i$ is an unavoidable feature of
DR and their precise definition is part of the renormalization scheme. In the present work, we adopt the evanescent basis in the HV scheme defined in Ref.~\cite{Naterop:2023dek}.

The renormalization scheme of Ref.~\cite{Naterop:2023dek} includes finite counterterms to physical operator coefficients that compensate the insertion of bare evanescent operators~\cite{Dugan:1990df,Herrlich:1994kh}. This implies that renormalized evanescent operators, i.e., the bare evanescent operators together with the counterterms, have vanishing physical matrix elements and hence the scheme avoids a RG mixing of the evanescent sector into physical operators at NLL (two-loop) accuracy~\cite{Naterop:2024cfx,Naterop:2025cwg,Naterop:2025lzc}.

The definition of the evanescent basis directly affects the one-loop matching of the physical operators. Given the separation of the evanescent sector, the evanescent matching coefficients $K_i$ are of no further interest at NLL and would only be relevant for the determination of the two-loop matching of the physical sector. Therefore, we will not report the one-loop matching coefficients of evanescent operators. The situation changes in the presence of tree-level matching contributions to evanescent operators:
in that case the finite evanescent-compensating counterterms affect the one-loop matching, in particular when the method of regions~\cite{Beneke:1997zp} is employed, as discussed in Ref.~\cite{Crosas:2023anw}.

\subsection{Restoration of chiral symmetry}

Although the HV scheme correctly reproduces the axial anomaly, it leads to spurious violations of
chiral symmetry in the regularized theory. In particular, the fermion gauge--kinetic terms induce
chirality-flipping contributions proportional to evanescent Dirac structures. When combined with
minimal subtraction, at one loop these effects result in finite chiral-symmetry-breaking terms in Green's
functions.

To disentangle these spurious effects from the physical breaking of chiral symmetry due to fermion
masses and higher-dimension operators, we follow Ref.~\cite{Naterop:2023dek} and restore chiral
symmetry by means of finite counterterms. To this end, the fermion mass matrices, the theta parameters, and the Wilson coefficients of higher-dimension operators are promoted to spurion fields with a definite transformation under global chiral symmetry, which renders the physical sector of the LEFT Lagrangian chirally invariant. The restoration of chiral symmetry in the spurion sense is then achieved by defining an appropriate
non-minimal subtraction scheme in which the renormalized parameters are related to the
MS ones by finite shifts.
%\begin{equation}
%  X_{\text{ren}} = X_{\overline{\text{MS}}}
%  + \frac{1}{16\pi^2}\,\Delta X ,
%\end{equation}
%where $X$ denotes gauge couplings, theta parameters, masses, and wave-function normalizations.
The explicit expressions for these finite counterterms are taken from
Ref.~\cite{Naterop:2023dek}. By construction, they cancel all spurious
chiral-symmetry-breaking terms arising from the HV regulator and ensure that the renormalized
effective action respects chiral spurion symmetry.

We emphasize that this scheme choice only affects intermediate quantities, as the broken chiral spurion symmetry is a global symmetry of the LEFT. Physical observables
and relations between them are scheme independent, hence in any scheme the symmetry-breaking terms need to cancel between matching contributions and the RG evolution. The scheme of Ref.~\cite{Naterop:2023dek} has the advantage that chiral symmetry is preserved in intermediate stages. As a result, the matching to flowed operators is manifestly free of unphysical chiral-symmetry-violating terms, and such scheme-dependent artifacts do not obscure the interpretation of the results.

We will perform the gradient-flow matching keeping the full flavor structure and in particular generic mass matrices. This allows us to directly identify symmetry-breaking terms in our calculation. The cancellation of these contributions with the finite counterterms of Ref.~\cite{Naterop:2023dek} provides a strong cross-check of both calculations and shows that the full LEFT renormalization can be extracted from the matching.

	% !TEX root = ../Paper-LEFT-GF.tex

\section{Gradient flow}
\label{sec:GradientFlow}

\subsection{Conventional gauge fixing}

In the present work, we are interested in the gradient flow as a bridge to lattice simulations that are used to obtain non-perturbative matrix elements. We consider the situation where only quarks and gluons are simulated as dynamical degrees of freedom on the lattice, whereas the remaining fields are treated as external static sources on the lattice and dealt with in continuum perturbation theory. Therefore, we only apply the gradient flow to gluon and quark fields. With conventional gauge fixing, the flow equations read~\cite{Luscher:2010iy,Luscher:2011bx,Luscher:2013cpa}
\begin{align}
	\label{eq:ConventionalFlowEquations}
	\p_t B_\mu &= D_\nu G_{\nu\mu} + \alpha_0 D_\mu \p_\nu B_\nu \, , \nn
	\p_t \chi &= D_\mu D_\mu \chi - \alpha_0 (\p_\mu B_\mu) \chi \, , \nn
	\p_t \bar\chi &= \bar\chi \overleftarrow{D\!}_\mu \overleftarrow{D\!}_\mu + \alpha_0 \bar\chi (\p_\mu B_\mu)  \, ,
\end{align}
where as a boundary condition the flowed fields agree with the fundamental fields at vanishing flow time
\begin{equation}
	\label{eq:BoundaryConditionsFullFields}
	B_\mu(x;t=0) = G_\mu(x) \, , \quad
	\chi(x;t=0) = q(x) \, , \quad
	\bar\chi(x;t=0) = \bar q(x) \, .
\end{equation}
In order not to break $U(1)_\mathrm{em}$ invariance, the flowed covariant derivative of the quark fields in Eq.~\eqref{eq:ConventionalFlowEquations} reads~\cite{Buhler:2023gsg}
\begin{equation}
	D_\mu \chi(x;t) = ( \p_\mu + B_\mu(x;t) - i Q A_\mu(x) ) \chi(x;t) \, ,
\end{equation}
with the photon field $A_\mu$, which can be treated as an unflowed static external source if $\O(\alpha_\mathrm{QED})$ corrections are neglected.

\subsection{Background-field method for the gradient flow}

The BFM has been formulated for the gradient flow in Ref.~\cite{Suzuki:2015bqa}. The flowed fields are split into background and quantum fields
\begin{equation}
	B_\mu(x;t) = \hat B_\mu(x;t) + \B_\mu(x;t) \, , \quad
	\chi(x;t) = \hat \chi(x;t) + \zeta(x;t) \, , \quad
	\bar\chi(x;t) = \hat{\bar\chi}(x;t) + \bar\zeta(x;t)
\end{equation}
and the flow equations in conventional gauge~\eqref{eq:ConventionalFlowEquations} are replaced by
\begin{align}
	\label{eq:FlowEquationsBFM}
	\p_t B_\mu &= D_\nu G_{\nu\mu} + \alpha_0 D_\mu \hat D_\nu \B_\nu \, , \nn
	\p_t \chi &= D_\mu D_\mu \chi - \alpha_0 (\hat D_\mu \B_\mu) \chi \, , \nn
	\p_t \bar\chi &= \bar\chi \overleftarrow{D\!}_\mu \overleftarrow{D\!}_\mu + \alpha_0 \bar\chi (\hat D_\mu \B_\mu)
\end{align}
in order to preserve background-gauge covariance. The background fields are flowed according to
\begin{equation}
	\label{eq:FlowEquationsBackground}
	\p_t \hat B_\mu = \hat D_\nu \hat G_{\nu\mu} \, , \quad
	\p_t \hat \chi = \hat D_\mu \hat D_\mu \hat \chi \, , \quad
	\p_t \hat {\bar\chi} = \hat{\bar\chi} {\,\hat{\!\overleftarrow{\,D}}}_\mu {\,\hat{\!\overleftarrow{\,D}}}_\mu
\end{equation}
and they fulfill the boundary conditions
\begin{equation}
	\label{eq:BoundaryConditionsBackground}
	\hat B_\mu(x;t=0) = \hat G_\mu(x) \, , \quad
	\hat \chi(x;t=0) = \hat q(x) \, , \quad
	\hat {\bar\chi}(x;t=0) = \hat {\bar q}(x) \, .
\end{equation}
The boundary conditions~\eqref{eq:BoundaryConditionsFullFields} and~\eqref{eq:BoundaryConditionsBackground} imply the boundary conditions for the quantum fields
\begin{equation}
	\B_\mu(x;t=0) = \G_\mu(x) \, , \quad
	\zeta(x;t=0) = \tilde q(x) \, , \quad
	\bar\zeta(x;t=0) = \tilde {\bar q}(x) \, .
\end{equation}
Their flow equations follow from Eqs.~\eqref{eq:FlowEquationsBFM} and~\eqref{eq:FlowEquationsBackground}.

For the perturbative solution of the flow equations, we rewrite them in integral form. For the background fields, we obtain
\begin{align}
	\label{eq:FlowIntegralEquationsBackground}
	\hat B_\mu(x;t) &= \int \dD y \left[ K^{(0)}_{\mu\nu}(x-y;t) \hat G_\nu(y) + \int_0^t ds \, K^{(0)}_{\mu\nu}(x-y;t-s) \hat R_\nu(y;s) \right] \, , \nn
	\hat \chi(x;t) &= \int \dD y \left[ J(x-y;t) \hat q(y) + \int_0^t ds \, J(x-y;t-s) \hat\Delta(y;s) \right] \, , \nn
	\hat {\bar\chi}(x;t) &= \int \dD y \left[ \hat{\bar q}(y) \bar J(x-y;t) + \int_0^t ds \, \hat{\bar\Delta}(y;s) \bar J(x-y;t-s) \right] \, ,
\end{align}
with the heat kernels
\begin{align}
	K^{(\alpha_0)}_{\mu\nu}(x;t) &= \int \frac{\dD p}{(2\pi)^D} \frac{e^{ip\cdot x}}{p^2} \left[ (\delta_{\mu\nu}p^2 - p_\mu p_\nu) e^{-tp^2} + p_\mu p_\nu e^{-\alpha_0 t p^2} \right] \, , \nn
	J(x;t) &= \bar J(x;t) = \int \frac{\dD p}{(2\pi)^D} e^{ip\cdot x} e^{-t p^2} = \frac{e^{-\frac{x^2}{4t}}}{(4\pi t)^{D/2}} \, ,
\end{align}
and interaction terms
\begin{align}
	\label{eq:FlowInteractionTermsBackground}
	\hat R_\mu &= 2 [ \hat B_\nu, \p_\nu \hat B_\mu] - [\hat B_\nu, \p_\mu \hat B_\nu] - [\hat B_\mu, \p_\nu \hat B_\nu] + [\hat B_\nu, [\hat B_\nu, \hat B_\mu]] \, , \nn
	\hat \Delta &= \left[ (\p_\nu \hat B_\nu) + 2 \hat B_\nu \p_\nu - i Q (\p_\nu \hat A_\nu) - 2i Q \hat A_\nu \p_\nu + \hat B_\nu \hat B_\nu - 2 i Q \hat A_\nu \hat B_\nu - Q^2 \hat A_\nu \hat A_\nu \right]  \hat\chi \, , \nn
	\hat{\bar \Delta} &= \hat{\bar\chi} \left[ - (\p_\nu \hat B_\nu) - 2 \leftp_\nu \hat B_\nu + i Q (\p_\nu \hat A_\nu) + 2 i Q \leftp_\nu \hat A_\nu + \hat B_\nu \hat B_\nu - 2 i Q \hat A_\nu \hat B_\nu - Q^2 \hat A_\nu \hat A_\nu \right] \, .
\end{align}
For the quantum fields, the integral form of the flow equations reads
\begin{align}
	\label{eq:FlowIntegralEquationsQuantum}
	\B_\mu(x;t) &= \int \dD y \left[ K^{(\alpha_0)}_{\mu\nu}(x-y;t) \G_\nu(y) + \int_0^t ds \, K^{(\alpha_0)}_{\mu\nu}(x-y;t-s) \R_\nu(y;s) \right] \, , \nn
	\zeta(x;t) &= \int \dD y \left[ J(x-y;t) \tilde q(y) + \int_0^t ds \, J(x-y;t-s) \tilde\Delta(y;s) \right] \, , \nn
	\bar\zeta(x;t) &= \int \dD y \left[ \tilde{\bar q}(y) \bar J(x-y;t) + \int_0^t ds \, \tilde{\bar\Delta}(y;s) \bar J(x-y;t-s) \right] \, ,
\end{align}
with interaction terms
\begin{align}
	\label{eq:FlowInteractionTermsQuantum}
	\R_\mu &= 2 [ \B_\nu, \p_\nu \B_\mu] - [ \B_\nu, \p_\mu \B_\nu] - (1-\alpha_0)[ \B_\mu, \p_\nu \B_\nu] \nn
		&\quad + 2 [\B_\nu, \p_\nu \hat B_\mu] - (1+\alpha_0)  [\B_\nu, \p_\mu \hat B_\nu] - [\B_\mu, \p_\nu \hat B_\nu] \nn
		&\quad + 2 [\hat B_\nu, \p_\nu \B_\mu] - (1-\alpha_0)  [\hat B_\nu, \p_\mu \B_\nu] - (1-\alpha_0) [\hat B_\mu, \p_\nu \B_\nu] \nn
		&\quad + [ \B_\nu, [ \B_\nu, \B_\mu ]] + [ \B_\nu, [ \B_\nu, \hat B_\mu ]] + (1+\alpha_0) [ \B_\nu, [ \hat B_\nu, \B_\mu ]] + (1-\alpha_0) [ \hat B_\nu, [ \B_\nu, \B_\mu ]] \nn
		&\quad + [ \hat B_\nu, [ \hat B_\nu, \B_\mu ]] + (1-\alpha_0) [ \hat B_\nu, [ \B_\nu, \hat B_\mu ]] + (1+\alpha_0) [ \B_\nu, [ \hat B_\nu, \hat B_\mu ]] \, , \nn
	\tilde \Delta &= \bigg[ (\p_\nu \hat B_\nu) + 2 \hat B_\nu \p_\nu - i Q (\p_\nu \hat A_\nu) - 2 i Q \hat A_\nu \p_\nu + \hat B_\nu \hat B_\nu - 2 i Q \hat A_\nu \hat B_\nu - Q^2 \hat A_\nu \hat A_\nu \bigg]  \zeta \nn
			&\quad + \begin{aligned}[t] &\bigg[ (1-\alpha_0) (\p_\nu \B_\nu) + 2 \B_\nu \p_\nu - i Q (\p_\nu \A_\nu) -  2 i Q \A_\nu \p_\nu + \B_\nu \B_\nu - 2 i Q \A_\nu \B_\nu - Q^2 \A_\nu \A_\nu \\
				& + (1 + \alpha_0) \B_\nu \hat B_\nu + (1 - \alpha_0) \hat B_\nu \B_\nu - 2 i Q \hat A_\nu \B_\nu - 2 i Q \A_\nu \hat B_\nu - 2 Q^2 \A_\nu \hat A_\nu \bigg] (\zeta + \hat\chi) \, , \end{aligned} \nn
	\tilde{\bar \Delta} &= \bar\zeta \bigg[ - (\p_\nu \hat B_\nu) - 2 \leftp_\nu \hat B_\nu + i Q (\p_\nu \hat A_\nu) + 2 i Q \leftp_\nu \hat A_\nu + \hat B_\nu \hat B_\nu - 2 i Q \hat A_\nu \hat B_\nu - Q^2 \hat A_\nu \hat A_\nu \bigg] \nn
			&\quad +  (\bar\zeta + \hat{\bar\chi}) \begin{aligned}[t] &\bigg[ (\alpha_0-1) (\p_\nu \B_\nu) - 2 \leftp_\nu \B_\nu + i Q (\p_\nu \A_\nu) + 2 i Q \leftp_\nu \A_\nu + \B_\nu \B_\nu - 2 i Q \A_\nu \B_\nu - Q^2 \A_\nu \A_\nu \\
				& + (1 + \alpha_0) \hat B_\nu \B_\nu + (1 - \alpha_0)  \B_\nu \hat B_\nu - 2 i Q \hat A_\nu \B_\nu - 2 i Q \A_\nu \hat B_\nu - 2 Q^2 \A_\nu \hat A_\nu \bigg] \, . \end{aligned} \nonumber\mytag
\end{align}
The perturbative solution of the flow equations is most conveniently expressed in terms of Feynman rules, which we provide in App.~\ref{sec:FeynmanRules}. The interaction terms due to the flow equations of the background fields only depend on background fields, whereas the interaction terms of quantum fields always involve at least one quantum field. This implies that the effective action with insertions of flowed operators is obtained from one-particle-irreducible (1PI) Feynman diagrams, where the external legs are given by background fields. Propagators and flow lines (representing insertions of the heat kernels) inside loops always correspond to quantum fields. The notion of 1PI only concerns the cutting of propagators, but not flow lines. Therefore, 1PI diagrams can contain two types of tree-level flow lines: if in the direction of decreasing flow time a tree-level flow line leads to a tree-level sub-diagram (consisting only of further flow lines and external legs), then the flow line has to be a background field. In contrast, if in the direction of decreasing flow time a flow line leads to a sub-diagram containing loops, it has to be a quantum-field flow line.

\subsection{Short-flow-time expansion for the LEFT}
\label{sec:SFTE}

When computing in the LEFT low-energy observables in a hadronic context, one encounters non-perturbative matrix elements of quark and gluon operators in the MS (or \msbar{}) scheme. In order to use lattice-QCD input for these non-perturbative quantities, we need to relate them (in perturbation theory) to non-perturbative matrix elements of renormalized operators that are defined in a regularization-independent (RI) scheme. Such a scheme is given by the gradient flow: composite operators of the flowed quark and gluon fields~\eqref{eq:FlowEquationsBFM} are UV finite in QCD up to a multiplicative renormalization of the flowed quark fields (provided that the QCD gauge coupling and quark masses are renormalized). For short flow times, these flowed operators can be related through an operator-product expansion to MS operators at zero flow time. In order to perform this short-flow-time expansion (SFTE) for the complete theory, we define the flowed LEFT Lagrangian
\begin{equation}
	\label{eq:FlowedLEFT}
	\L_\text{LEFT}^t(x;t) = \L_\text{QCD+QED}(x) + \sum_i L_i^{t,(3)} \O_i^{(3)}(x;t) + \sum_i L_i^{t,(4)} \O_i^{(4)}(x;t) + \sum_i L_i^t \O_i(x;t) \, ,
\end{equation}
where the flowed (physical) operators $\O_i(x;t)$ are obtained by replacing quark and gluon fields in the operators $\O_i$ by their flowed counterparts at flow time $t$. In the case of quark fields, we use ``ringed fields''~\cite{Makino:2014taa,Makino:2014wca}, which are related to the flowed quark fields in Eq.~\eqref{eq:FlowEquationsBFM} by a multiplicative renormalization factor fixed by the condition
\begin{equation}
	\< 0 | \mathring{\bar\chi}(x;t) \slashed D \mathring\chi(x;t) | 0 \> \Big|_{M_q = 0} = - \frac{N_c}{(4\pi)^2 t^2}
\end{equation}
for each quark flavor. In addition to the flowed higher-dimension operators $\O_i(x;t)$, we also include in Eq.~\eqref{eq:FlowedLEFT} flowed operators of dimension three and four,
\begin{align}
	\label{eq:FlowedDim34Ops}
	\O^{(3)}_{\substack{M\chi\\pr}}(x;t) &= \mathring{\bar\chi}_{Rp}(x;t) \mathring\chi_{Lr}(x;t) \, , \nn
	\O^{(3)\dagger}_{\substack{M\chi\\pr}}(x;t) &= \mathring{\bar\chi}_{Lp}(x;t) \mathring\chi_{Rr}(x;t) \, , \nn
	\O^{L(4)}_{\substack{\chi D\\pr}}(x;t) &= \mathring{\bar\chi}_{Lp}(x;t) \slashed D \mathring\chi_{Lr}(x;t) \, , \nn
	\O^{R(4)}_{\substack{\chi D\\pr}}(x;t) &= \mathring{\bar\chi}_{Rp}(x;t) \slashed D \mathring\chi_{Rr}(x;t) \, , \nn
	\O^{(4)}_G(x;t) &= G_{\mu\nu}^A(x;t) G_{\mu\nu}^A(x;t) \, , \nn
	\O^{(4)}_\theta(x;t) &= i G_{\mu\nu}^A(x;t) \widetilde G_{\mu\nu}^A(x;t) \, ,
\end{align}
where $p$, $r$ denote flavor indices. All the flowed operators $\O(x;t)$ have UV-finite QCD matrix elements (we neglect QED corrections) and they can be written in terms of MS operators in the SFTE
\begin{equation}
	\O_i(x;t) = C_{ij}(t,\mu) \O^\mathrm{MS}_j(x;\mu) \, .
\end{equation}
Equivalently, the SFTE can be expressed in terms of Wilson coefficients as
\begin{equation}
	L_i^t \, \O_i(x;t) = L_i^t \, C_{ij}(t,\mu) \O^\mathrm{MS}_j(x;\mu) = L_j^r(\mu) \O^\mathrm{MS}_j(x;\mu) \, , \quad L_i^r(\mu) = L_j^t C_{ji}(t,\mu) \, .
\end{equation}
We find it convenient to rewrite the matching coefficients as follows. Starting from the unflowed bare LEFT Lagrangian~\eqref{eq:LEFTLagrangian}, we renormalize the quark fields
\begin{equation}
	q_L = Z_{q,L}^{1/2} q_L^r \, , \quad q_R = Z_{q,R}^{1/2} q_R^r
\end{equation}
and all the parameters including gauge coupling, mass matrices, and Wilson coefficients,
\begin{equation}
	X_i = \mu^{n_i \varepsilon} (X_i^r(\mu) + X_i^\mathrm{ct} ) \, , \quad X_i = g, \, M_q, \, L_i,\, L_i^\mathrm{red} ,\, K_i \, ,
\end{equation}
where the appropriate powers $n_i$ of the renormalization scale ensure integer mass dimensions of the renormalized parameters in $D$ dimensions.
The counterterms are taken from Ref.~\cite{Naterop:2023dek} and in addition to the one-loop divergences contain finite symmetry-restoring and finite evanescent-compensating counterterms, see Sect.~\ref{sec:Scheme},
\begin{equation}
	X_i^\mathrm{ct} = \frac{1}{\varepsilon} \frac{1}{16\pi^2} X_i^{(1,1)} + \frac{1}{16\pi^2} X_{i,\chi}^{(1,0)} + \frac{1}{16\pi^2} X_{i,\text{ev}}^{(1,0)} \, .
\end{equation}
The matching amounts to expressing the renormalized quantities in terms of the parameters of the flowed theory. To this end, we write
\begin{equation}
	X_i^r(\mu) = X_i^t + \Delta_0(X_i) + \frac{1}{16\pi^2} \Delta_1(X_i) \, ,
\end{equation}
where $X_i^t$ is the trivial (diagonal) tree-level matching, $\Delta_0$ denotes the off-diagonal part of the tree-level matching, and $\Delta_1$ is the one-loop matching contribution. For the wave-function contribution, we proceed analogously and write
\begin{equation}
	Z_{q,L/R}^{1/2} = 1 + \Delta_0(Z_{q,L/R}^{1/2}) + \frac{1}{16\pi^2} \Delta_1(Z_{q,L/R}^{1/2}) + (Z_{q,L/R}^{1/2})^\mathrm{ct} \, .
\end{equation}
With these conventions, the field redefinition that canonically normalizes the quark kinetic terms simply absorbs the wave-function matching contributions into the quark fields without changing the matching contributions to the other parameters.

Due to the addition of flowed operators at mass dimension three and four~\eqref{eq:FlowedDim34Ops}, a non-trivial tree-level matching contribution arises, which is given by
\begin{align}
	\Delta_0(Z_{u,L}^{1/2}) &= \frac{1}{2} \lwc{uD}{L(4)}[][] \, , \nn
	\Delta_0(Z_{d,L}^{1/2}) &= \frac{1}{2} \lwc{dD}{L(4)}[][] \, , \nn
	\Delta_0(Z_{u,R}^{1/2}) &= \frac{1}{2} \lwc{uD}{R(4)}[][] \, , \nn
	\Delta_0(Z_{d,R}^{1/2}) &= \frac{1}{2} \lwc{dD}{R(4)}[][] \, , \nn
	\Delta_0(K_{uD}) &= -\frac{1}{2} \left( \lwc{uD}{L(4)}[][] + \lwc{uD}{R(4)}[][] \right) \, , \nn
	\Delta_0(K_{dD}) &= -\frac{1}{2} \left( \lwc{dD}{L(4)}[][] + \lwc{dD}{R(4)}[][] \right) \, , \nn
	\Delta_0(M_u) &= \lwc{Mu}{(3)}[][] -\frac{1}{2} \left( M_u \lwc{uD}{L(4)}[][] + \lwc{uD}{R(4)}[][] M_u \right) \, , \nn
	\Delta_0(M_d) &= \lwc{Md}{(3)}[][] -\frac{1}{2} \left( M_d \lwc{dD}{L(4)}[][] + \lwc{dD}{R(4)}[][] M_d \right) \, , \nn
	\Delta_0(g) &= - 2 g^3 L_G^{(4)} \, , \nn
	\Delta_0(\theta_\mathrm{QCD}) &= 32\pi^2 L_\theta^{(4)} \, ,
\end{align}
where the parameters on the right-hand side of these equations are the flowed ones and we drop for notational simplicity the superscript $^t$. We note that there is a tree-level matching onto the evanescent dimension-4 operators
\begin{equation}
	\EOp{qD}{}[][pr] = \bar q_{Lp} \hat{\slashed D} q_{Rr} \, ,
\end{equation}
which in the determination of the one-loop matching enters through the evanescent-compensating counterterms $X_{i,\text{ev}}^{(1,0)}$.

We include the matching onto the cosmological constant $\Lambda$, which in Ref.~\cite{Naterop:2023dek} was not considered. The missing one-loop counterterms for $\Lambda$ are given (in our Euclidean conventions) by
\begin{align}
	\Lambda^{(1,1)} &= - N_c \left( \< M_u M_u^\dagger M_u M_u^\dagger \> + \< M_d M_d^\dagger M_d M_d^\dagger \> \right) \, , \nn
	\Lambda^{(1,0)}_\chi &= \frac{N_c}{12} \begin{aligned}[t]
		&\Big( \< M_u M_u M_u M_u \> + \< M_d M_d M_d M_d \> + \< M_u^\dagger M_u^\dagger M_u^\dagger M_u^\dagger \> + \< M_d^\dagger M_d^\dagger M_d^\dagger M_d^\dagger \> \\
		&+ 8 \< M_u M_u M_u M_u^\dagger \> + 8 \< M_d M_d M_d M_d^\dagger \> + 8 \< M_u M_u^\dagger M_u^\dagger M_u^\dagger \> + 8 \< M_d M_d^\dagger M_d^\dagger M_d^\dagger \> \\
		&- 8 \< M_u M_u M_u^\dagger M_u^\dagger \> - 8 \< M_d M_d M_d^\dagger M_d^\dagger \> \Big) \, , \end{aligned} \nn
	\Lambda^{(1,0)}_\mathrm{ev} &= \frac{N_c}{6} \begin{aligned}[t]
		&\Big( \< K_{uD} M_u M_u M_u M_u \> - \< K_{uD} M_u M_u M_u M_u^\dagger \> + 3 \< K_{uD} M_u M_u M_u^\dagger M_u \> \\
		&+ \< K_{uD} M_u M_u M_u^\dagger M_u^\dagger \> + 3 \< K_{uD} M_u M_u^\dagger M_u M_u \> - 3 \< K_{uD} M_u M_u^\dagger M_u M_u^\dagger \> \\
		&- 3 \< K_{uD} M_u M_u^\dagger M_u^\dagger M_u \> - \< K_{uD} M_u M_u^\dagger M_u^\dagger M_u^\dagger \> - \< K_{uD} M_u^\dagger M_u M_u M_u \> \\
		&+ \< K_{uD} M_u^\dagger M_u M_u M_u^\dagger \> - 3 \< K_{uD} M_u^\dagger M_u M_u^\dagger M_u \> - \< K_{uD} M_u^\dagger M_u M_u^\dagger M_u^\dagger \> \\
		&+ \< K_{uD} M_u^\dagger M_u^\dagger M_u M_u \> - \< K_{uD} M_u^\dagger M_u^\dagger M_u M_u^\dagger \> - \< K_{uD} M_u^\dagger M_u^\dagger M_u^\dagger M_u \> \\
		&- \< K_{uD} M_u^\dagger M_u^\dagger M_u^\dagger M_u^\dagger \> \Big) + \left( u \leftrightarrow d \right) + \hc \, . \end{aligned}
\end{align}
Note that in the present work, we will only compute single insertions of flowed operators.

	% !TEX root = ../Paper-LEFT-GF.tex

\section{Computation}
\label{sec:Computation}

\subsection{Off-shell matching and method of regions}

The one-loop matching is performed most efficiently by using the method of regions~\cite{Beneke:1997zp}. We will match the off-shell 1PI effective action by diagrammatically computing all relevant Green's functions with single insertions of flowed operators. The only hard scale in the matching is given by the flow time $t$. Expanding the integrals before integration in the masses and external momenta turns all unflowed loop integrals into vanishing scaleless integrals. On the unflowed side of the matching equation, this leaves only the tree-level insertions of counterterms and matching contributions. On the flowed side, we are left with single-scale flowed loop integrals. Apart from QCD renormalization and the multiplicative renormalization of flowed quark fields, no UV-renormalization is required on the flowed side of the matching equations. Upon expansion of the integrands, IR singularities are generated that are cancelled by the UV counterterms of the unflowed theory. The method of regions allows us to keep generic off-diagonal mass matrices, which can be treated as spurions under chiral transformations. Therefore, we are able to independently extract and cross-check the finite symmetry-restoring counterterms $X_{i,\chi}^{(1,0)}$ of Ref.~\cite{Naterop:2023dek}. Similarly, the insertion of flowed evanescent operators would allow us to extract the finite counterterms $X_{i,\mathrm{ev}}^{(1,0)}$. This illustrates that the gradient flow can be used as an efficient tool to renormalize the theory and extract both divergent and finite counterterms: the method of regions trades off the UV singularities of the unflowed theory against IR singularities in the expanded flowed loop integrals and the gradient flow as a UV regulator replaces the need for an infrared rearrangement.

By matching the off-shell 1PI effective action, we obtain the results in the off-shell basis of the LEFT~\cite{Naterop:2023dek}. As usual, in a subsequent step we remove redundant operators with the appropriate field redefinitions in order to arrive at the matching results in the physical on-shell basis.

\subsection{Algorithm for the calculation of one-loop flow-time integrals}

After applying the method of regions, we are left with rather simple flowed one-loop integrals, which can be easily computed with a general closed algorithm. The most general integral that one obtains
when computing correlation functions of a flowed operator at flow time $t$ is
\begin{align}
	&I\left[\vec{B},\alpha,\{m_n,\dots,m_1\},\{c_n,\dots,c_1\}\right] \nn
	&:= \int_0^{B_{n}(t)}dt_n\dotsb \int_0^{B_{1}(t_2,t_3,\dots,t_n,t)}dt_1 \int_k\left(k^2\right)^\alpha t_n^{m_n}\dotsb t_1^{m_1} \exp\left(-\sum_{i=1}^n c_i k^2 t_i - c k^2 t\right) \, ,
\end{align}
where we use the short-hand notation $\int_k := \int d^Dk/(2\pi)^D$ and define $B_i$ as
\begin{equation}
	B_i\left(t_{i+1},\dots,t_n,t\right)=a_i \, t+\sum_{j=i+1}^n a_{ij}t_j \, ,
\end{equation}
where in the sets $\{a_i, a_{i\,i+1}, \ldots, a_{in} \}_i$ only one element is equal to $1$ and the other elements are zero for each $i$.
We will compute the innermost flow-time integral, reducing our integral to an integral with one less flow-time integration. Therefore, the only integral that we have to consider is
\begin{equation}
	\int_0^{B_1(t_2,\dots,t_n)}dt_1 t_1^{m_1} \exp\left(-c_1k^2 t_1\right) \, ,
\end{equation}
for which we distinguish three cases:
\begin{align}
    \text{case 1: } c_1=0 &\implies \int_0^{B_1}dt_1 t_1^{m_1}=\frac{B_1^{m_1+1}}{m_1+1} \, , \nn
    \text{case 2: } c_1\neq 0, m_1=0 &\implies \int_0^{B_1}dt_1e^{-c_1 k^2 t_1}=\frac{1}{c_1 k^2}\left(1-e^{-c_1 k^2 B_1}\right) \, , \nn
    \text{case 3: } c_1\neq 0,m_1\neq 0 &\implies \int_0^{B_1}dt_1t_1^{m_1}e^{-c_1 k^2 t_1} \nn
    	&\qquad =\frac{1}{c_1k^2}\left(m_1\int_0^{B_1}dt_1t_1^{m_1-1}e^{-c_1k^2 t_1}-B_1^{m_1}e^{-c_1k^2B_1}\right) \, ,
\end{align}
where case 3 is applied recursively until $m_1=0$. 

Therefore our algorithm recursively reduces all integrals to an integral without any flow-time integration
\begin{equation}
	I[\alpha] := \int_k\left(k^2\right)^\alpha \exp\left(-ck^2t\right) \, ,
\end{equation}
which we further reduce to $\alpha=0$ using the integration-by-parts relation 
\begin{equation}
	(D+2\alpha)I[\alpha]-2ct I[\alpha+1] = 0 \, ,
\end{equation}
allowing us to write everything in terms of the simple Gaussian integral
\begin{equation}
	I[0] = \int_k \exp\left(-ck^2t\right)=\left(4\pi t c\right)^{-D/2} \, .
\end{equation}

\subsection{Implementation and checks}

For the one-loop matching, we automate the calculation to a large extent by making use of a computer-algebra tool chain. We treat the gradient flow diagrammatically and we generate all diagrams with single-operator insertions using \texttt{qgraf}~\cite{Nogueira:1991ex}. We insert the Feynman rules in the background-field formulation given in App.~\ref{sec:FeynmanRules} and we perform color, Dirac, and Lorentz algebra as well as the integral reduction using our own routines written in \texttt{Mathematica}, \texttt{FORM}~\cite{Vermaseren:2000nd,Ruijl:2017dtg}, and \texttt{Symbolica}. We are relying on two completely independent implementations of the whole calculation. In one of them, we make use of projectors to directly compute the matching contributions, in the other implementation we compute the full Green's functions by applying tensor reductions.

In addition, the following cross-checks are used to validate our results. We check the cancellation of $1/\varepsilon$ divergences with the counterterms of Ref.~\cite{Naterop:2023dek}. We further check that all contributions that violate spurion chiral symmetry cancel with the finite counterterms. In addition, we keep fully generic gauge parameters $\xi_g$ and $\alpha_0$ in the flow equations, which increases the computational cost significantly but provides a strong check: our results are indeed independent of $\xi_g$ and $\alpha_0$ already before performing the field redefinitions that remove redundant operators.

	% !TEX root = ../Paper-LEFT-GF.tex

\section{Results and comparison to the literature}
\label{sec:Results}

The explicit results for the one-loop matching of the LEFT to the QCD gradient flow are provided in App.~\ref{sec:MatchingEquations}. For convenience, we provide both the off-shell results before field redefinitions, as well as the final results in the non-redundant operator basis after field redefinitions. In addition to the most interesting matching coefficients at the same mass dimension, we compute power-divergent matchings onto lower-dimension operators: although these contributions receive important non-perturbative corrections, the results can be of interest to constrain lattice computations in the weak-coupling limit~\cite{Kim:2021qae}. We do not compute power corrections to higher-dimension operators, which are suppressed by positive powers of the flow time $t$.

By construction of the scheme, our results respect chiral symmetry in the spurion sense. Another interesting property is holomorphy~\cite{Alonso:2014rga}, as discussed also in Refs.~\cite{Jenkins:2017dyc,Naterop:2024cfx,Naterop:2025cwg}. Considering the linear combination of couplings
\begin{equation}
	\tau_\mathrm{QED} = i \frac{4\pi}{e^2} + \frac{\theta_\mathrm{QED}}{2\pi} \, , \quad \tau_\mathrm{QCD} = i \frac{4\pi}{g^2} + \frac{\theta_\mathrm{QCD}}{2\pi} \, ,
\end{equation}
we find that the one-loop matching contribution $\Delta_1(\tau_\mathrm{QED})$ only depends on the self-dual dipole-operator coefficients $L_{q\gamma}$ but not on $L_{q\gamma}^\dagger$. Also $\Delta_1(\tau_\mathrm{QCD})$ respects holomorphy with respect to dipole-operator coefficients and three-gluon-operator coefficients, i.e., it only depends on the self-dual operator coefficients $L_{qG}$ and $L_G - i L_{\widetilde G}$, but not on the anti-self-dual operator coefficients $L_{qG}^\dagger$ or $L_G + i L_{\widetilde G}$. Holomorphy is also respected by the matching contributions of the three-gluon operators to the dipole-operator coefficients and to the three-gluon-operator coefficients themselves. Holomorphy in the matching of dipoles to dipoles is respected: it is a direct consequence of chiral symmetry. We do find a violation of holomorphy in the matching of dipole operators to the quark masses, i.e., the quark-mass matrices $M_q$ receive matching contributions proportional to the dipole-operator coefficients $L_{qG}$, see Eqs.~\eqref{eq:DeltaMuOnShell} and~\eqref{eq:DeltaMdOnShell}. Since this dependence is polynomial, it could be removed by another (chirally symmetric) finite renormalization. We also note that the dependence of $\Delta_1(\tau_\mathrm{QCD})$ on the dimension-four operator coefficients $L_G^{(4)}$ and $L_\theta^{(4)}$ is not holomorphic, which is a consequence of the necessity to define the gauge-kinetic term in $D$ dimensions, whereas the theta terms are intrinsically four-dimensional. Yet, the breaking of holomorphy is again polynomial and could be removed by a finite renormalization of the gauge coupling.

After field redefinitions, we find polynomial matching contributions from $CP$-even three-gluon-operator insertions into four-quark vector-type operators, see Eqs.~\eqref{eq:DeltaLVLLuuOnShell}--\eqref{eq:DeltaLV8LRduOnShell}. There is no such logarithmic matching contribution, since the corresponding RG mixing only starts at two-loop order~\cite{Jenkins:2017dyc,Naterop:2025cwg}. Also, a similar contribution from the $CP$-odd three-gluon operator is prohibited by $CP$ symmetry, as there are no $CP$-odd flavor-conserving vector-type four-quark operators~\cite{Khatsimovsky:1987fr,Cirigliano:2020msr,Buhler:2023gsg}.

All our results are given in the HV scheme of Ref.~\cite{Naterop:2023dek}. They can be compared to partial existing results in the MS HV scheme by removing possible finite renormalizations, provided that the operator definitions agree. As we are defining all higher-dimension physical operators in four space-time dimensions, a scheme change to operators in $D$ dimension involves a finite one-loop shift due to the evanescent difference of the operator definitions. Whenever possible, we have performed these checks, as explained in the following.

The contribution to the cosmological constant $\Delta_1(\Lambda)$ due to $L_G^{(4)}$ was found in Ref.~\cite{Artz:2019bpr}. Concerning the contribution to $\Delta_1(\Lambda)$ due to $L_{Mu}^{(3)}$, Ref.~\cite{Artz:2019bpr} computed the flavor-diagonal contribution without mass corrections, while in the present work we provide the full flavor dependence and the mass corrections. The contributions to $\Delta_1(\Lambda)$ due to the kinetic operators in the massless limit have to reproduce the ringed-field conditions~\cite{Makino:2014taa,Makino:2014wca}. The $\O(t^{-1})$ mass corrections in the flavor-diagonal case can be compared to Ref.~\cite{Lange:2021vqg}, while, to our knowledge, the $\O(t^0)$ corrections are new.

The contributions to $\Delta_1(g)$ due to dimension-four operators are in agreement with the results obtained from the SFTE of the energy-momentum tensor~\cite{Harlander:2018zpi}.

Regarding the SFTE of the $CP$-odd three-gluon operator, we have checked for agreement with Ref.~\cite{Crosas:2023anw} after field redefinitions, using the existing results for the scheme in which operators are kept in four dimensions.

The SFTE of semileptonic operators is given, up to QED corrections that we do not compute, by the SFTE of the quark bilinears. Therefore, for all the semileptonic operators we could check for agreement with Ref.~\cite{Hieda:2016lly}.

Finally, after removing finite chiral-symmetry-restoring counterterms, we also checked that our results agree with the dipole-operator matching results of Ref.~\cite{Mereghetti:2021nkt} and the matching of $CP$-odd flavor-conserving four-quark operators of Ref.~\cite{Buhler:2023gsg}.

In the following, we illustrate our results, in particular the peculiarities of the HV scheme and the effect of the removal of redundant operators, for an example at dimension five: the insertion of the flowed version of the gluonic quark-dipole operators
\begin{equation}
	\O_{qG} = i \bar q_L \sigma_{\bar\mu\bar\nu} T^A q_R G^A_{\bar\mu\bar\nu} \, , \quad \O_{qG}^\dagger = i \bar q_R \sigma_{\bar\mu\bar\nu} T^A q_L G^A_{\bar\mu\bar\nu}
\end{equation}
into a quark two-point function generates matching contributions to both the mass term and the redundant operator $\O_{qD}^{(5)}$ (as well as the quark wave function). The matching result for the bare parameters before field redefinitions reads
\begin{align}
	\label{eq:BareMatching}
	L_{qD}^{(5)} &= \frac{g^2 C_F}{16\pi^2} \left[ -\frac{1}{2} \left(5-12\lt - \frac{12}{\varepsilon}\right)L_{qG} \right] \, , \nn
	M_q &= \frac{g^2 C_F}{16\pi^2} \biggl[ \begin{aligned}[t]
			& - \frac{6}{t} L_{qG}^\dagger \\
			& -\left(5+3\lt + \frac{3}{\varepsilon}\right)M_qL_{qG}M_q - \frac{3}{2}\left( 9 + 5 \lt + \frac{5}{\varepsilon} \right) \left( L_{qG}^\dagger M_q^\dagger M_q + M_q M_q^\dagger L_{qG}^\dagger \right) \\
			& + M_q L_{qG} M_q^\dagger + M_q^\dagger L_{qG} M_q + 3 \left( M_q M_q L_{qG}^\dagger + L_{qG}^\dagger M_q M_q \right) \\
			& + 2 \left( L_{qG}^\dagger M_q^\dagger M_q^\dagger + M_q^\dagger M_q^\dagger L_{qG}^\dagger - L_{qG}^\dagger M_q M_q^\dagger - M_q^\dagger M_q L_{qG}^\dagger \right) \biggr] \, , \end{aligned} \nn[-0.75cm]
\end{align}
where the logarithmic dependence on the MS scale enters through $\lt := \log(8\pi\mu^2 t)$. The Wilson coefficient on the right-hand side of these equations is the flowed one, but for notational simplicity we again drop the superscript $^t$.
As usual for matching results to the gradient flow, we express the results in terms of the MS scale $\mu$ (and not the \msbar{} scale $\bar\mu$), as this leads to simpler expressions. 

In the matching result for the bare mass matrix $M_q$ in the second equation of Eq.~\eqref{eq:BareMatching}, the first line contains the power divergence, the second line contains divergent, logarithmic and polynomial matching contributions that preserve chiral spurion symmetry, whereas the last two lines contain terms that break chiral spurion symmetry. As expected, these are purely local terms and they arise due to the use of the HV scheme.

In order to remove the redundant operator $\mathcal{O}_{qD}^{(5)}$, we perform the field redefinition 
\begin{align}
	q_{L,R}&\mapsto q_{L,R}+A_{L,R}q_{L,R}+B_{L,R}\bar{\slashed{D}}q_{R,L} \, , \nn
	\bar{q}_{L,R}&\mapsto \bar{q}_{L,R}+\bar{q}_{L,R}A_{L,R}^\dagger-\bar{q}_{R,L}\overleftarrow{\bar{\slashed{D}}}B_{L,R}^\dagger \, ,
\end{align}
where $A_{L,R}$ and $B_{L,R}$ are matrices in flavor space. We then fix 
\begin{align}
	A_L+A_L^\dagger &= -B_R^\dagger M_q -M_q^\dagger B_R \, , \quad A_L - A_L^\dagger = 0 \, , \nn
	A_R+A_R^\dagger &= -B_L^\dagger M_q^\dagger -M_q B_L \, , \quad A_R - A_R^\dagger = 0 \, , \nn
	B_L+B_R^\dagger &= -L_{qD}^{(5)} \, , \quad B_L - B_R^\dagger = 0 \, , \nn
	B_L^\dagger+B_R &= -L_{qD}^{(5)\,\dagger} \, , \quad B_L^\dagger - B_R = 0 \, ,
\end{align}
which in turn induces the shift
\begin{equation}
	M_q \mapsto M_q + \frac{1}{4}\left(M_qM_q^\dagger L_{qD}^{(5)\,\dagger} + 2M_q L_{qD}^{(5)} M_q + L_{qD}^{(5)\,\dagger} M_q^\dagger M_q \right) \, ,
\end{equation}
leading to the bare matching result after field redefinitions
\begin{align}
	\label{eq:BareMatchingAfterFR}
	M_q &= \frac{g^2 C_F}{16\pi^2} \biggl[ \begin{aligned}[t]
			& - \frac{6}{t} L_{qG}^\dagger \\
			& -\frac{25}{4} M_qL_{qG}M_q - \frac{1}{8}\left( 113 + 48 \lt + \frac{48}{\varepsilon} \right) \left( L_{qG}^\dagger M_q^\dagger M_q + M_q M_q^\dagger L_{qG}^\dagger \right) \\
			& + M_q L_{qG} M_q^\dagger + M_q^\dagger L_{qG} M_q + 3 \left( M_q M_q L_{qG}^\dagger + L_{qG}^\dagger M_q M_q \right) \\
			& + 2 \left( L_{qG}^\dagger M_q^\dagger M_q^\dagger + M_q^\dagger M_q^\dagger L_{qG}^\dagger - L_{qG}^\dagger M_q M_q^\dagger - M_q^\dagger M_q L_{qG}^\dagger \right) \biggr] \, . \end{aligned} \nn[-0.75cm]
\end{align}
As described in Sect.~\ref{sec:SFTE}, we now split the bare result into divergent counterterms, finite counterterms, as well as the (renormalized) finite matching contribution,
\begin{equation}
	M_q = \frac{1}{16\pi^2} \left( \frac{1}{\varepsilon} M_q^{(1,1)} + M_{q,\chi}^{(1,0)} + \Delta_1(M_q) \right) \, ,
\end{equation}
where the divergent and finite counterterms $M_q^{(1,1)}$ and $M_{q,\chi}^{(1,0)}$ are already known from Ref.~\cite{Naterop:2023dek}: they exactly cancel the divergent and symmetry-breaking contributions to the bare matching result, leading to
\begin{equation}
	\label{eq:RenormMatchingAfterFR}
	\Delta_1(M_q) = g^2 C_F \biggl[ - \frac{6}{t} L_{qG}^\dagger  -\frac{25}{4} M_qL_{qG}M_q - \frac{1}{8}\left( 113 + 48 \lt \right) \left( L_{qG}^\dagger M_q^\dagger M_q + M_q M_q^\dagger L_{qG}^\dagger \right) \biggr] \, ,
\end{equation}
as reported in App.~\ref{sec:MatchingEquations}. The final matching result is finite and respects chiral spurion symmetry.

	% !TEX root = ../Paper-LEFT-GF.tex

\section{Summary and conclusions}
\label{sec:Conclusions}

In this work, we have performed the one-loop matching of the LEFT to the QCD gradient flow, providing a unified and algebraically consistent treatment of the full baryon- and lepton-number-conserving operator basis up to mass dimension six. Working in Euclidean space-time and employing the background-field formulation of the gradient flow, we derived the short-flow-time expansion of all relevant flowed operators and extracted the corresponding matching coefficients in dimensional regularization.

The matching has been carried out in the ’t~Hooft--Veltman scheme as defined in Ref.~\cite{Naterop:2023dek}, including a systematic treatment of evanescent operators and the finite counterterms required to restore chiral symmetry in the spurion sense. By keeping generic flavor structures and non-diagonal mass matrices throughout the calculation, we were able to directly verify the cancellation of spurious chiral-symmetry-violating contributions and independently reproduce the finite symmetry-restoring counterterms. This provides a non-trivial cross-check of both the continuum renormalization scheme and the gradient-flow matching procedure.

A key technical ingredient of our calculation is the off-shell matching of the one-particle-irreducible effective action using the method of regions. In this framework, the gradient flow acts as a gauge-invariant ultraviolet regulator, allowing us to trade ultraviolet divergences of the unflowed theory against infrared singularities of expanded flowed loop integrals. This approach avoids the need for infrared rearrangements and enables an efficient and transparent extraction of both divergent and finite counterterms, in addition to the matching contributions. The background-field formulation ensures that gauge invariance is preserved at every step and that no gauge-variant off-shell operators are required.

Our results are presented both in the off-shell operator basis, prior to field redefinitions, and in the final non-redundant on-shell basis commonly used in phenomenological applications. Besides the matching of operators at the same mass dimension, we have also determined power-divergent mixings into lower-dimensional operators, which may serve as useful perturbative constraints for lattice simulations at small flow times. All matching coefficients are given in a scheme that is particularly well suited for consistent renormalization-group evolution beyond leading-logarithmic accuracy.

The results obtained in this work constitute an essential building block for the systematic use of lattice-QCD input in low-energy precision studies based on the LEFT. They enable a perturbative connection between continuum EFT calculations performed in dimensional regularization and non-perturbative matrix elements computed in gradient-flow-based schemes, thereby facilitating reliable phenomenological predictions for a wide class of low-energy observables.

Several extensions of this work are left for future studies. The inclusion of QED corrections and of baryon- and lepton-number-violating operators will complete the matching of the full LEFT. Moreover, the present framework provides a promising starting point for higher-order calculations, in particular for extending the matching to two-loop order or for exploring the renormalization properties of the LEFT using the gradient flow as a regulator. We expect that the methods developed here will find further applications in precision flavor physics and in the study of $CP$-violating observables.

% ============================================================================
	
	\section*{Acknowledgements}
	\addcontentsline{toc}{section}{\numberline{}Acknowledgements}

	We thank Fabian Lange and Luca Naterop for useful discussions.
	\`OLC is supported by a UZH Candoc Grant (Grant No. [FK-25-094]).
	Financial support by the Swiss National Science Foundation (Project No.~PCEFP2\_194272) is gratefully acknowledged.

% ============================================================================
	
	\appendix
	
	% !TEX root = ../Paper-LEFT-GF.tex

\section{Conventions}
\label{sec:Conventions}

We largely use the same conventions as Ref.~\cite{Mereghetti:2021nkt}, which for completeness are reproduced in the following and related to the conventions of Ref.~\cite{Jenkins:2017jig}.

\subsection[$SU(3)_c$]{\boldmath $SU(3)_c$}

In contrast to Ref.~\cite{Jenkins:2017jig}, for $SU(3)_c$ we use traceless and anti-Hermitian generators
\begin{equation}
	T^A = -i \frac{\lambda^A}{2} = -i T^A_\text{\cite{Jenkins:2017jig}} \, ,
\end{equation}
where $\lambda^A$ are the Gell-Mann matrices. The generators fulfill
\begin{align}
	[ T^A, T^B ] &= f^{ABC} T^C \, , \quad  \{ T^A, T^B \} = - \frac{1}{3} \delta^{AB} - i d^{ABC} T^C \, , \nn
	\tr[T^A T^B] &= -\frac{1}{2} \delta^{AB} \, , \quad T^A T^A = - C_F \, ,
\end{align}
with the quadratic Casimir invariant $C_F = (N_c^2-1)/(2N_c)$.

\subsection{Minkowski vs. Euclidean space}

In four dimensions, we relate Euclidean space-time and momentum vectors to the ones in Minkowski space-time by
\begin{align}
	x_\mu^E &= ( \vec x, i t ) \, , \quad x_M^\mu = ( t, \vec x ) \, , \quad x_\mu^M = ( t, - \vec x ) \, , \nn
	p_\mu^E &= ( \vec p, i E ) \, , \quad p_M^\mu = (E, \vec p) \, , \quad p^M_\mu = (E, -\vec p) \, ,
\end{align}
where only here $t$ denotes ordinary time (not to be confused with the flow time). Scalar products are related by $p^E \cdot x^E = \delta_{\mu\nu} p^E_\mu x^E_\nu = - p_M \cdot x^M = - g_{\mu\nu} p_M^\mu x_M^\nu$. Partial space-time derivatives are related by
\begin{equation}
	\p_\mu^E = \left( \vec\nabla, -i \p_t \right) \, , \quad \p_\mu^M = \left( \p_t , \vec\nabla \right) \, , \quad \p^\mu_M = \left( \p_t , -\vec\nabla \right) \, , \quad \vec\nabla = \left( \p_x, \p_y, \p_z \right) \, ,
\end{equation}
thus
\begin{equation}
	\Box = \p_\mu^M \p^\mu_M = \p_t^2 - \vec\nabla^2 = - \p_\mu^E \p_\mu^E \, .
\end{equation}
Vector fields are defined in Minkowski and Euclidean space as
\begin{equation}
	V_\mu^E = ( \vec V , i V^0 ) \, , \quad V_M^\mu = ( V^0, \vec V ) \, , \quad V^M_\mu = ( V^0 , - \vec V ) \, ,
\end{equation}
hence
\begin{equation}
	\p_\mu^E V^E_\mu = \p_\mu^M V_M^\mu
\end{equation}
and for the (Abelian) field-strength tensor $V_{\mu\nu} = \p_\mu V_\nu - \p_\nu V_\mu$, one obtains
\begin{align}
	V_{\mu\nu}^E V_{\mu\nu}^E = V_{\mu\nu}^M V^{\mu\nu}_M \, .
\end{align}
The generating functional and actions in Euclidean space are
\begin{equation}
	Z_E = \int \!\D\phi \, e^{-S_E[\phi]} \, , \quad S_E = \int d^4x \, \L_E \, ,
\end{equation}
where the Euclidean Lagrangian is related to the one in Minkowski space by $\L_E = - \L_M$, and the Wick rotation gives $S_M = i S_E$.

Compared to Ref.~\cite{Jenkins:2017jig}, we choose a different normalization of the gauge fields:
\begin{equation}
	G_\mu^A = g ( \vec G^A_\text{\cite{Jenkins:2017jig}}, i G^{A,0}_\text{\cite{Jenkins:2017jig}} ) \, , \quad A_\mu = e ( \vec A_\text{\cite{Jenkins:2017jig}}, i A^{0}_\text{\cite{Jenkins:2017jig}} ) \, ,
\end{equation}
which leads to the (Euclidean) gauge-covariant derivative
\begin{equation}
	D_\mu = \p_\mu + G_\mu - i Q A_\mu = \p_\mu + T^A G_\mu^A - i Q A_\mu \, .
\end{equation}
The field-strength tensors are defined as
\begin{equation}
	G_{\mu\nu}^A = \p_\mu G_\nu^A - \p_\nu G_\mu^A + f^{ABC} G_\mu^B G_\nu^C  \, , \quad F_{\mu\nu}^A = \p_\mu A_\nu - \p_\nu A_\mu \, .
\end{equation}

\subsection{Dirac algebra and dimensional regularization}

The Dirac matrices in $D=4-2\varepsilon$ Euclidean space-time dimensions are chosen Hermitian and fulfill the algebra
\begin{equation}
	\{ \gamma_\mu, \gamma_\nu \} = 2 \delta_{\mu\nu} \, .
\end{equation}
They are related to the Dirac matrices used in Ref.~\cite{Jenkins:2017jig} by
\begin{equation}
	\gamma_4 = \gamma_0^\text{\cite{Jenkins:2017jig}} \, , \quad \gamma_i = -i \gamma^i_\text{\cite{Jenkins:2017jig}} = i \gamma_i^\text{\cite{Jenkins:2017jig}} \quad \text{for $i\neq4$} \, .
\end{equation}
%The charge-conjugation matrix $C$ satisfies
%\begin{equation}
%	C \gamma_\mu C^{-1} = - \gamma_\mu^T \, , \quad C = -C^{-1} = -C^\dagger = - C^T \, ,
%\end{equation}
%and in the chiral basis it is given by $C=-\gamma_2\gamma_4$, i.e., it is identical to the one of Ref.~\cite{Jenkins:2017jig}, $C = C_\text{\cite{Jenkins:2017jig}}$.

In the HV scheme, we perform as usual a dimensional split of the metric into a four-dimensional part and an evanescent $-2\varepsilon$-dimensional part,
\begin{equation}
	\delta_{\mu\nu} = \delta_{\bar\mu\bar\nu} + \delta_{\hat\mu\hat\nu} \, ,
\end{equation}
with $\delta_{\bar\mu\bar\mu} = 4$, $\delta_{\hat\mu\hat\mu} = -2\varepsilon$. The respective metric tensors are used to project onto the four-dimensional and evanescent sub-spaces.

The fifth gamma matrix is defined by\footnote{This implies $\gamma_5 = - \gamma_5^\text{\cite{Jenkins:2017jig}}$ and explains the sign of the theta terms in Eq.~\eqref{eq:QCDQEDLagrangian} or the relation for the $CP$-odd three-gluon operator coefficient in Eq.~\eqref{eq:translationLEFTcoeffs}. When mapping the conventions, we apply a parity conjugation to compensate this change of conventions, so that the meaning of left- and right-chiral fields $\psi_{L,R}$ remains unchanged.}
\begin{equation}
	\gamma_5 = \frac{1}{4!} \epsilon_{\mu\nu\lambda\sigma} \gamma_\mu \gamma_\nu \gamma_\lambda \gamma_\sigma = \gamma_1 \gamma_2 \gamma_3 \gamma_4 \, ,
\end{equation}
where the Levi-Civita symbol is defined strictly in four dimensions with $\epsilon_{1234} = +1$. It is Hermitian, $\gamma_5^\dagger = \gamma_5$, and fulfills
\begin{equation}
	\{ \gamma_5, \gamma_{\bar\mu} \} = [ \gamma_5, \gamma_{\hat\mu} ] = 0 \, .
\end{equation}
The Dirac matrices in the four-dimensional sub-space fulfill the Chisholm identity
\begin{equation}
	\gamma_{\bar\mu} \gamma_{\bar\nu} \gamma_{\bar\lambda} = \gamma_{\bar\mu} \delta_{\bar\nu\bar\lambda} + \gamma_{\bar\lambda} \delta_{\bar\mu\bar\nu} - \gamma_{\bar\nu} \delta_{\bar\mu\bar\lambda} - \gamma_{\sigma} \gamma_5 \epsilon_{\mu\nu\lambda\sigma} \, .
\end{equation}
Finally, we use the definition
\begin{equation}
	\sigma_{\mu\nu} = \frac{i}{2} [ \gamma_\mu, \gamma_\nu ] \, .
\end{equation}

	% !TEX root = ../Paper-LEFT-GF.tex

\clearpage

\section{LEFT operator basis}
\label{sec:LEFTBasis}

\subsection{LEFT in Euclidean conventions}
\label{sec:LEFTEuclideanBasis}

The following list of LEFT operators up to dimension six is reproduced from~\cite{Jenkins:2017jig}. We define the operators to have a similar form as in Ref.~\cite{Jenkins:2017jig}. Due to our use of a different space-time signature, $SU(3)$ algebra, and normalization of fields, the Wilson coefficients are rescaled compared to Ref.~\cite{Jenkins:2017jig}, see App.~\ref{sec:TranslationCoefficients}. As in Ref.~\cite{Naterop:2023dek}, we use the HV scheme and define the basis of higher-dimension physical operators with all Lorentz indices summed over only the four-dimensional sub-space. Operators violating baryon or lepton number will be considered in a forthcoming work~\cite{Crosas:2026}.

%%
% --- START TABLE
%%%
\begin{table}[H]
\capstart
\vspace{-0.5cm}
\begin{adjustbox}{width=0.425\textwidth,center}
%\begin{minipage}[t]{3cm}
%\renewcommand{\arraystretch}{1.51}
%\small
%\begin{align*}
%\begin{array}[t]{c|c}
%\multicolumn{2}{c}{\boldsymbol{\nu \nu+\hc}} \\
%\hline
%\O_{\nu} & (\nu_{Lp}^T C \nu_{Lr})  \\
%%
%\end{array}
%\end{align*}
%\end{minipage}
%%%
% --- END TABLE
%%
%
%%
% --- START TABLE
%%
%\begin{minipage}[t]{3cm}
%\renewcommand{\arraystretch}{1.51}
%\small
%\begin{align*}
%\begin{array}[t]{c|c}
%\multicolumn{2}{c}{\boldsymbol{(\nu \nu) X+\hc}} \\
%\hline
%\O_{\nu \gamma} & (\nu_{Lp}^T C   \sigma_{\bar\mu\bar\nu}  \nu_{Lr})  F_{\bar\mu\bar\nu}  \\
%\end{array}
%\end{align*}
%\end{minipage}
%%%
% --- END TABLE
%%%
%%
% --- START TABLE
%%
\begin{minipage}[t]{3cm}
\renewcommand{\arraystretch}{1.51}
\small
\begin{align*}
\begin{array}[t]{c|c}
\multicolumn{2}{c}{\boldsymbol{(\overline L R ) X+\hc}} \\
\hline
\O_{e \gamma} & \bar e_{Lp}   \sigma_{\bar\mu\bar\nu} e_{Rr}\, F_{\bar\mu\bar\nu}  \\
\O_{u \gamma} & \bar u_{Lp}   \sigma_{\bar\mu\bar\nu}  u_{Rr}\, F_{\bar\mu\bar\nu}   \\
\O_{d \gamma} & \bar d_{Lp}  \sigma_{\bar\mu\bar\nu} d_{Rr}\, F_{\bar\mu\bar\nu}  \\
\O_{u G} & i \bar u_{Lp}   \sigma_{\bar\mu\bar\nu}  T^A u_{Rr}\,  G_{\bar\mu\bar\nu}^A  \\
\O_{d G} & i \bar d_{Lp}   \sigma_{\bar\mu\bar\nu} T^A d_{Rr}\,  G_{\bar\mu\bar\nu}^A \\
\end{array}
\end{align*}
\end{minipage}
\begin{minipage}[t]{3cm}
\renewcommand{\arraystretch}{1.51}
\small
\begin{align*}
\begin{array}[t]{c|c}
\multicolumn{2}{c}{\boldsymbol{X^3}} \\
\hline
\O_G     & f^{ABC} G_{\bar\mu\bar\nu}^{A} G_{\bar\nu\bar\rho}^{B} G_{\bar\rho\bar\mu}^{C}  \\
\O_{\widetilde G} & i f^{ABC} \widetilde G_{\bar\mu\bar\nu}^{A} G_{\bar\nu\bar\rho}^{B} G_{\bar\rho\bar\mu}^{C}   \\
\end{array}
\end{align*}
\end{minipage}
\end{adjustbox}
%

%
%%
% --- START TABLE
%%
\mbox{}\\[-1cm]

\begin{adjustbox}{width=1.05\textwidth,center}
%%%
% --- END TABLE
%%%
\begin{minipage}[t]{3cm}
\renewcommand{\arraystretch}{1.51}
\small
\begin{align*}
\begin{array}[t]{c|c}
\multicolumn{2}{c}{\boldsymbol{(\overline L L)(\overline L  L)}} \\
\hline
\op{\nu\nu}{V}{LL} & (\bar \nu_{Lp}  \gamma_{\bar\mu} \nu_{Lr} )(\bar \nu_{Ls} \gamma_{\bar\mu} \nu_{Lt})   \\
\op{ee}{V}{LL}       & (\bar e_{Lp}  \gamma_{\bar\mu} e_{Lr})(\bar e_{Ls} \gamma_{\bar\mu} e_{Lt})   \\
\op{\nu e}{V}{LL}       & (\bar \nu_{Lp} \gamma_{\bar\mu} \nu_{Lr})(\bar e_{Ls}  \gamma_{\bar\mu} e_{Lt})  \\
\op{\nu u}{V}{LL}       & (\bar \nu_{Lp} \gamma_{\bar\mu} \nu_{Lr}) (\bar u_{Ls}  \gamma_{\bar\mu} u_{Lt})  \\
\op{\nu d}{V}{LL}       & (\bar \nu_{Lp} \gamma_{\bar\mu} \nu_{Lr})(\bar d_{Ls} \gamma_{\bar\mu} d_{Lt})     \\
\op{eu}{V}{LL}      & (\bar e_{Lp}  \gamma_{\bar\mu} e_{Lr})(\bar u_{Ls} \gamma_{\bar\mu} u_{Lt})   \\
\op{ed}{V}{LL}       & (\bar e_{Lp}  \gamma_{\bar\mu} e_{Lr})(\bar d_{Ls} \gamma_{\bar\mu} d_{Lt})  \\
\op{\nu edu}{V}{LL}      & (\bar \nu_{Lp} \gamma_{\bar\mu} e_{Lr}) (\bar d_{Ls} \gamma_{\bar\mu} u_{Lt})  + \hc   \\
\op{uu}{V}{LL}        & (\bar u_{Lp} \gamma_{\bar\mu} u_{Lr})(\bar u_{Ls} \gamma_{\bar\mu} u_{Lt})    \\
\op{dd}{V}{LL}   & (\bar d_{Lp} \gamma_{\bar\mu} d_{Lr})(\bar d_{Ls} \gamma_{\bar\mu} d_{Lt})    \\
\op{ud}{V1}{LL}     & (\bar u_{Lp} \gamma_{\bar\mu} u_{Lr}) (\bar d_{Ls} \gamma_{\bar\mu} d_{Lt})  \\
\op{ud}{V8}{LL}     & (\bar u_{Lp} \gamma_{\bar\mu} T^A u_{Lr}) (\bar d_{Ls} \gamma_{\bar\mu} T^A d_{Lt})   \\[-0.5cm]
\end{array}
\end{align*}
%\end{minipage}
%\begin{minipage}[t]{3cm}
\renewcommand{\arraystretch}{1.51}
\small
\begin{align*}
\begin{array}[t]{c|c}
\multicolumn{2}{c}{\boldsymbol{(\overline R  R)(\overline R R)}} \\
\hline
\op{ee}{V}{RR}     & (\bar e_{Rp} \gamma_{\bar\mu} e_{Rr})(\bar e_{Rs} \gamma_{\bar\mu} e_{Rt})  \\
\op{eu}{V}{RR}       & (\bar e_{Rp}  \gamma_{\bar\mu} e_{Rr})(\bar u_{Rs} \gamma_{\bar\mu} u_{Rt})   \\
\op{ed}{V}{RR}     & (\bar e_{Rp} \gamma_{\bar\mu} e_{Rr})  (\bar d_{Rs} \gamma_{\bar\mu} d_{Rt})   \\
\op{uu}{V}{RR}      & (\bar u_{Rp} \gamma_{\bar\mu} u_{Rr})(\bar u_{Rs} \gamma_{\bar\mu} u_{Rt})  \\
\op{dd}{V}{RR}      & (\bar d_{Rp} \gamma_{\bar\mu} d_{Rr})(\bar d_{Rs} \gamma_{\bar\mu} d_{Rt})    \\
\op{ud}{V1}{RR}       & (\bar u_{Rp} \gamma_{\bar\mu} u_{Rr}) (\bar d_{Rs} \gamma_{\bar\mu} d_{Rt})  \\
\op{ud}{V8}{RR}    & (\bar u_{Rp} \gamma_{\bar\mu} T^A u_{Rr}) (\bar d_{Rs} \gamma_{\bar\mu} T^A d_{Rt})  \\
\end{array}
\end{align*}
\end{minipage}
%
%\hspace{-1.5cm}
%
\begin{minipage}[t]{3cm}
\renewcommand{\arraystretch}{1.51}
\small
\begin{align*}
\begin{array}[t]{c|c}
\multicolumn{2}{c}{\boldsymbol{(\overline L  L)(\overline R  R)}} \\
\hline
\op{\nu e}{V}{LR}     & (\bar \nu_{Lp} \gamma_{\bar\mu} \nu_{Lr})(\bar e_{Rs}  \gamma_{\bar\mu} e_{Rt})  \\
\op{ee}{V}{LR}       & (\bar e_{Lp}  \gamma_{\bar\mu} e_{Lr})(\bar e_{Rs} \gamma_{\bar\mu} e_{Rt}) \\
\op{\nu u}{V}{LR}         & (\bar \nu_{Lp} \gamma_{\bar\mu} \nu_{Lr})(\bar u_{Rs}  \gamma_{\bar\mu} u_{Rt})    \\
\op{\nu d}{V}{LR}         & (\bar \nu_{Lp} \gamma_{\bar\mu} \nu_{Lr})(\bar d_{Rs} \gamma_{\bar\mu} d_{Rt})   \\
\op{eu}{V}{LR}        & (\bar e_{Lp}  \gamma_{\bar\mu} e_{Lr})(\bar u_{Rs} \gamma_{\bar\mu} u_{Rt})   \\
\op{ed}{V}{LR}        & (\bar e_{Lp}  \gamma_{\bar\mu} e_{Lr})(\bar d_{Rs} \gamma_{\bar\mu} d_{Rt})   \\
\op{ue}{V}{LR}        & (\bar u_{Lp} \gamma_{\bar\mu} u_{Lr})(\bar e_{Rs}  \gamma_{\bar\mu} e_{Rt})   \\
\op{de}{V}{LR}         & (\bar d_{Lp} \gamma_{\bar\mu} d_{Lr}) (\bar e_{Rs} \gamma_{\bar\mu} e_{Rt})   \\
\op{\nu edu}{V}{LR}        & (\bar \nu_{Lp} \gamma_{\bar\mu} e_{Lr})(\bar d_{Rs} \gamma_{\bar\mu} u_{Rt})  +\hc \\
\op{uu}{V1}{LR}        & (\bar u_{Lp} \gamma_{\bar\mu} u_{Lr})(\bar u_{Rs} \gamma_{\bar\mu} u_{Rt})   \\
\op{uu}{V8}{LR}       & (\bar u_{Lp} \gamma_{\bar\mu} T^A u_{Lr})(\bar u_{Rs} \gamma_{\bar\mu} T^A u_{Rt})    \\ 
\op{ud}{V1}{LR}       & (\bar u_{Lp} \gamma_{\bar\mu} u_{Lr}) (\bar d_{Rs} \gamma_{\bar\mu} d_{Rt})  \\
\op{ud}{V8}{LR}       & (\bar u_{Lp} \gamma_{\bar\mu} T^A u_{Lr})  (\bar d_{Rs} \gamma_{\bar\mu} T^A d_{Rt})  \\
\op{du}{V1}{LR}       & (\bar d_{Lp} \gamma_{\bar\mu} d_{Lr})(\bar u_{Rs} \gamma_{\bar\mu} u_{Rt})   \\
\op{du}{V8}{LR}       & (\bar d_{Lp} \gamma_{\bar\mu} T^A d_{Lr})(\bar u_{Rs} \gamma_{\bar\mu} T^A u_{Rt}) \\
\op{dd}{V1}{LR}      & (\bar d_{Lp} \gamma_{\bar\mu} d_{Lr})(\bar d_{Rs} \gamma_{\bar\mu} d_{Rt})  \\
\op{dd}{V8}{LR}   & (\bar d_{Lp} \gamma_{\bar\mu} T^A d_{Lr})(\bar d_{Rs} \gamma_{\bar\mu} T^A d_{Rt}) \\
\op{uddu}{V1}{LR}   & (\bar u_{Lp} \gamma_{\bar\mu} d_{Lr})(\bar d_{Rs} \gamma_{\bar\mu} u_{Rt})  + \hc  \\
\op{uddu}{V8}{LR}      & (\bar u_{Lp} \gamma_{\bar\mu} T^A d_{Lr})(\bar d_{Rs} \gamma_{\bar\mu} T^A  u_{Rt})  + \hc \\
\end{array}
\end{align*}
\end{minipage}

\begin{minipage}[t]{3cm}
\renewcommand{\arraystretch}{1.51}
\small
\begin{align*}
\begin{array}[t]{c|c}
\multicolumn{2}{c}{\boldsymbol{(\overline L R)(\overline L R)+\hc}} \\
\hline
\op{ee}{S}{RR} 		& (\bar e_{Lp}   e_{Rr}) (\bar e_{Ls} e_{Rt})   \\
\op{eu}{S}{RR}  & (\bar e_{Lp}   e_{Rr}) (\bar u_{Ls} u_{Rt})   \\
\op{eu}{T}{RR} & (\bar e_{Lp}   \sigma_{\bar\mu\bar\nu}   e_{Rr}) (\bar u_{Ls}  \sigma_{\bar\mu\bar\nu}  u_{Rt})  \\
\op{ed}{S}{RR}  & (\bar e_{Lp} e_{Rr})(\bar d_{Ls} d_{Rt})  \\
\op{ed}{T}{RR} & (\bar e_{Lp} \sigma_{\bar\mu\bar\nu} e_{Rr}) (\bar d_{Ls} \sigma_{\bar\mu\bar\nu} d_{Rt})   \\
\op{\nu edu}{S}{RR} & (\bar   \nu_{Lp} e_{Rr})  (\bar d_{Ls} u_{Rt} ) \\
\op{\nu edu}{T}{RR} &  (\bar  \nu_{Lp}  \sigma_{\bar\mu\bar\nu} e_{Rr} )  (\bar  d_{Ls}  \sigma_{\bar\mu\bar\nu} u_{Rt} )   \\
\op{uu}{S1}{RR}  & (\bar u_{Lp}   u_{Rr}) (\bar u_{Ls} u_{Rt})  \\
\op{uu}{S8}{RR}   & (\bar u_{Lp}   T^A u_{Rr}) (\bar u_{Ls} T^A u_{Rt})  \\
\op{ud}{S1}{RR}   & (\bar u_{Lp} u_{Rr})  (\bar d_{Ls} d_{Rt})   \\
\op{ud}{S8}{RR}  & (\bar u_{Lp} T^A u_{Rr})  (\bar d_{Ls} T^A d_{Rt})  \\
\op{dd}{S1}{RR}   & (\bar d_{Lp} d_{Rr}) (\bar d_{Ls} d_{Rt}) \\
\op{dd}{S8}{RR}  & (\bar d_{Lp} T^A d_{Rr}) (\bar d_{Ls} T^A d_{Rt})  \\
\op{uddu}{S1}{RR} &  (\bar u_{Lp} d_{Rr}) (\bar d_{Ls}  u_{Rt})   \\
\op{uddu}{S8}{RR}  &  (\bar u_{Lp} T^A d_{Rr}) (\bar d_{Ls}  T^A u_{Rt})  \\[-0.5cm]
\end{array}
\end{align*}
%\end{minipage}
%\hspace{2cm}
%\begin{minipage}[t]{3cm}
\renewcommand{\arraystretch}{1.51}
\small
\begin{align*}
\begin{array}[t]{c|c}
\multicolumn{2}{c}{\boldsymbol{(\overline L R)(\overline R L) +\hc}} \\
\hline
\op{eu}{S}{RL}  & (\bar e_{Lp} e_{Rr}) (\bar u_{Rs}  u_{Lt})  \\
\op{ed}{S}{RL} & (\bar e_{Lp} e_{Rr}) (\bar d_{Rs} d_{Lt}) \\
\op{\nu edu}{S}{RL}  & (\bar \nu_{Lp} e_{Rr}) (\bar d_{Rs}  u_{Lt})  \\
\end{array}
\end{align*}
\end{minipage}
\end{adjustbox}
%%%
% --- END TABLE
%%%
\setlength{\abovecaptionskip}{0.75cm}
\setlength{\belowcaptionskip}{-2cm}
\caption{LEFT operators of dimension five and six that conserve baryon and lepton number, reproduced from Ref.~\cite{Jenkins:2017jig}. For each operator with ${}+\hc$, there is an additional operator in the basis, which in Minkowski space-time is related by Hermitian conjugation.}
\label{tab:oplist1}
\end{table}

\clearpage

%%
% --- START TABLE
%%%
\begin{table}[H]
\capstart
\centering
%\vspace{-0.75cm}
%%
% --- START TABLE
%%
\begin{minipage}[t]{3cm}
\renewcommand{\arraystretch}{1.51}
\small
\begin{align*}
\begin{array}[t]{c|c}
\multicolumn{2}{c}{\boldsymbol{(\overline L R)D^2 + \hc}} \\
\hline
\O_{eD}^{(5)} & \bar e_{Lp} \bar{\slashed D}^2 e_{Rr} \\
\O_{uD}^{(5)} & \bar u_{Lp} \bar{\slashed D}^2 u_{Rr} \\
\O_{dD}^{(5)} & \bar d_{Lp} \bar{\slashed D}^2 d_{Rr} \\
\end{array}
\end{align*}
\end{minipage}
%%%
% --- END TABLE
%%%
%%
% --- START TABLE
%%
\begin{minipage}[t]{3cm}
\renewcommand{\arraystretch}{1.51}
\small
\begin{align*}
\begin{array}[t]{c|c}
\multicolumn{2}{c}{\boldsymbol{X^2D^2}} \\
\hline
\O_{\gamma D} & (\p_{\bar\mu} F_{\bar\mu\bar\nu})(\p_{\bar\lambda} F_{\bar\lambda\bar\nu}) \\
\O_{GD} & (D_{\bar\mu} G_{\bar\mu\bar\nu})^A (D_{\bar\lambda} G_{\bar\lambda\bar\nu})^A \\
\end{array}
\end{align*}
\end{minipage}
%%%
% --- END TABLE
%%%
\\
\begin{adjustbox}{width=\textwidth,center}
%%
% --- START TABLE
%%
\begin{minipage}[t]{3cm}
\renewcommand{\arraystretch}{1.51}
\small
\begin{align*}
\begin{array}[t]{c|c}
\multicolumn{2}{c}{\boldsymbol{(\overline L L)D^3}} \\
\hline
\O_{\nu D}^{L} & \bar \nu_{Lp} \bar{\slashed \p}^3 \nu_{Lr} \\
\O_{eD}^{L} & \bar e_{Lp} \bar{\slashed D}^3 e_{Lr} \\
\O_{uD}^{L} & \bar u_{Lp} \bar{\slashed D}^3 u_{Lr} \\
\O_{dD}^{L} & \bar d_{Lp} \bar{\slashed D}^3 d_{Lr} \\
\end{array}
\end{align*}
\end{minipage}
%%%
% --- END TABLE
%%%
%%
% --- START TABLE
%%
\begin{minipage}[t]{3cm}
\renewcommand{\arraystretch}{1.51}
\small
\begin{align*}
\begin{array}[t]{c|c}
\multicolumn{2}{c}{\boldsymbol{(\overline R R)D^3}} \\
\hline
\O_{eD}^{R} & \bar e_{Rp} \bar{\slashed D}^3 e_{Rr} \\
\O_{uD}^{R} & \bar u_{Rp} \bar{\slashed D}^3 u_{Rr} \\
\O_{dD}^{R} & \bar d_{Rp} \bar{\slashed D}^3 d_{Rr} \\
\end{array}
\end{align*}
\end{minipage}
%%%
% --- END TABLE
%%%
%%
% --- START TABLE
%%
\begin{minipage}[t]{3cm}
\renewcommand{\arraystretch}{1.51}
\small
\begin{align*}
\begin{array}[t]{c|c}
\multicolumn{2}{c}{\boldsymbol{(\overline L L)XD}} \\
\hline
\O_{D\nu\gamma}^{L} & (\bar \nu_{Lp} \overleftarrow{\bar{\slashed \p}} \sigma_{\bar\mu\bar\nu} \nu_{Lr})  F_{\bar\mu\bar\nu}  \\
\O_{\nu D\gamma}^{L} & (\bar \nu_{Lp} \sigma_{\bar\mu\bar\nu} \bar{\slashed \p} \nu_{Lr})  F_{\bar\mu\bar\nu}  \\
\O_{\nu\gamma D}^{L} & i (\bar \nu_{Lp} \gamma_{\bar\nu} \nu_{Lr})  (\p_{\bar\mu} F_{\bar\mu\bar\nu})  \\
\O_{De\gamma}^{L} & (\bar e_{Lp} \overleftarrow{\bar{\slashed D}} \sigma_{\bar\mu\bar\nu} e_{Lr})  F_{\bar\mu\bar\nu}  \\
\O_{eD\gamma}^{L} & (\bar e_{Lp} \sigma_{\bar\mu\bar\nu} \bar{\slashed D} e_{Lr})  F_{\bar\mu\bar\nu}  \\
\O_{e\gamma D}^{L} & i (\bar e_{Lp} \gamma_{\bar\nu} e_{Lr})  (\p_{\bar\mu} F_{\bar\mu\bar\nu})  \\
\O_{Du\gamma}^{L} & (\bar u_{Lp} \overleftarrow{\bar{\slashed D}} \sigma_{\bar\mu\bar\nu} u_{Lr})  F_{\bar\mu\bar\nu}  \\
\O_{uD\gamma}^{L} & (\bar u_{Lp} \sigma_{\bar\mu\bar\nu} \bar{\slashed D} u_{Lr})  F_{\bar\mu\bar\nu}  \\
\O_{u\gamma D}^{L} & i (\bar u_{Lp} \gamma_{\bar\nu} u_{Lr})  (\p_{\bar\mu} F_{\bar\mu\bar\nu})  \\
\O_{Dd\gamma}^{L} & (\bar d_{Lp} \overleftarrow{\bar{\slashed D}} \sigma_{\bar\mu\bar\nu} d_{Lr})  F_{\bar\mu\bar\nu}  \\
\O_{dD\gamma}^{L} & (\bar d_{Lp} \sigma_{\bar\mu\bar\nu} \bar{\slashed D} d_{Lr})  F_{\bar\mu\bar\nu}  \\
\O_{d\gamma D}^{L} & i (\bar d_{Lp} \gamma_{\bar\nu} d_{Lr})  (\p_{\bar\mu} F_{\bar\mu\bar\nu})  \\
\O_{DuG}^{L} & i (\bar u_{Lp} \overleftarrow{\bar{\slashed D}} \sigma_{\bar\mu\bar\nu} T^A u_{Lr})  G^A_{\bar\mu\bar\nu}  \\
\O_{uDG}^{L} & i (\bar u_{Lp} \sigma_{\bar\mu\bar\nu} T^A \bar{\slashed D} u_{Lr})  G^A_{\bar\mu\bar\nu}  \\
\O_{uGD}^{L} & (\bar u_{Lp} \gamma_{\bar\nu} T^A u_{Lr})  (D_{\bar\mu} G_{\bar\mu\bar\nu})^A  \\
\O_{DdG}^{L} & i (\bar d_{Lp} \overleftarrow{\bar{\slashed D}} \sigma_{\bar\mu\bar\nu} T^A d_{Lr})  G^A_{\bar\mu\bar\nu}  \\
\O_{dDG}^{L} & i (\bar d_{Lp} \sigma_{\bar\mu\bar\nu} T^A \bar{\slashed D} d_{Lr})  G^A_{\bar\mu\bar\nu}  \\
\O_{dGD}^{L} & (\bar d_{Lp} \gamma_{\bar\nu} T^A d_{Lr})  (D_{\bar\mu} G_{\bar\mu\bar\nu})^A  \\
\end{array}
\end{align*}
\end{minipage}
%%%
% --- END TABLE
%%%
%%
% --- START TABLE
%%
\begin{minipage}[t]{3cm}
\renewcommand{\arraystretch}{1.51}
\small
\begin{align*}
\begin{array}[t]{c|c}
\multicolumn{2}{c}{\boldsymbol{(\overline R R)XD}} \\
\hline
\O_{De\gamma}^{R} & (\bar e_{Rp} \overleftarrow{\bar{\slashed D}} \sigma_{\bar\mu\bar\nu} e_{Rr})  F_{\bar\mu\bar\nu}  \\
\O_{eD\gamma}^{R} & (\bar e_{Rp} \sigma_{\bar\mu\bar\nu} \bar{\slashed D} e_{Rr})  F_{\bar\mu\bar\nu}  \\
\O_{e\gamma D}^{R} & i (\bar e_{Rp} \gamma_{\bar\nu} e_{Rr})  (\p_{\bar\mu} F_{\bar\mu\bar\nu})  \\
\O_{Du\gamma}^{R} & (\bar u_{Rp} \overleftarrow{\bar{\slashed D}} \sigma_{\bar\mu\bar\nu} u_{Rr})  F_{\bar\mu\bar\nu}  \\
\O_{uD\gamma}^{R} & (\bar u_{Rp} \sigma_{\bar\mu\bar\nu} \bar{\slashed D} u_{Rr})  F_{\bar\mu\bar\nu}  \\
\O_{u\gamma D}^{R} & i (\bar u_{Rp} \gamma_{\bar\nu} u_{Rr})  (\p_{\bar\mu} F_{\bar\mu\bar\nu})  \\
\O_{Dd\gamma}^{R} & (\bar d_{Rp} \overleftarrow{\bar{\slashed D}} \sigma_{\bar\mu\bar\nu} d_{Rr})  F_{\bar\mu\bar\nu}  \\
\O_{dD\gamma}^{R} & (\bar d_{Rp} \sigma_{\bar\mu\bar\nu} \bar{\slashed D} d_{Rr})  F_{\bar\mu\bar\nu}  \\
\O_{d\gamma D}^{R} & i (\bar d_{Rp} \gamma_{\bar\nu} d_{Rr})  (\p_{\bar\mu} F_{\bar\mu\bar\nu})  \\
\O_{DuG}^{R} & i (\bar u_{Rp} \overleftarrow{\bar{\slashed D}} \sigma_{\bar\mu\bar\nu} T^A u_{Rr})  G^A_{\bar\mu\bar\nu}  \\
\O_{uDG}^{R} & i (\bar u_{Rp} \sigma_{\bar\mu\bar\nu} T^A \bar{\slashed D} u_{Rr})  G^A_{\bar\mu\bar\nu}  \\
\O_{uGD}^{R} & (\bar u_{Rp} \gamma_{\bar\nu} T^A u_{Rr})  (D_{\bar\mu} G_{\bar\mu\bar\nu})^A  \\
\O_{DdG}^{R} & i (\bar d_{Rp} \overleftarrow{\bar{\slashed D}} \sigma_{\bar\mu\bar\nu} T^A d_{Rr})  G^A_{\bar\mu\bar\nu}  \\
\O_{dDG}^{R} & i (\bar d_{Rp} \sigma_{\bar\mu\bar\nu} T^A \bar{\slashed D} d_{Rr})  G^A_{\bar\mu\bar\nu}  \\
\O_{dGD}^{R} & (\bar d_{Rp} \gamma_{\bar\nu} T^A d_{Rr})  (D_{\bar\mu} G_{\bar\mu\bar\nu})^A  \\
\end{array}
\end{align*}
\end{minipage}
%%%
% --- END TABLE
%%%
\end{adjustbox}

\caption{Redundant LEFT operators of dimension five and six that conserve lepton and baryon number, reproduced from Ref.~\cite{Naterop:2023dek}. These operators can be removed by field redefinitions but are required to renormalize off-shell Green's functions.}
\label{tab:EOMRedundantOperators}
\end{table}

\clearpage

\subsection{Translation between Euclidean and Minkowskian conventions}
\label{sec:TranslationCoefficients}

Due to different conventions for space-time signature, $SU(3)$ algebra, and normalization of fields compared to Ref.~\cite{Jenkins:2017jig,Naterop:2023dek}, some of our Wilson coefficients are rescaled compared to the ones in the HV scheme of Ref.~\cite{Naterop:2023dek} by powers of the couplings and overall signs. For the coefficients of the on-shell basis, we find
\begin{align}
	\label{eq:translationLEFTcoeffs}
	L_{\psi\gamma} &= - \frac{1}{e} L_{\psi\gamma}^\text{\cite{Naterop:2023dek}} \, , \quad
	L_{\psi G} = - \frac{1}{g} L_{\psi G}^\text{\cite{Naterop:2023dek}} \, , \quad
	L_{G} = - \frac{1}{g^3} L_{G}^\text{\cite{Naterop:2023dek}} \, , \quad
	L_{\widetilde G} = \frac{1}{g^3} L_{\widetilde G}^\text{\cite{Naterop:2023dek}} \, , \nn
	L^{V,LL} &= - L^{V,LL}_{\text{\cite{Naterop:2023dek}}} \, , \quad
	L^{V,RR} = - L^{V,RR}_{\text{\cite{Naterop:2023dek}}} \, , \quad
	L^{V,LR} = - L^{V,LR}_{\text{\cite{Naterop:2023dek}}} \, , \nn
	L^{V1,LL} &= - L^{V1,LL}_{\text{\cite{Naterop:2023dek}}} \, , \quad
	L^{V1,RR} = - L^{V1,RR}_{\text{\cite{Naterop:2023dek}}} \, , \quad
	L^{V1,LR} = - L^{V1,LR}_{\text{\cite{Naterop:2023dek}}} \, , \nn
	L^{V8,LL} &= L^{V8,LL}_{\text{\cite{Naterop:2023dek}}} \, , \quad
	L^{V8,RR} = L^{V8,RR}_{\text{\cite{Naterop:2023dek}}} \, , \quad
	L^{V8,LR} = L^{V8,LR}_{\text{\cite{Naterop:2023dek}}} \, , \nn
	L^{S,RR} &= -L^{S,RR}_{\text{\cite{Naterop:2023dek}}} \, , \quad
	L^{S,RL} = -L^{S,RL}_{\text{\cite{Naterop:2023dek}}} \, , \quad
	L^{S1,RR} = -L^{S1,RR}_{\text{\cite{Naterop:2023dek}}} \, , \nn
	L^{S8,RR} &= L^{S8,RR}_{\text{\cite{Naterop:2023dek}}} \, , \quad
	L^{T,RR} = -L^{T,RR}_{\text{\cite{Naterop:2023dek}}} \, ,
\end{align}
whereas the relations for the coefficients of redundant operators read
\begin{align}
	L_{\psi D}^{(5)} &= -L_{\psi D}^{(5)\,\text{\cite{Naterop:2023dek}}} \, , \quad
%%%%
	L_{\gamma D} = \frac{1}{e^2} L_{\gamma D}^\text{\cite{Naterop:2023dek}} \, , \quad
	L_{GD} = \frac{1}{g^2} L_{GD}^\text{\cite{Naterop:2023dek}} \, , \nn
%%%%
	L_{\psi D}^L &= L_{\psi D}^{L\,\text{\cite{Naterop:2023dek}}} \, , \quad
	L_{\psi D}^R = L_{\psi D}^{R\,\text{\cite{Naterop:2023dek}}} \, , \nn
%%%%
	L_{D\psi\gamma}^L &= \frac{1}{e} L_{D\psi\gamma}^{L\,\text{\cite{Naterop:2023dek}}} \, , \quad
	L_{\psi D\gamma}^L = \frac{1}{e} L_{\psi D\gamma}^{L\,\text{\cite{Naterop:2023dek}}} \, , \quad
	L_{\psi\gamma D}^L = - \frac{1}{e} L_{\psi\gamma D}^{L\,\text{\cite{Naterop:2023dek}}} \, , \nn
	L_{D\psi G}^L &= \frac{1}{g} L_{D\psi G}^{L\,\text{\cite{Naterop:2023dek}}} \, , \quad
	L_{\psi DG}^L = \frac{1}{g} L_{\psi DG}^{L\,\text{\cite{Naterop:2023dek}}} \, , \quad
	L_{\psi GD}^L = \frac{1}{g} L_{\psi GD}^{L\,\text{\cite{Naterop:2023dek}}} \, , \nn
%%%%
	L_{D\psi\gamma}^R &= \frac{1}{e} L_{D\psi\gamma}^{R\,\text{\cite{Naterop:2023dek}}} \, , \quad
	L_{\psi D\gamma}^R = \frac{1}{e} L_{\psi D\gamma}^{R\,\text{\cite{Naterop:2023dek}}} \, , \quad
	L_{\psi\gamma D}^R = - \frac{1}{e} L_{\psi\gamma D}^{R\,\text{\cite{Naterop:2023dek}}} \, , \nn
	L_{D\psi G}^R &= \frac{1}{g} L_{D\psi G}^{R\,\text{\cite{Naterop:2023dek}}} \, , \quad
	L_{\psi DG}^R = \frac{1}{g} L_{\psi DG}^{R\,\text{\cite{Naterop:2023dek}}} \, , \quad
	L_{\psi GD}^R = \frac{1}{g} L_{\psi GD}^{R\,\text{\cite{Naterop:2023dek}}} \, .
\end{align}

	% !TEX root = ../Paper-LEFT-GF.tex

\section{Feynman rules}
\label{sec:FeynmanRules}

In the following, all momenta are understood as outgoing. Ordinary external lines denote quantum fields, whereas lines ending in a crossed dot denote background fields. Lines ending in a colored crossed dot stand for either quantum or background fields.

The QCD and QED interaction vertices in the background-field gauge are~\cite{Abbott:1980hw}
\begin{align}
	\nn[-0.2cm]
	\quad\qquad\begin{gathered}
		\scalebox{0.8}{\begin{fmfgraph*}(60,50)
			\fmftop{t1} \fmfbottom{b1,b2}
			\fmf{quark,tension=2}{b1,v1}
			\fmf{quark,tension=2}{v1,b2}
			\fmf{gluon,tension=3}{t1,v1}
			\fmflabel{$\mu,A$}{t1}
			\fmfv{decor.shape=otimes, decor.filled=empty, decor.size=(3mm), foreground=(.9,,.4,,.4)}{t1,b1,b2}
		\end{fmfgraph*}}
	\end{gathered}
	\qquad &= - \gamma_\mu t^A \, , \qquad\qquad
	\begin{gathered}
		\scalebox{0.8}{\begin{fmfgraph*}(60,50)
			\fmftop{t1} \fmfbottom{b1,b2}
			\fmf{quark,tension=2}{b1,v1}
			\fmf{quark,tension=2}{v1,b2}
			\fmf{photon,tension=3}{t1,v1}
			\fmflabel{$\mu$}{t1}
			\fmfv{decor.shape=otimes, decor.filled=empty, decor.size=(3mm), foreground=(.9,,.4,,.4)}{t1,b1,b2}
		\end{fmfgraph*}}
	\end{gathered}
	\qquad = i \gamma_\mu \q_\psi \, , \nn[0.5cm]
	\quad\qquad\begin{gathered}
		\scalebox{0.8}{\begin{fmfgraph*}(60,50)
			\fmftop{t1} \fmfbottom{b1,b2}
			\fmf{ghost,tension=2}{b1,v1}
			\fmf{ghost,label.side=left,tension=2}{v1,b2}
			\fmf{gluon,tension=3}{t1,v1}
			\fmflabel{$\mu,A$}{t1}
			\fmflabel{$B$}{b1}
			\fmflabel{$p,C$}{b2}
		\end{fmfgraph*}}
	\end{gathered}
	\qquad &= i p_\mu f^{ABC} \, , \qquad\qquad
	\begin{gathered}
		\scalebox{0.8}{\begin{fmfgraph*}(60,50)
			\fmftop{t1} \fmfbottom{b1,b2}
			\fmf{ghost,tension=2}{b1,v1}
			\fmf{ghost,label.side=left,tension=2}{v1,b2}
			\fmf{gluon,tension=3}{t1,v1}
			\fmflabel{$\mu,A$}{t1}
			\fmflabel{$p_1,B$}{b1}
			\fmflabel{$p_2,C$}{b2}
			\fmfv{decor.shape=otimes, decor.filled=empty, decor.size=(3mm)}{t1}
		\end{fmfgraph*}}
	\end{gathered}
	\qquad = i ({p_2}_\mu-{p_1}_\mu) f^{ABC} \, , \nn[1cm]
%%%%%%%%%%%%%%%%%%
	\quad\qquad\begin{gathered}
		\scalebox{0.8}{\begin{fmfgraph*}(50,50)
			\fmftop{t1,t2} \fmfbottom{b1,b2}
			\fmf{ghost,tension=2}{b1,v1}
			\fmf{ghost,label.side=left,tension=2}{v1,b2}
			\fmf{gluon,tension=3}{t1,v1,t2}
			\fmflabel{$\mu,A$}{t1}
			\fmflabel{$\nu,B$}{t2}
			\fmflabel{$C$}{b1}
			\fmflabel{$D$}{b2}
			\fmfv{decor.shape=otimes, decor.filled=empty, decor.size=(3mm)}{t2}
		\end{fmfgraph*}}
	\end{gathered}
	\qquad &= - \delta_{\mu\nu} f^{ADE} f^{BCE} \, , \nn[1cm]
	\quad\qquad\begin{gathered}
		\scalebox{0.8}{\begin{fmfgraph*}(50,50)
			\fmftop{t1,t2} \fmfbottom{b1,b2}
			\fmf{ghost,tension=2}{b1,v1}
			\fmf{ghost,label.side=left,tension=2}{v1,b2}
			\fmf{gluon,tension=3}{t1,v1,t2}
			\fmflabel{$\mu,A$}{t1}
			\fmflabel{$\nu,B$}{t2}
			\fmflabel{$C$}{b1}
			\fmflabel{$D$}{b2}
			\fmfv{decor.shape=otimes, decor.filled=empty, decor.size=(3mm)}{t1,t2}
		\end{fmfgraph*}}
	\end{gathered}
	\qquad &= - \delta_{\mu\nu} \left( f^{ADE} f^{BCE} + f^{ACE} f^{BDE} \right) \, , \nn[1cm]
%%%%%%%%%%%%%%%%%%
	\quad\qquad\begin{gathered}
		\scalebox{0.8}{\begin{fmfgraph*}(60,50)
			\fmftop{t1} \fmfbottom{b1,b2}
			\fmf{gluon,tension=3}{t1,v1}
			\fmf{gluon,tension=2}{b1,v1}
			\fmf{gluon,tension=2}{v1,b2}
			\fmflabel{$p_1,\mu,A$}{t1}
			\fmflabel{$p_2,\nu,B$}{b1}
			\fmflabel{$p_3,\rho,C$}{b2}
			\fmfv{decor.shape=otimes, decor.filled=empty, decor.size=(3mm), foreground=(.9,,.4,,.4)}{t1}
			\fmfv{decor.shape=otimes, decor.filled=empty, decor.size=(3mm)}{b1,b2}
		\end{fmfgraph*}}
	\end{gathered}
	\qquad &= 
	\qquad\qquad\begin{gathered}
		\scalebox{0.8}{\begin{fmfgraph*}(60,50)
			\fmftop{t1} \fmfbottom{b1,b2}
			\fmf{gluon,tension=3}{t1,v1}
			\fmf{gluon,tension=2}{b1,v1}
			\fmf{gluon,tension=2}{v1,b2}
			\fmflabel{$p_1,\mu,A$}{t1}
			\fmflabel{$p_2,\nu,B$}{b1}
			\fmflabel{$p_3,\rho,C$}{b2}
		\end{fmfgraph*}}
	\end{gathered} \nn[0.5cm]
	&=  -\frac{i f^{ABC}}{g_0^2} \Bigl[ \delta_{\mu\nu} ( p_1 - p_2)_\rho + \delta_{\nu\rho} ( p_2 - p_3 )_\mu + \delta_{\mu\rho} ( p_3 - p_1 )_\nu \Bigr] \, , \nn[0.75cm]
	\quad\qquad\begin{gathered}
		\scalebox{0.8}{\begin{fmfgraph*}(60,50)
			\fmftop{t1} \fmfbottom{b1,b2}
			\fmf{gluon,tension=3}{t1,v1}
			\fmf{gluon,tension=2}{b1,v1}
			\fmf{gluon,tension=2}{v1,b2}
			\fmflabel{$p_1,\mu,A$}{t1}
			\fmflabel{$p_2,\nu,B$}{b1}
			\fmflabel{$p_3,\rho,C$}{b2}
			\fmfv{decor.shape=otimes, decor.filled=empty, decor.size=(3mm)}{t1}
		\end{fmfgraph*}}
	\end{gathered}
	\qquad &=  -\frac{i f^{ABC}}{g_0^2} \Bigl[ \delta_{\mu\nu} \Bigl( p_1 - p_2 - \frac{p_3}{\xi_g}\Bigr)_\rho + \delta_{\nu\rho} ( p_2 - p_3 )_\mu + \delta_{\mu\rho} \Bigl( p_3 - p_1 + \frac{p_2}{\xi_g} \Bigr)_\nu \Bigr] \, , \nn[0.75cm]
%%%%%%%%%%%%%%%%%%
%             TO DO
%%%%%%%%%%%%%%%%%%
	\quad\qquad\begin{gathered}
		\scalebox{0.8}{\begin{fmfgraph*}(60,50)
			\fmftop{t1,t2} \fmfbottom{b1,b2}
			\fmf{gluon}{t1,v1}
			\fmf{gluon}{v1,t2}
			\fmf{gluon}{b1,v1}
			\fmf{gluon}{v1,b2}
			\fmflabel{$p_1,\mu,A$}{t1}
			\fmflabel{$p_2,\nu,B$}{t2}
			\fmflabel{$p_3,\rho,C$}{b1}
			\fmflabel{$p_4,\sigma,D$}{b2}
			\fmfv{decor.shape=otimes, decor.filled=empty, decor.size=(3mm)}{t2,b1,b2}
			\fmfv{decor.shape=otimes, decor.filled=empty, decor.size=(3mm), foreground=(.9,,.4,,.4)}{t1}
		\end{fmfgraph*}}
	\end{gathered}
	\qquad &= 
	\qquad\qquad\begin{gathered}
		\scalebox{0.8}{\begin{fmfgraph*}(60,50)
			\fmftop{t1,t2} \fmfbottom{b1,b2}
			\fmf{gluon}{t1,v1}
			\fmf{gluon}{v1,t2}
			\fmf{gluon}{b1,v1}
			\fmf{gluon}{v1,b2}
			\fmflabel{$p_1,\mu,A$}{t1}
			\fmflabel{$p_2,\nu,B$}{t2}
			\fmflabel{$p_3,\rho,C$}{b1}
			\fmflabel{$p_4,\sigma,D$}{b2}
			\fmfv{decor.shape=otimes, decor.filled=empty, decor.size=(3mm), foreground=(.9,,.4,,.4)}{t1}
		\end{fmfgraph*}}
	\end{gathered}
	\qquad = - \frac{1}{g_0^2} \begin{aligned}[t] &\Bigl[ f^{ADE} f^{BCE} \left(\delta_{\mu \nu } \delta_{\rho \sigma }-\delta_{\mu \rho } \delta_{\nu \sigma }\right) \nn
		& + f^{ACE} f^{BDE} \left(\delta_{\mu \nu } \delta_{\rho\sigma }-\delta_{\mu \sigma } \delta_{\nu \rho }\right) \nn
		& + f^{ABE} f^{CDE} \left(\delta_{\mu \rho } \delta_{\nu \sigma }-\delta_{\mu \sigma } \delta_{\nu \rho}\right) \Bigr]  \, , \end{aligned} \nn[1cm]
	\quad\qquad\begin{gathered}
		\scalebox{0.8}{\begin{fmfgraph*}(60,50)
			\fmftop{t1,t2} \fmfbottom{b1,b2}
			\fmf{gluon}{t1,v1}
			\fmf{gluon}{v1,t2}
			\fmf{gluon}{b1,v1}
			\fmf{gluon}{v1,b2}
			\fmflabel{$p_1,\mu,A$}{t1}
			\fmflabel{$p_2,\nu,B$}{t2}
			\fmflabel{$p_3,\rho,C$}{b1}
			\fmflabel{$p_4,\sigma,D$}{b2}
			\fmfv{decor.shape=otimes, decor.filled=empty, decor.size=(3mm)}{t1,t2}
		\end{fmfgraph*}}
	\end{gathered}
	\qquad &= - \frac{1}{g_0^2} \begin{aligned}[t] &\Bigl[ f^{ADE} f^{BCE} \Bigl(\delta_{\mu \nu } \delta_{\rho \sigma }-\delta_{\mu \rho } \delta_{\nu \sigma } + \frac{\delta_{\mu \sigma } \delta_{\nu \rho }}{\xi_g}\Bigr) \nn
		& + f^{ACE} f^{BDE} \Bigl(\delta_{\mu \nu } \delta_{\rho\sigma }-\delta_{\mu \sigma } \delta_{\nu \rho }  + \frac{\delta_{\mu \rho } \delta_{\nu \sigma }}{\xi_g} \Bigr) \nn
		& + f^{ABE} f^{CDE} \Bigl(\delta_{\mu \rho } \delta_{\nu \sigma }-\delta_{\mu \sigma } \delta_{\nu \rho}\Bigr) \Bigr]  \, . \end{aligned} \mytag
\end{align}
The flowed propagators of the quantum fields are given by
\begin{align}
	\begin{gathered}
		\scalebox{0.8}{\begin{fmfgraph*}(60,50)
			\fmfleft{l1} \fmfright{r1}
			\fmf{gluon,label=$p$,label.dist=-15}{l1,r1}
			\fmflabel{$s,\nu,B$}{l1}
			\fmflabel{$t,\mu,A$}{r1}
		\end{fmfgraph*}}
	\end{gathered} \qquad\quad\;
		&= g_0^2 \delta^{AB} \frac{1}{p^2} \left[ \left( \delta_{\mu\nu} - \frac{p_\mu p_\nu}{p^2} \right) e^{-(s+t)p^2} + \xi \frac{p_\mu p_\nu}{p^2} e^{-\alpha_0 (s+t) p^2} \right] \, , \nn
	\begin{gathered}
		\scalebox{0.8}{\begin{fmfgraph*}(60,50)
			\fmfleft{l1} \fmfright{r1}
			\fmf{quark,label=$p$,label.dist=-15}{l1,r1}
			\fmflabel{$s$}{l1}
			\fmflabel{$t$}{r1}
		\end{fmfgraph*}}
	\end{gathered} \quad\;
	&= \Bigl[ i \slashed p + M P_L + M^\dagger P_R \Bigr]^{-1} e^{-(s+t)p^2} \, .
\end{align}
Insertions of the kernels of the flow equations are represented as flow lines of both quantum and background fields:
\begin{align}
	\qquad\begin{gathered}
		\scalebox{0.8}{\begin{fmfgraph*}(60,50)
			\fmfleft{l1} \fmfright{r1}
			\fmf{gluon,label=$p$,label.dist=-15}{l1,r1}
			\fmf{marrowd,tension=0}{l1,r1}
			\fmflabel{$s,\nu,B$}{l1}
			\fmflabel{$t,\mu,A$}{r1}
		\end{fmfgraph*}}
	\end{gathered} \qquad\quad\;
	&= \delta^{AB} \theta(t-s) \frac{1}{p^2} \left[ (\delta_{\mu\nu}p^2 - p_\mu p_\nu) e^{-(t-s)p^2} + p_\mu p_\nu e^{-\alpha_0 (t-s) p^2} \right] \, , \nn
	\qquad\begin{gathered}
		\scalebox{0.8}{\begin{fmfgraph*}(60,50)
			\fmfleft{l1} \fmfright{r1}
			\fmf{gluon,label=$p$,label.dist=-15}{l1,r1}
			\fmf{marrowd,tension=0}{l1,r1}
			\fmflabel{$s,\nu,B$}{l1}
			\fmflabel{$t,\mu,A$}{r1}
			\fmfv{decor.shape=otimes, decor.filled=empty, decor.size=(3mm)}{l1,r1}
		\end{fmfgraph*}}
	\end{gathered} \qquad\quad\;
	&= \delta^{AB} \theta(t-s) \frac{1}{p^2} \left[ (\delta_{\mu\nu}p^2 - p_\mu p_\nu) e^{-(t-s)p^2} + p_\mu p_\nu \right] \, ,
\end{align}
where the adjacent arrow points into the direction of increasing flow time. The quark flow lines are
\begin{align}
	\begin{gathered}
		\scalebox{0.8}{\begin{fmfgraph*}(60,50)
			\fmfleft{l1} \fmfright{r1}
			\fmf{quark,label=$p$,label.dist=-15}{l1,r1}
			\fmf{marrowd,tension=0}{l1,r1}
			\fmflabel{$s$}{l1}
			\fmflabel{$t$}{r1}
		\end{fmfgraph*}}
	\end{gathered} \quad\;
	&= \quad\;
	\begin{gathered}
		\scalebox{0.8}{\begin{fmfgraph*}(60,50)
			\fmfleft{l1} \fmfright{r1}
			\fmf{quark,label=$p$,label.dist=-15}{l1,r1}
			\fmf{marrowd,tension=0}{l1,r1}
			\fmflabel{$s$}{l1}
			\fmflabel{$t$}{r1}
			\fmfv{decor.shape=otimes, decor.filled=empty, decor.size=(3mm)}{l1,r1}
		\end{fmfgraph*}}
	\end{gathered} \quad\;
	= \theta(t-s) e^{-(t-s)p^2} \, , \nn
	\begin{gathered}
		\scalebox{0.8}{\begin{fmfgraph*}(60,50)
			\fmfleft{l1} \fmfright{r1}
			\fmf{quark,label=$p$,label.dist=10}{r1,l1}
			\fmf{marrowd,tension=0}{l1,r1}
			\fmflabel{$s$}{l1}
			\fmflabel{$t$}{r1}
		\end{fmfgraph*}}
	\end{gathered} \quad\;
	&= \quad\;
	\begin{gathered}
		\scalebox{0.8}{\begin{fmfgraph*}(60,50)
			\fmfleft{l1} \fmfright{r1}
			\fmf{quark,label=$p$,label.dist=10}{r1,l1}
			\fmf{marrowd,tension=0}{l1,r1}
			\fmflabel{$s$}{l1}
			\fmflabel{$t$}{r1}
			\fmfv{decor.shape=otimes, decor.filled=empty, decor.size=(3mm)}{l1,r1}
		\end{fmfgraph*}}
	\end{gathered} \quad\;
	= \theta(t-s) e^{-(t-s)p^2} \, .
\end{align}
As we are considering the 1PI effective action, we do not require any explicit background-field propagators. However, since the amputation happens at vanishing flow time one has to keep track of the evolution to the flow time of the vertex to which the leg is attached: the rules for amputated external legs therefore correspond to flow lines from zero flow time to $t$. These external-leg factors are relevant in particular in the case of power divergences:
\begin{align}
	\begin{gathered}
		\scalebox{0.8}{\begin{fmfgraph*}(40,50)
			\fmfleft{l1} \fmfright{r1}
			\fmf{gluon,label=$\;\;p$,label.dist=-15}{l1,r1}
			\fmflabel{$\nu,B\quad\;{\bigg/}\hspace{-0.7cm}$}{l1}
			\fmflabel{\;\;$t,\mu,A$}{r1}
			\fmfv{decor.shape=otimes, decor.filled=empty, decor.size=(3mm)}{l1}
			\fmfv{decor.shape=circle, decor.filled=hatched, decor.size=(5mm)}{r1}
		\end{fmfgraph*}}
	\end{gathered} \qquad\quad\;
		&= \delta^{AB} \frac{1}{p^2} \left[ (\delta_{\mu\nu} p^2 - p_\mu p_\nu ) e^{-t p^2} + p_\mu p_\nu \right] \, , \nn[-0.1cm]
	\begin{gathered}
		\scalebox{0.8}{\begin{fmfgraph*}(40,50)
			\fmfleft{l1} \fmfright{r1}
			\fmf{quark,label=$p$,label.dist=-15}{l1,r1}
			\fmflabel{$\Big/\hspace{-0.65cm}$}{l1}
			\fmflabel{\;\;$t$}{r1}
			\fmfv{decor.shape=otimes, decor.filled=empty, decor.size=(3mm)}{l1}
			\fmfv{decor.shape=circle, decor.filled=hatched, decor.size=(5mm)}{r1}
		\end{fmfgraph*}}
	\end{gathered} \quad\quad
	&= \qquad \begin{gathered}
		\scalebox{0.8}{\begin{fmfgraph*}(40,50)
			\fmfleft{l1} \fmfright{r1}
			\fmf{quark,label=$p$,label.dist=-15}{l1,r1}
			\fmflabel{$t$\;\;}{l1}
			\fmflabel{$\hspace{-0.65cm}\Big/$}{r1}
			\fmfv{decor.shape=otimes, decor.filled=empty, decor.size=(3mm)}{r1}
			\fmfv{decor.shape=circle, decor.filled=hatched, decor.size=(5mm)}{l1}
		\end{fmfgraph*}}
	\end{gathered} \quad
	= e^{-t p^2} \, .
\end{align}

The flow vertices involving only quantum gauge fields correspond to the ones of conventional gauge fixing, whereas the flow vertices involving only background gauge fields correspond to the same expressions with $\alpha_0 = 0$:
\begin{align}
	\nn
	\begin{gathered}
		\scalebox{0.8}{\begin{fmfgraph*}(70,60)
			\fmfleft{l1,l2} \fmfright{r1}
			\fmf{gluon,label=$\hspace{-1.5cm}$,label.side=right,label.dist=35,tension=2}{l1,v1}
			\fmf{gluon,label=$\hspace{-1.5cm}\quad\quad$,label.dist=-50,label.side=left,tension=2}{v1,l2}
			\fmf{gluon,label=$t\qquad\quad$,label.dist=5,label.side=right,label.side=left,tension=2}{v1,r1}
			\fmfv{decor.shape=circle, decor.filled=empty, decor.size=(4mm)}{v1}
			\fmf{marrowd,tension=0}{v1,r1}
			\fmf{darrowl,tension=0}{l1,v1}
			\fmf{darrowl,tension=0}{l2,v1}
			\fmflabel{$p_1,\mu,A$}{r1}
			\fmflabel{$p_2,\nu,B$}{l2}
			\fmflabel{$p_3,\rho,C$}{l1}
		\end{fmfgraph*}}
	\end{gathered} \qquad\quad
	\; &=  - i f^{ABC} \int_0^\infty dt \, \begin{aligned}[t]
				& \Big[ \delta_{\nu\rho} ( p_2 - p_3)_\mu + 2 \delta_{\mu\rho} {p_3}_\nu - 2 \delta_{\mu\nu} {p_2}_\rho \\
				&+ (\alpha_0-1) (  \delta_{\mu\nu} {p_3}_\rho - \delta_{\mu\rho} {p_2}_\nu ) \Big] \, , \end{aligned} \nn[0.75cm]
	\begin{gathered}
		\scalebox{0.8}{\begin{fmfgraph*}(70,60)
			\fmfleft{l1,l2} \fmfright{r1}
			\fmf{gluon,label=$\hspace{-1.5cm}$,label.side=right,label.dist=35,tension=2}{l1,v1}
			\fmf{gluon,label=$\hspace{-1.5cm}\quad\quad$,label.dist=-50,label.side=left,tension=2}{v1,l2}
			\fmf{gluon,label=$t\qquad\quad$,label.dist=5,label.side=right,label.side=left,tension=2}{v1,r1}
			\fmfv{decor.shape=circle, decor.filled=empty, decor.size=(4mm)}{v1}
			\fmf{marrowd,tension=0}{v1,r1}
			\fmf{darrowl,tension=0}{l1,v1}
			\fmf{darrowl,tension=0}{l2,v1}
			\fmflabel{$\;p_1,\mu,A$}{r1}
			\fmflabel{$p_2,\nu,B$}{l2}
			\fmflabel{$p_3,\rho,C$}{l1}
			\fmfv{decor.shape=otimes, decor.filled=empty, decor.size=(3mm)}{r1,l1,l2}
		\end{fmfgraph*}}
	\end{gathered} \qquad\quad\;
	\; &=  - i f^{ABC} \int_0^\infty dt \, \begin{aligned}[t]
				& \Big[ \delta_{\mu\rho} (2 {p_3} + p_2 )_\nu - \delta_{\mu\nu} (2 {p_2} + p_3 )_\rho \\
				& + \delta_{\nu\rho} ( p_2 - p_3)_\mu \Big] \, , \end{aligned} \nn[0.75cm]
	\begin{gathered}
		\scalebox{0.8}{\begin{fmfgraph*}(80,70)
			\fmfleft{l1} \fmftop{t1} \fmfbottom{b1} \fmfright{r1}
			\fmf{gluon}{v1,b1}
			\fmf{gluon}{l1,v1}
			\fmf{gluon}{t1,v1}
			\fmf{gluon,label=$t\qquad\quad$,label.dist=5,label.side=right,label.side=left}{v1,r1}
			\fmfv{decor.shape=circle, decor.filled=empty, decor.size=(4mm)}{v1}
			\fmf{marrowd,tension=0}{v1,r1}
			\fmf{darrowd,tension=0}{l1,v1}
			\fmf{darrowl,tension=0}{t1,v1}
			\fmf{darrowl,tension=0}{b1,v1}
			\fmflabel{$p_1,\mu,A$}{r1}
			\fmflabel{$p_2,\nu,B$}{t1}
			\fmflabel{$p_3,\rho,C$}{l1}
			\fmflabel{$p_4,\sigma,D$}{b1}
		\end{fmfgraph*}}
	\end{gathered} \qquad\quad\;
	&= \qquad\quad\quad \begin{gathered}
		\scalebox{0.8}{\begin{fmfgraph*}(80,70)
			\fmfleft{l1} \fmftop{t1} \fmfbottom{b1} \fmfright{r1}
			\fmf{gluon}{v1,b1}
			\fmf{gluon}{l1,v1}
			\fmf{gluon}{t1,v1}
			\fmf{gluon,label=$t\qquad\quad$,label.dist=5,label.side=right,label.side=left}{v1,r1}
			\fmfv{decor.shape=circle, decor.filled=empty, decor.size=(4mm)}{v1}
			\fmf{marrowd,tension=0}{v1,r1}
			\fmf{darrowd,tension=0}{l1,v1}
			\fmf{darrowl,tension=0}{t1,v1}
			\fmf{darrowl,tension=0}{b1,v1}
			\fmflabel{$\;p_1,\mu,A$}{r1}
			\fmflabel{$p_2,\nu,B$}{t1}
			\fmflabel{$p_3,\rho,C\;$}{l1}
			\fmflabel{$p_4,\sigma,D$}{b1}
			\fmfv{decor.shape=otimes, decor.filled=empty, decor.size=(3mm)}{r1,l1,t1,b1}
		\end{fmfgraph*}}
	\end{gathered} \nn[0.5cm]
	\; &= - \int_0^\infty dt \, \begin{aligned}[t]
				& \Big[ f^{ABE} f^{CDE} (\delta_{\mu\rho}\delta_{\nu\sigma} -  \delta_{\mu\sigma}\delta_{\rho\nu} )  \nn
				& + f^{ACE} f^{BDE} ( \delta_{\mu\nu}\delta_{\rho\sigma} - \delta_{\mu\sigma}\delta_{\nu\rho} )  \nn
				& + f^{ADE} f^{BCE} ( \delta_{\mu\nu}\delta_{\rho\sigma} - \delta_{\mu\rho}\delta_{\nu\sigma} )  \Big] \, , \end{aligned} \mytag
\end{align}
where lines with a dashed adjacent arrow can be either a propagator or a flow line. The flow vertices involving both quantum and background gauge fields are
\begin{align}
	\nn
	\begin{gathered}
		\scalebox{0.8}{\begin{fmfgraph*}(70,60)
			\fmfleft{l1,l2} \fmfright{r1}
			\fmf{gluon,label=$\hspace{-1.5cm}$,label.side=right,label.dist=35,tension=2}{l1,v1}
			\fmf{gluon,label=$\hspace{-1.5cm}\quad\quad$,label.dist=-50,label.side=left,tension=2}{v1,l2}
			\fmf{gluon,label=$t\qquad\quad$,label.dist=5,label.side=right,label.side=left,tension=2}{v1,r1}
			\fmfv{decor.shape=circle, decor.filled=empty, decor.size=(4mm)}{v1}
			\fmf{marrowd,tension=0}{v1,r1}
			\fmf{darrowl,tension=0}{l1,v1}
			\fmf{darrowl,tension=0}{l2,v1}
			\fmflabel{$\;p_1,\mu,A$}{r1}
			\fmflabel{$p_2,\nu,B$}{l2}
			\fmflabel{$p_3,\rho,C$}{l1}
			\fmfv{decor.shape=otimes, decor.filled=empty, decor.size=(3mm)}{l1}
		\end{fmfgraph*}}
	\end{gathered} \qquad\quad\;
	\; &=  - i f^{ABC} \int_0^\infty dt \, \begin{aligned}[t]
				& \Big[ \delta_{\mu\rho} \bigl(2 {p_3} + (1-\alpha_0) p_2 \bigr)_\nu - \delta_{\mu\nu} \bigl(2 {p_2} + p_3 \bigr)_\rho \\
				& + \delta_{\nu\rho} \bigl( (1-\alpha_0) p_2 - (1+\alpha_0)p_3 \bigr)_\mu \Big] \, , \end{aligned} \nn[0.75cm]
	\begin{gathered}
		\scalebox{0.8}{\begin{fmfgraph*}(80,70)
			\fmfleft{l1} \fmftop{t1} \fmfbottom{b1} \fmfright{r1}
			\fmf{gluon}{v1,b1}
			\fmf{gluon}{l1,v1}
			\fmf{gluon}{t1,v1}
			\fmf{gluon,label=$t\qquad\quad$,label.dist=5,label.side=right,label.side=left}{v1,r1}
			\fmfv{decor.shape=circle, decor.filled=empty, decor.size=(4mm)}{v1}
			\fmf{marrowd,tension=0}{v1,r1}
			\fmf{darrowd,tension=0}{l1,v1}
			\fmf{darrowl,tension=0}{t1,v1}
			\fmf{darrowl,tension=0}{b1,v1}
			\fmflabel{$p_1,\mu,A$}{r1}
			\fmflabel{$p_2,\nu,B$}{t1}
			\fmflabel{$p_3,\rho,C$}{l1}
			\fmflabel{$p_4,\sigma,D$}{b1}
			\fmfv{decor.shape=otimes, decor.filled=empty, decor.size=(3mm)}{b1}
		\end{fmfgraph*}}
	\end{gathered} \qquad\quad
	\; &= - \int_0^\infty dt \, \begin{aligned}[t]
				& \Big[ f^{ABE} f^{CDE} \bigl( (1+\alpha_0) \delta_{\mu\rho}\delta_{\nu\sigma} -  \delta_{\mu\sigma}\delta_{\rho\nu} \bigr)  \nn
				& + f^{ACE} f^{BDE} \bigl( (1+\alpha_0) \delta_{\mu\nu}\delta_{\rho\sigma} - \delta_{\mu\sigma}\delta_{\nu\rho} \bigr)  \nn
				& + f^{ADE} f^{BCE} (1-\alpha_0) \bigl( \delta_{\mu\nu}\delta_{\rho\sigma} - \delta_{\mu\rho}\delta_{\nu\sigma} \bigr)  \Big] \, , \end{aligned} \\[0.75cm]
	\begin{gathered}
		\scalebox{0.8}{\begin{fmfgraph*}(80,70)
			\fmfleft{l1} \fmftop{t1} \fmfbottom{b1} \fmfright{r1}
			\fmf{gluon}{v1,b1}
			\fmf{gluon}{l1,v1}
			\fmf{gluon}{t1,v1}
			\fmf{gluon,label=$t\qquad\quad$,label.dist=5,label.side=right,label.side=left}{v1,r1}
			\fmfv{decor.shape=circle, decor.filled=empty, decor.size=(4mm)}{v1}
			\fmf{marrowd,tension=0}{v1,r1}
			\fmf{darrowd,tension=0}{l1,v1}
			\fmf{darrowl,tension=0}{t1,v1}
			\fmf{darrowl,tension=0}{b1,v1}
			\fmflabel{$p_1,\mu,A$}{r1}
			\fmflabel{$p_2,\nu,B$}{t1}
			\fmflabel{$p_3,\rho,C$}{l1}
			\fmflabel{$p_4,\sigma,D$}{b1}
			\fmfv{decor.shape=otimes, decor.filled=empty, decor.size=(3mm)}{l1,b1}
		\end{fmfgraph*}}
	\end{gathered} \qquad\quad
	\; &= - \int_0^\infty dt \, \begin{aligned}[t]
				& \Big[ f^{ABE} f^{CDE} (1+\alpha_0) \bigl(\delta_{\mu\rho}\delta_{\nu\sigma} -  \delta_{\mu\sigma}\delta_{\rho\nu} \bigr)  \nn
				& + f^{ACE} f^{BDE} \bigl( \delta_{\mu\nu}\delta_{\rho\sigma} - (1-\alpha_0)\delta_{\mu\sigma}\delta_{\nu\rho} \bigr)  \nn
				& + f^{ADE} f^{BCE} \bigl( \delta_{\mu\nu}\delta_{\rho\sigma} - (1-\alpha_0)\delta_{\mu\rho}\delta_{\nu\sigma} \bigr)  \Big] \, . \end{aligned} \mytag
\end{align}

Also in the case of fermionic vertices, the flow vertices involving only quantum fields agree with conventional gauge fixing, whereas the vertices involving only background fields correspond to the same expressions with $\alpha_0 = 0$. The complete set of vertex rules is as follows:
\begin{align}
	\nn
	\qquad \begin{gathered}
		\scalebox{0.8}{\begin{fmfgraph*}(70,60)
			\fmfleft{l1,l2} \fmfright{r1}
			\fmf{quark}{l1,v1}
			\fmf{gluon}{v1,l2}
			\fmf{quark,label=$t\qquad$,label.dist=5,label.side=right,label.side=left}{v1,r1}
			\fmfv{decor.shape=circle, decor.filled=empty, decor.size=(4mm)}{v1}
			\fmf{marrowd,tension=0}{v1,r1}
			\fmf{darrowl,tension=0}{l1,v1}
			\fmf{darrowl,tension=0}{l2,v1}
			\fmflabel{$p_1,\mu,A$}{l2}
			\fmflabel{$p_2$}{l1}
			\fmfv{decor.shape=otimes, decor.filled=empty, decor.size=(3mm), foreground=(.9,,.4,,.4)}{l1}
		\end{fmfgraph*}}
	\end{gathered}
	\quad &=  - i t^A \int_0^\infty dt \, \begin{aligned}[t]
				& \Bigl[ (1-\alpha_0) {p_1}_\mu + 2 {p_2}_\mu \Bigr] \, , \end{aligned} \nn[1cm]
	\qquad \begin{gathered}
		\scalebox{0.8}{\begin{fmfgraph*}(70,60)
			\fmfleft{l1,l2} \fmfright{r1}
			\fmf{quark}{l1,v1}
			\fmf{gluon}{v1,l2}
			\fmf{quark,label=$t\qquad\quad$,label.dist=5,label.side=right,label.side=left}{v1,r1}
			\fmfv{decor.shape=circle, decor.filled=empty, decor.size=(4mm)}{v1}
			\fmf{darrowd,tension=0}{r1,v1}
			\fmf{marrowl,tension=0}{v1,l1}
			\fmf{darrowl,tension=0}{l2,v1}
			\fmflabel{$p_1,\mu,A$}{l2}
			\fmflabel{$\;p_2$}{r1}
			\fmfv{decor.shape=otimes, decor.filled=empty, decor.size=(3mm), foreground=(.9,,.4,,.4)}{r1}
		\end{fmfgraph*}}
	\end{gathered} \quad\;
	\quad &=  i t^A \int_0^\infty dt \, \begin{aligned}[t]
				& \Bigl[ (1-\alpha_0) {p_1}_\mu + 2 {p_2}_\mu \Bigr] \, , \end{aligned} \nn[0.75cm]
	\qquad \begin{gathered}
		\scalebox{0.8}{\begin{fmfgraph*}(70,60)
			\fmfleft{l1,l2} \fmfright{r1}
			\fmf{quark}{l1,v1}
			\fmf{gluon}{v1,l2}
			\fmf{quark,label=$t\qquad$,label.dist=5,label.side=right,label.side=left}{v1,r1}
			\fmfv{decor.shape=circle, decor.filled=empty, decor.size=(4mm)}{v1}
			\fmf{marrowd,tension=0}{v1,r1}
			\fmf{darrowl,tension=0}{l1,v1}
			\fmf{darrowl,tension=0}{l2,v1}
			\fmflabel{$p_1,\mu,A$}{l2}
			\fmflabel{$p_2$}{l1}
			\fmfv{decor.shape=otimes, decor.filled=empty, decor.size=(3mm)}{l2}
		\end{fmfgraph*}}
	\end{gathered} \quad &= \quad
	\begin{gathered}
		\scalebox{0.8}{\begin{fmfgraph*}(70,60)
			\fmfleft{l1,l2} \fmfright{r1}
			\fmf{quark}{l1,v1}
			\fmf{gluon}{v1,l2}
			\fmf{quark,label=$t\qquad$,label.dist=5,label.side=right,label.side=left}{v1,r1}
			\fmfv{decor.shape=circle, decor.filled=empty, decor.size=(4mm)}{v1}
			\fmf{marrowd,tension=0}{v1,r1}
			\fmf{darrowl,tension=0}{l1,v1}
			\fmf{darrowl,tension=0}{l2,v1}
			\fmflabel{$p_1,\mu,A$}{l2}
			\fmflabel{$p_2$}{l1}
			\fmfv{decor.shape=otimes, decor.filled=empty, decor.size=(3mm)}{l1,l2,r1}
		\end{fmfgraph*}}
	\end{gathered}
	\quad =  - i t^A \int_0^\infty dt \, \begin{aligned}[t]
				& \Big( {p_1}_\mu + 2 {p_2}_\mu \Big) \, , \end{aligned} \nn[1cm]
	\qquad \begin{gathered}
		\scalebox{0.8}{\begin{fmfgraph*}(70,60)
			\fmfleft{l1,l2} \fmfright{r1}
			\fmf{quark}{l1,v1}
			\fmf{gluon}{v1,l2}
			\fmf{quark,label=$t\qquad\quad$,label.dist=5,label.side=right,label.side=left}{v1,r1}
			\fmfv{decor.shape=circle, decor.filled=empty, decor.size=(4mm)}{v1}
			\fmf{darrowd,tension=0}{r1,v1}
			\fmf{marrowl,tension=0}{v1,l1}
			\fmf{darrowl,tension=0}{l2,v1}
			\fmflabel{$p_1,\mu,A$}{l2}
			\fmflabel{$\;p_2$}{r1}
			\fmfv{decor.shape=otimes, decor.filled=empty, decor.size=(3mm)}{l2}
		\end{fmfgraph*}}
	\end{gathered} \qquad &= \quad
	\begin{gathered}
		\scalebox{0.8}{\begin{fmfgraph*}(70,60)
			\fmfleft{l1,l2} \fmfright{r1}
			\fmf{quark}{l1,v1}
			\fmf{gluon}{v1,l2}
			\fmf{quark,label=$t\qquad\quad$,label.dist=5,label.side=right,label.side=left}{v1,r1}
			\fmfv{decor.shape=circle, decor.filled=empty, decor.size=(4mm)}{v1}
			\fmf{darrowd,tension=0}{r1,v1}
			\fmf{marrowl,tension=0}{v1,l1}
			\fmf{darrowl,tension=0}{l2,v1}
			\fmflabel{$p_1,\mu,A$}{l2}
			\fmflabel{$\;p_2$}{r1}
			\fmfv{decor.shape=otimes, decor.filled=empty, decor.size=(3mm)}{l1,l2,r1}
		\end{fmfgraph*}}
	\end{gathered} \quad
	\quad =  i t^A \int_0^\infty dt \, \begin{aligned}[t]
				& \Big( {p_1}_\mu + 2 {p_2}_\mu \Big) \, , \end{aligned} \nn[0.75cm]
%%%%%%%%%%%%%%%%%%%%%%%%
	\qquad \begin{gathered}
		\scalebox{0.8}{\begin{fmfgraph*}(70,60)
			\fmfleft{l1,l2} \fmfright{r1}
			\fmf{quark}{l1,v1}
			\fmf{photon}{v1,l2}
			\fmf{quark,label=$t\qquad$,label.dist=5,label.side=right,label.side=left}{v1,r1}
			\fmfv{decor.shape=circle, decor.filled=empty, decor.size=(4mm)}{v1}
			\fmf{marrowd,tension=0}{v1,r1}
			\fmf{darrowl,tension=0}{l1,v1}
			\fmflabel{$p_1,\mu$}{l2}
			\fmflabel{$p_2$}{l1}
			\fmfv{decor.shape=otimes, decor.filled=empty, decor.size=(3mm), foreground=(.9,,.4,,.4)}{l1}
		\end{fmfgraph*}}
	\end{gathered} \quad &= \quad 
	\begin{gathered}
		\scalebox{0.8}{\begin{fmfgraph*}(70,60)
			\fmfleft{l1,l2} \fmfright{r1}
			\fmf{quark}{l1,v1}
			\fmf{photon}{v1,l2}
			\fmf{quark,label=$t\qquad$,label.dist=5,label.side=right,label.side=left}{v1,r1}
			\fmfv{decor.shape=circle, decor.filled=empty, decor.size=(4mm)}{v1}
			\fmf{marrowd,tension=0}{v1,r1}
			\fmf{darrowl,tension=0}{l1,v1}
			\fmflabel{$p_1,\mu$}{l2}
			\fmflabel{$p_2$}{l1}
			\fmfv{decor.shape=otimes, decor.filled=empty, decor.size=(3mm)}{l2}
		\end{fmfgraph*}}
	\end{gathered} \quad = \quad 
	\begin{gathered}
		\scalebox{0.8}{\begin{fmfgraph*}(70,60)
			\fmfleft{l1,l2} \fmfright{r1}
			\fmf{quark}{l1,v1}
			\fmf{photon}{v1,l2}
			\fmf{quark,label=$t\qquad$,label.dist=5,label.side=right,label.side=left}{v1,r1}
			\fmfv{decor.shape=circle, decor.filled=empty, decor.size=(4mm)}{v1}
			\fmf{marrowd,tension=0}{v1,r1}
			\fmf{darrowl,tension=0}{l1,v1}
			\fmflabel{$p_1,\mu$}{l2}
			\fmflabel{$p_2$}{l1}
			\fmfv{decor.shape=otimes, decor.filled=empty, decor.size=(3mm)}{l1,l2,r1}
		\end{fmfgraph*}}
	\end{gathered}
	\quad =  - \q_\psi \int_0^\infty dt \, \begin{aligned}[t]
				& \Big( {p_1}_\mu + 2 {p_2}_\mu \Big) \, , \end{aligned} \nn[1cm]
	\qquad \begin{gathered}
		\scalebox{0.8}{\begin{fmfgraph*}(70,60)
			\fmfleft{l1,l2} \fmfright{r1}
			\fmf{quark}{l1,v1}
			\fmf{photon}{v1,l2}
			\fmf{quark,label=$t\qquad\quad$,label.dist=5,label.side=right,label.side=left}{v1,r1}
			\fmfv{decor.shape=circle, decor.filled=empty, decor.size=(4mm)}{v1}
			\fmf{darrowd,tension=0}{r1,v1}
			\fmf{marrowl,tension=0}{v1,l1}
			\fmflabel{$p_1,\mu$}{l2}
			\fmflabel{$\;p_2$}{r1}
			\fmfv{decor.shape=otimes, decor.filled=empty, decor.size=(3mm), foreground=(.9,,.4,,.4)}{r1}
		\end{fmfgraph*}}
	\end{gathered} \qquad &= \quad
	\begin{gathered}
		\scalebox{0.8}{\begin{fmfgraph*}(70,60)
			\fmfleft{l1,l2} \fmfright{r1}
			\fmf{quark}{l1,v1}
			\fmf{photon}{v1,l2}
			\fmf{quark,label=$t\qquad\quad$,label.dist=5,label.side=right,label.side=left}{v1,r1}
			\fmfv{decor.shape=circle, decor.filled=empty, decor.size=(4mm)}{v1}
			\fmf{darrowd,tension=0}{r1,v1}
			\fmf{marrowl,tension=0}{v1,l1}
			\fmflabel{$p_1,\mu$}{l2}
			\fmflabel{$\;p_2$}{r1}
			\fmfv{decor.shape=otimes, decor.filled=empty, decor.size=(3mm)}{l2}
		\end{fmfgraph*}}
	\end{gathered} \qquad = \quad
	\begin{gathered}
		\scalebox{0.8}{\begin{fmfgraph*}(70,60)
			\fmfleft{l1,l2} \fmfright{r1}
			\fmf{quark}{l1,v1}
			\fmf{photon}{v1,l2}
			\fmf{quark,label=$t\qquad\quad$,label.dist=5,label.side=right,label.side=left}{v1,r1}
			\fmfv{decor.shape=circle, decor.filled=empty, decor.size=(4mm)}{v1}
			\fmf{darrowd,tension=0}{r1,v1}
			\fmf{marrowl,tension=0}{v1,l1}
			\fmflabel{$p_1,\mu$}{l2}
			\fmflabel{$\;p_2$}{r1}
			\fmfv{decor.shape=otimes, decor.filled=empty, decor.size=(3mm)}{l1,l2,r1}
		\end{fmfgraph*}}
	\end{gathered}
	\qquad =  \q_\psi \int_0^\infty dt \, \begin{aligned}[t]
				& \Big( {p_1}_\mu + 2 {p_2}_\mu \Big) \, , \end{aligned} \nn[0.75cm]
%%%%%%%%%%%%%%%%%%%%%%%%
%%%%%%%%%%%%%%%%%%%%%%%%
%%%%%%%%%%%%%%%%%%%%%%%%
	\qquad \begin{gathered}
		\scalebox{0.8}{\begin{fmfgraph*}(80,70)
			\fmfleft{l1} \fmftop{t1} \fmfbottom{b1} \fmfright{r1}
			\fmf{gluon}{v1,b1}
			\fmf{quark}{l1,v1}
			\fmf{gluon}{t1,v1}
			\fmf{quark,label=$\! t\qquad$,label.dist=5,label.side=right,label.side=left}{v1,r1}
			\fmfv{decor.shape=circle, decor.filled=empty, decor.size=(4mm)}{v1}
			\fmf{marrowd,tension=0}{v1,r1}
			\fmf{darrowd,tension=0}{l1,v1}
			\fmf{darrowl,tension=0}{t1,v1}
			\fmf{darrowl,tension=0}{b1,v1}
			\fmflabel{$\mu,A$}{t1}
			\fmflabel{$\nu,B$}{b1}
			\fmfv{decor.shape=otimes, decor.filled=empty, decor.size=(3mm), foreground=(.9,,.4,,.4)}{l1}
		\end{fmfgraph*}}
	\end{gathered} \quad &= \quad
	\begin{gathered}
		\scalebox{0.8}{\begin{fmfgraph*}(80,70)
			\fmfleft{l1} \fmftop{t1} \fmfbottom{b1} \fmfright{r1}
			\fmf{gluon}{v1,b1}
			\fmf{quark}{l1,v1}
			\fmf{gluon}{t1,v1}
			\fmf{quark,label=$\! t\qquad$,label.dist=5,label.side=right,label.side=left}{v1,r1}
			\fmfv{decor.shape=circle, decor.filled=empty, decor.size=(4mm)}{v1}
			\fmf{marrowd,tension=0}{v1,r1}
			\fmf{darrowd,tension=0}{l1,v1}
			\fmf{darrowl,tension=0}{t1,v1}
			\fmf{darrowl,tension=0}{b1,v1}
			\fmflabel{$\mu,A$}{t1}
			\fmflabel{$\nu,B$}{b1}
			\fmfv{decor.shape=otimes, decor.filled=empty, decor.size=(3mm)}{b1,t1}
		\end{fmfgraph*}}
	\end{gathered} \quad = \quad
	\begin{gathered}
		\scalebox{0.8}{\begin{fmfgraph*}(80,70)
			\fmfleft{l1} \fmftop{t1} \fmfbottom{b1} \fmfright{r1}
			\fmf{gluon}{v1,b1}
			\fmf{quark}{l1,v1}
			\fmf{gluon}{t1,v1}
			\fmf{quark,label=$\! t\qquad$,label.dist=5,label.side=right,label.side=left}{v1,r1}
			\fmfv{decor.shape=circle, decor.filled=empty, decor.size=(4mm)}{v1}
			\fmf{marrowd,tension=0}{v1,r1}
			\fmf{darrowd,tension=0}{l1,v1}
			\fmf{darrowl,tension=0}{t1,v1}
			\fmf{darrowl,tension=0}{b1,v1}
			\fmflabel{$\mu,A$}{t1}
			\fmflabel{$\nu,B$}{b1}
			\fmfv{decor.shape=otimes, decor.filled=empty, decor.size=(3mm)}{l1,r1,t1,b1}
		\end{fmfgraph*}}
	\end{gathered} \nn[0.8cm]
	= \quad
	\begin{gathered}
		\scalebox{0.8}{\begin{fmfgraph*}(80,70)
			\fmfleft{l1} \fmftop{t1} \fmfbottom{b1} \fmfright{r1}
			\fmf{gluon}{v1,b1}
			\fmf{quark}{l1,v1}
			\fmf{gluon}{t1,v1}
			\fmf{quark,label=$\! t\qquad$,label.dist=5,label.side=right,label.side=left}{v1,r1}
			\fmfv{decor.shape=circle, decor.filled=empty, decor.size=(4mm)}{v1}
			\fmf{darrowd,tension=0}{r1,v1}
			\fmf{marrowd,tension=0}{v1,l1}
			\fmf{darrowl,tension=0}{t1,v1}
			\fmf{darrowl,tension=0}{b1,v1}
			\fmflabel{$\mu,A$}{t1}
			\fmflabel{$\nu,B$}{b1}
			\fmfv{decor.shape=otimes, decor.filled=empty, decor.size=(3mm), foreground=(.9,,.4,,.4)}{r1}
		\end{fmfgraph*}}
	\end{gathered} \quad &= \quad
	\begin{gathered}
		\scalebox{0.8}{\begin{fmfgraph*}(80,70)
			\fmfleft{l1} \fmftop{t1} \fmfbottom{b1} \fmfright{r1}
			\fmf{gluon}{v1,b1}
			\fmf{quark}{l1,v1}
			\fmf{gluon}{t1,v1}
			\fmf{quark,label=$\! t\qquad$,label.dist=5,label.side=right,label.side=left}{v1,r1}
			\fmfv{decor.shape=circle, decor.filled=empty, decor.size=(4mm)}{v1}
			\fmf{darrowd,tension=0}{r1,v1}
			\fmf{marrowd,tension=0}{v1,l1}
			\fmf{darrowl,tension=0}{t1,v1}
			\fmf{darrowl,tension=0}{b1,v1}
			\fmflabel{$\mu,A$}{t1}
			\fmflabel{$\nu,B$}{b1}
			\fmfv{decor.shape=otimes, decor.filled=empty, decor.size=(3mm)}{b1,t1}
		\end{fmfgraph*}}
	\end{gathered} \quad = \quad
	\begin{gathered}
		\scalebox{0.8}{\begin{fmfgraph*}(80,70)
			\fmfleft{l1} \fmftop{t1} \fmfbottom{b1} \fmfright{r1}
			\fmf{gluon}{v1,b1}
			\fmf{quark}{l1,v1}
			\fmf{gluon}{t1,v1}
			\fmf{quark,label=$\! t\qquad$,label.dist=5,label.side=right,label.side=left}{v1,r1}
			\fmfv{decor.shape=circle, decor.filled=empty, decor.size=(4mm)}{v1}
			\fmf{darrowd,tension=0}{r1,v1}
			\fmf{marrowd,tension=0}{v1,l1}
			\fmf{darrowl,tension=0}{t1,v1}
			\fmf{darrowl,tension=0}{b1,v1}
			\fmflabel{$\mu,A$}{t1}
			\fmflabel{$\nu,B$}{b1}
			\fmfv{decor.shape=otimes, decor.filled=empty, decor.size=(3mm)}{l1,r1,t1,b1}
		\end{fmfgraph*}}
	\end{gathered}
	\quad =  \delta_{\mu\nu}  \{ t^A, t^B \} \int_0^\infty dt \, , \nn[1cm]
	\qquad \begin{gathered}
		\scalebox{0.8}{\begin{fmfgraph*}(80,70)
			\fmfleft{l1} \fmftop{t1} \fmfbottom{b1} \fmfright{r1}
			\fmf{gluon}{v1,b1}
			\fmf{quark}{l1,v1}
			\fmf{gluon}{t1,v1}
			\fmf{quark,label=$\! t\qquad$,label.dist=5,label.side=right,label.side=left}{v1,r1}
			\fmfv{decor.shape=circle, decor.filled=empty, decor.size=(4mm)}{v1}
			\fmf{marrowd,tension=0}{v1,r1}
			\fmf{darrowd,tension=0}{l1,v1}
			\fmf{darrowl,tension=0}{t1,v1}
			\fmf{darrowl,tension=0}{b1,v1}
			\fmflabel{$\mu,A$}{t1}
			\fmflabel{$\nu,B$}{b1}
			\fmfv{decor.shape=otimes, decor.filled=empty, decor.size=(3mm), foreground=(.9,,.4,,.4)}{l1}
			\fmfv{decor.shape=otimes, decor.filled=empty, decor.size=(3mm)}{t1}
		\end{fmfgraph*}}
	\end{gathered} \quad
	&= \delta_{\mu\nu} \Bigl[ (1+\alpha_0) t^B t^A + (1-\alpha_0) t^A t^B \Bigr] \int_0^\infty dt \, , \nn[1cm]
	\qquad \begin{gathered}
		\scalebox{0.8}{\begin{fmfgraph*}(80,70)
			\fmfleft{l1} \fmftop{t1} \fmfbottom{b1} \fmfright{r1}
			\fmf{gluon}{v1,b1}
			\fmf{quark}{l1,v1}
			\fmf{gluon}{t1,v1}
			\fmf{quark,label=$\! t\qquad$,label.dist=5,label.side=right,label.side=left}{v1,r1}
			\fmfv{decor.shape=circle, decor.filled=empty, decor.size=(4mm)}{v1}
			\fmf{darrowd,tension=0}{r1,v1}
			\fmf{marrowd,tension=0}{v1,l1}
			\fmf{darrowl,tension=0}{t1,v1}
			\fmf{darrowl,tension=0}{b1,v1}
			\fmflabel{$\mu,A$}{t1}
			\fmflabel{$\nu,B$}{b1}
			\fmfv{decor.shape=otimes, decor.filled=empty, decor.size=(3mm), foreground=(.9,,.4,,.4)}{r1}
			\fmfv{decor.shape=otimes, decor.filled=empty, decor.size=(3mm)}{t1}
		\end{fmfgraph*}}
	\end{gathered} \quad
	& =  \delta_{\mu\nu} \Bigl[ (1+\alpha_0) t^A t^B + (1-\alpha_0) t^B t^A \Bigr] \int_0^\infty dt \, , \nn[1cm]
%%%%%%%%%%%%%%%%%%%%%%%%%%%
%%%%%%%%%%%%%%%%%%%%%%%%%%%
	\qquad \begin{gathered}
		\scalebox{0.8}{\begin{fmfgraph*}(80,70)
			\fmfleft{l1} \fmftop{t1} \fmfbottom{b1} \fmfright{r1}
			\fmf{photon}{v1,b1}
			\fmf{quark}{l1,v1}
			\fmf{gluon}{t1,v1}
			\fmf{quark,label=$\! t\qquad$,label.dist=5,label.side=right,label.side=left}{v1,r1}
			\fmfv{decor.shape=circle, decor.filled=empty, decor.size=(4mm)}{v1}
			\fmf{marrowd,tension=0}{v1,r1}
			\fmf{darrowd,tension=0}{l1,v1}
			\fmf{darrowl,tension=0}{t1,v1}
			\fmflabel{$\mu,A$}{t1}
			\fmflabel{$\nu$}{b1}
			\fmfv{decor.shape=otimes, decor.filled=empty, decor.size=(3mm), foreground=(.9,,.4,,.4)}{l1,t1}
		\end{fmfgraph*}}
	\end{gathered} \quad &= \quad
	\begin{gathered}
		\scalebox{0.8}{\begin{fmfgraph*}(80,70)
			\fmfleft{l1} \fmftop{t1} \fmfbottom{b1} \fmfright{r1}
			\fmf{photon}{v1,b1}
			\fmf{quark}{l1,v1}
			\fmf{gluon}{t1,v1}
			\fmf{quark,label=$\! t\qquad$,label.dist=5,label.side=right,label.side=left}{v1,r1}
			\fmfv{decor.shape=circle, decor.filled=empty, decor.size=(4mm)}{v1}
			\fmf{marrowd,tension=0}{v1,r1}
			\fmf{darrowd,tension=0}{l1,v1}
			\fmf{darrowl,tension=0}{t1,v1}
			\fmflabel{$\mu,A$}{t1}
			\fmflabel{$\nu$}{b1}
			\fmfv{decor.shape=otimes, decor.filled=empty, decor.size=(3mm), foreground=(.9,,.4,,.4)}{l1}
			\fmfv{decor.shape=otimes, decor.filled=empty, decor.size=(3mm)}{b1}
		\end{fmfgraph*}}
	\end{gathered} \quad = \quad
	\begin{gathered}
		\scalebox{0.8}{\begin{fmfgraph*}(80,70)
			\fmfleft{l1} \fmftop{t1} \fmfbottom{b1} \fmfright{r1}
			\fmf{photon}{v1,b1}
			\fmf{quark}{l1,v1}
			\fmf{gluon}{t1,v1}
			\fmf{quark,label=$\! t\qquad$,label.dist=5,label.side=right,label.side=left}{v1,r1}
			\fmfv{decor.shape=circle, decor.filled=empty, decor.size=(4mm)}{v1}
			\fmf{marrowd,tension=0}{v1,r1}
			\fmf{darrowd,tension=0}{l1,v1}
			\fmf{darrowl,tension=0}{t1,v1}
			\fmflabel{$\mu,A$}{t1}
			\fmflabel{$\nu$}{b1}
			\fmfv{decor.shape=otimes, decor.filled=empty, decor.size=(3mm)}{b1,t1}
		\end{fmfgraph*}}
	\end{gathered} \quad = \quad
	\begin{gathered}
		\scalebox{0.8}{\begin{fmfgraph*}(80,70)
			\fmfleft{l1} \fmftop{t1} \fmfbottom{b1} \fmfright{r1}
			\fmf{photon}{v1,b1}
			\fmf{quark}{l1,v1}
			\fmf{gluon}{t1,v1}
			\fmf{quark,label=$\! t\qquad$,label.dist=5,label.side=right,label.side=left}{v1,r1}
			\fmfv{decor.shape=circle, decor.filled=empty, decor.size=(4mm)}{v1}
			\fmf{marrowd,tension=0}{v1,r1}
			\fmf{darrowd,tension=0}{l1,v1}
			\fmf{darrowl,tension=0}{t1,v1}
			\fmflabel{$\mu,A$}{t1}
			\fmflabel{$\nu$}{b1}
			\fmfv{decor.shape=otimes, decor.filled=empty, decor.size=(3mm)}{l1,r1,t1,b1}
		\end{fmfgraph*}}
	\end{gathered} \nn[0.8cm]
	= \quad
	\begin{gathered}
		\scalebox{0.8}{\begin{fmfgraph*}(80,70)
			\fmfleft{l1} \fmftop{t1} \fmfbottom{b1} \fmfright{r1}
			\fmf{photon}{v1,b1}
			\fmf{quark}{l1,v1}
			\fmf{gluon}{t1,v1}
			\fmf{quark,label=$\! t\qquad$,label.dist=5,label.side=right,label.side=left}{v1,r1}
			\fmfv{decor.shape=circle, decor.filled=empty, decor.size=(4mm)}{v1}
			\fmf{darrowd,tension=0}{r1,v1}
			\fmf{marrowd,tension=0}{v1,l1}
			\fmf{darrowl,tension=0}{t1,v1}
			\fmflabel{$\mu,A$}{t1}
			\fmflabel{$\nu$}{b1}
			\fmfv{decor.shape=otimes, decor.filled=empty, decor.size=(3mm), foreground=(.9,,.4,,.4)}{r1,t1}
		\end{fmfgraph*}}
	\end{gathered} \quad &= \quad
	\begin{gathered}
		\scalebox{0.8}{\begin{fmfgraph*}(80,70)
			\fmfleft{l1} \fmftop{t1} \fmfbottom{b1} \fmfright{r1}
			\fmf{photon}{v1,b1}
			\fmf{quark}{l1,v1}
			\fmf{gluon}{t1,v1}
			\fmf{quark,label=$\! t\qquad$,label.dist=5,label.side=right,label.side=left}{v1,r1}
			\fmfv{decor.shape=circle, decor.filled=empty, decor.size=(4mm)}{v1}
			\fmf{darrowd,tension=0}{r1,v1}
			\fmf{marrowd,tension=0}{v1,l1}
			\fmf{darrowl,tension=0}{t1,v1}
			\fmflabel{$\mu,A$}{t1}
			\fmflabel{$\nu$}{b1}
			\fmfv{decor.shape=otimes, decor.filled=empty, decor.size=(3mm), foreground=(.9,,.4,,.4)}{r1}
			\fmfv{decor.shape=otimes, decor.filled=empty, decor.size=(3mm)}{b1}
		\end{fmfgraph*}}
	\end{gathered} \quad = \quad
	\begin{gathered}
		\scalebox{0.8}{\begin{fmfgraph*}(80,70)
			\fmfleft{l1} \fmftop{t1} \fmfbottom{b1} \fmfright{r1}
			\fmf{photon}{v1,b1}
			\fmf{quark}{l1,v1}
			\fmf{gluon}{t1,v1}
			\fmf{quark,label=$\! t\qquad$,label.dist=5,label.side=right,label.side=left}{v1,r1}
			\fmfv{decor.shape=circle, decor.filled=empty, decor.size=(4mm)}{v1}
			\fmf{darrowd,tension=0}{r1,v1}
			\fmf{marrowd,tension=0}{v1,l1}
			\fmf{darrowl,tension=0}{t1,v1}
			\fmflabel{$\mu,A$}{t1}
			\fmflabel{$\nu$}{b1}
			\fmfv{decor.shape=otimes, decor.filled=empty, decor.size=(3mm)}{b1,t1}
		\end{fmfgraph*}}
	\end{gathered} \quad = \quad
	\begin{gathered}
		\scalebox{0.8}{\begin{fmfgraph*}(80,70)
			\fmfleft{l1} \fmftop{t1} \fmfbottom{b1} \fmfright{r1}
			\fmf{photon}{v1,b1}
			\fmf{quark}{l1,v1}
			\fmf{gluon}{t1,v1}
			\fmf{quark,label=$\! t\qquad$,label.dist=5,label.side=right,label.side=left}{v1,r1}
			\fmfv{decor.shape=circle, decor.filled=empty, decor.size=(4mm)}{v1}
			\fmf{darrowd,tension=0}{r1,v1}
			\fmf{marrowd,tension=0}{v1,l1}
			\fmf{darrowl,tension=0}{t1,v1}
			\fmflabel{$\mu,A$}{t1}
			\fmflabel{$\nu$}{b1}
			\fmfv{decor.shape=otimes, decor.filled=empty, decor.size=(3mm)}{l1,r1,t1,b1}
		\end{fmfgraph*}}
	\end{gathered} \nn[0.5cm]
	&= -2i \q_\psi \delta_{\mu\nu}  t^A \int_0^\infty dt \, , \nn[1cm]
%%%%%%%%%%%%%%%%%%%%%%%%%%%
	\qquad \begin{gathered}
		\scalebox{0.8}{\begin{fmfgraph*}(80,70)
			\fmfleft{l1} \fmftop{t1} \fmfbottom{b1} \fmfright{r1}
			\fmf{photon}{v1,b1}
			\fmf{quark}{l1,v1}
			\fmf{photon}{t1,v1}
			\fmf{quark,label=$\! t\qquad$,label.dist=5,label.side=right,label.side=left}{v1,r1}
			\fmfv{decor.shape=circle, decor.filled=empty, decor.size=(4mm)}{v1}
			\fmf{marrowd,tension=0}{v1,r1}
			\fmf{darrowd,tension=0}{l1,v1}
			\fmflabel{$\mu$}{t1}
			\fmflabel{$\nu$}{b1}
			\fmfv{decor.shape=otimes, decor.filled=empty, decor.size=(3mm), foreground=(.9,,.4,,.4)}{l1,t1}
		\end{fmfgraph*}}
	\end{gathered} \quad &= \quad
	\begin{gathered}
		\scalebox{0.8}{\begin{fmfgraph*}(80,70)
			\fmfleft{l1} \fmftop{t1} \fmfbottom{b1} \fmfright{r1}
			\fmf{photon}{v1,b1}
			\fmf{quark}{l1,v1}
			\fmf{photon}{t1,v1}
			\fmf{quark,label=$\! t\qquad$,label.dist=5,label.side=right,label.side=left}{v1,r1}
			\fmfv{decor.shape=circle, decor.filled=empty, decor.size=(4mm)}{v1}
			\fmf{marrowd,tension=0}{v1,r1}
			\fmf{darrowd,tension=0}{l1,v1}
			\fmflabel{$\mu$}{t1}
			\fmflabel{$\nu$}{b1}
			\fmfv{decor.shape=otimes, decor.filled=empty, decor.size=(3mm)}{t1,b1}
		\end{fmfgraph*}}
	\end{gathered} \quad = \quad
	\begin{gathered}
		\scalebox{0.8}{\begin{fmfgraph*}(80,70)
			\fmfleft{l1} \fmftop{t1} \fmfbottom{b1} \fmfright{r1}
			\fmf{photon}{v1,b1}
			\fmf{quark}{l1,v1}
			\fmf{photon}{t1,v1}
			\fmf{quark,label=$\! t\qquad$,label.dist=5,label.side=right,label.side=left}{v1,r1}
			\fmfv{decor.shape=circle, decor.filled=empty, decor.size=(4mm)}{v1}
			\fmf{marrowd,tension=0}{v1,r1}
			\fmf{darrowd,tension=0}{l1,v1}
			\fmflabel{$\mu$}{t1}
			\fmflabel{$\nu$}{b1}
			\fmfv{decor.shape=otimes, decor.filled=empty, decor.size=(3mm)}{l1,r1,t1,b1}
		\end{fmfgraph*}}
	\end{gathered} \nn[0.8cm]
	= \quad \begin{gathered}
		\scalebox{0.8}{\begin{fmfgraph*}(80,70)
			\fmfleft{l1} \fmftop{t1} \fmfbottom{b1} \fmfright{r1}
			\fmf{photon}{v1,b1}
			\fmf{quark}{l1,v1}
			\fmf{photon}{t1,v1}
			\fmf{quark,label=$\! t\qquad$,label.dist=5,label.side=right,label.side=left}{v1,r1}
			\fmfv{decor.shape=circle, decor.filled=empty, decor.size=(4mm)}{v1}
			\fmf{darrowd,tension=0}{r1,v1}
			\fmf{marrowd,tension=0}{v1,l1}
			\fmflabel{$\mu$}{t1}
			\fmflabel{$\nu$}{b1}
			\fmfv{decor.shape=otimes, decor.filled=empty, decor.size=(3mm), foreground=(.9,,.4,,.4)}{r1,t1}
		\end{fmfgraph*}}
	\end{gathered}
	\quad &= \quad
	\begin{gathered}
		\scalebox{0.8}{\begin{fmfgraph*}(80,70)
			\fmfleft{l1} \fmftop{t1} \fmfbottom{b1} \fmfright{r1}
			\fmf{photon}{v1,b1}
			\fmf{quark}{l1,v1}
			\fmf{photon}{t1,v1}
			\fmf{quark,label=$\! t\qquad$,label.dist=5,label.side=right,label.side=left}{v1,r1}
			\fmfv{decor.shape=circle, decor.filled=empty, decor.size=(4mm)}{v1}
			\fmf{darrowd,tension=0}{r1,v1}
			\fmf{marrowd,tension=0}{v1,l1}
			\fmflabel{$\mu$}{t1}
			\fmflabel{$\nu$}{b1}
			\fmfv{decor.shape=otimes, decor.filled=empty, decor.size=(3mm)}{t1,b1}
		\end{fmfgraph*}}
	\end{gathered}
	\quad = \quad
	\begin{gathered}
		\scalebox{0.8}{\begin{fmfgraph*}(80,70)
			\fmfleft{l1} \fmftop{t1} \fmfbottom{b1} \fmfright{r1}
			\fmf{photon}{v1,b1}
			\fmf{quark}{l1,v1}
			\fmf{photon}{t1,v1}
			\fmf{quark,label=$\! t\qquad$,label.dist=5,label.side=right,label.side=left}{v1,r1}
			\fmfv{decor.shape=circle, decor.filled=empty, decor.size=(4mm)}{v1}
			\fmf{darrowd,tension=0}{r1,v1}
			\fmf{marrowd,tension=0}{v1,l1}
			\fmflabel{$\mu$}{t1}
			\fmflabel{$\nu$}{b1}
			\fmfv{decor.shape=otimes, decor.filled=empty, decor.size=(3mm)}{l1,r1,t1,b1}
		\end{fmfgraph*}}
	\end{gathered}
	\quad = -2 \q_\psi^2 \delta_{\mu\nu} \int_0^\infty dt \, .
\end{align}

	% !TEX root = ../Paper-LEFT-GF.tex

\section{Matching equations}
\label{sec:MatchingEquations}

The one-loop matching results are computed in the chirally symmetric HV scheme of Ref.~\cite{Naterop:2023dek}. We provide the off-shell matching results to the redundant operator basis in Sect.~\ref{sec:OffShellResults}, whereas in Sect.~\ref{sec:OnShellResults} we give the results in the non-redundant operator basis after field redefinitions.

The logarithmic dependence on the MS scale entering through $\lt = \log(8\pi\mu^2 t)$ is in direct correspondence to the one-loop RG equations~\cite{Jenkins:2017dyc,Naterop:2023dek}.

On the right-hand side of all matching equations, the Wilson coefficients denote the coefficients of the flowed operators, although for notational simplicity we drop a superscript $^t$.

\subsection{Before field redefinitions}
\label{sec:OffShellResults}

From the calculation of vacuum diagrams, we obtain the matching for the cosmological constant.
\begin{align}
	\label{eq:DeltaLambda}
	\Delta_1(\Lambda) &= \frac{N_c}{t^2} \left( 3 g^2 C_F L_G^{(4)} - \frac{1}{2} \left( \< \lwc{uD}{L(4)}[][] \> + \< \lwc{uD}{R(4)}[][] \> + \< \lwc{dD}{L(4)}[][] \> + \< \lwc{dD}{R(4)}[][] \> \right) \right) \nn
		&\quad + \frac{N_c}{t} \begin{aligned}[t]
			&\Big( \< \lwc{uD}{L(4)}[][] M_u^\dagger M_u \> + \< \lwc{uD}{R(4)}[][] M_u M_u^\dagger \> + \lwc{dD}{L(4)}[][] M_d^\dagger M_d \> + \< \lwc{dD}{R(4)}[][] M_d M_d^\dagger \> \\
			&- \< L_{Mu}^{(3)} M_u^\dagger \> - \< L_{Mu}^{(3)\dagger} M_u \> - \< L_{Md}^{(3)} M_d^\dagger \> - \< L_{Md}^{(3)\dagger} M_d \> \Big) \end{aligned} \nn
		&\quad + \frac{N_c}{3} \left( 11 + 6 \lt \right) \begin{aligned}[t]
			&\Big( \< \lwc{uD}{L(4)}[][] M_u^\dagger M_u M_u^\dagger M_u \> + \< \lwc{uD}{R(4)}[][] M_u M_u^\dagger M_u M_u^\dagger \> \\
			&+ \< \lwc{dD}{L(4)}[][] M_d^\dagger M_d M_d^\dagger M_d \> + \< \lwc{dD}{R(4)}[][] M_d M_d^\dagger M_d M_d^\dagger \> \\
			&- \< L_{Mu}^{(3)} M_u^\dagger M_u M_u^\dagger \> - \< L_{Mu}^{(3)\dagger} M_u M_u^\dagger M_u \> \\
			&- \< L_{Md}^{(3)} M_d^\dagger M_d M_d^\dagger \> - \< L_{Md}^{(3)\dagger} M_d M_d^\dagger M_d \> \Big) \, . \end{aligned}
\end{align}
The gauge-boson two-point function allows us to match the gauge couplings and theta parameters (computed with momentum insertion into the operator~\cite{Georgi:1980cn})
\begin{align}
	\label{eq:Deltae}
	\Delta_1(e) &= 4 e^3 N_c \left(2 + \lt \right) \left( \q_u \< L_{u\gamma} M_u \> + \q_u \< L_{u\gamma}^\dagger M_u^\dagger\> + \q_d \< L_{d\gamma} M_d \> + \q_d \< L_{d\gamma}^\dagger M_d^\dagger\> \right) \nn
		&\quad - \frac{5N_c}{6} e^3 \left( \q_u^2 \left( \< \lwc{uD}{L(4)}[][] \> + \< \lwc{uD}{R(4)}[][] \> \right) + \q_d^2 \left( \< \lwc{dD}{L(4)}[][] \> + \< \lwc{dD}{R(4)}[][] \> \right) \right) \, , \\
	\label{eq:Deltag}
	\Delta_1(g) &= 2 g^3 \left(2 + \lt \right) \left( \< L_{uG} M_u \> + \< L_{uG}^\dagger M_u^\dagger\> + \< L_{dG} M_d \> + \< L_{dG}^\dagger M_d^\dagger\> \right) + \frac{9 g^5 N_c}{t} L_G \nn
		&\quad - \frac{5}{12} g^3 \left( \< \lwc{uD}{L(4)}[][] \> + \< \lwc{uD}{R(4)}[][] \> + \< \lwc{dD}{L(4)}[][] \> + \< \lwc{dD}{R(4)}[][] \> \right) - 7 N_c g^5 \lwc{G}{(4)}[][] \, , \\
	\label{eq:DeltathetaQED}
	\Delta_1(\theta_\mathrm{QED}) &= -64 i \pi^2 N_c \left(2 + \lt \right) \left( \q_u \< L_{u\gamma} M_u \> - \q_u \< L_{u\gamma}^\dagger M_u^\dagger\> + \q_d \< L_{d\gamma} M_d \> - \q_d \< L_{d\gamma}^\dagger M_d^\dagger\> \right) \, , \\
	\label{eq:DeltathetaQCD}
	\Delta_1(\theta_\mathrm{QCD}) &= -32 i \pi^2 \left(2 + \lt \right) \left( \< L_{uG} M_u \> - \< L_{uG}^\dagger M_u^\dagger\> + \< L_{dG} M_d \> - \< L_{dG}^\dagger M_d^\dagger\> \right) \nn
		&\quad - \frac{144\pi^2 g^2 N_c}{t} L_{\widetilde G} \, ,
\end{align}
where $\< \cdot \>$ denotes the trace in flavor space. A finite contribution proportional to $\lwc{G}{(4)}$ arises (in contrast to the vanishing matching of $\lwc{\theta}{(4)}$ in the theta term) because the kinetic term needs to be defined in $D$ dimensions in order to dimensionally regulate the loop integrals. This coefficient agrees with Eq.~(C.5a) in Ref.~\cite{Harlander:2020duo}, where also the two-loop matching was computed. The gluon two-point function also allows us to extract the matching onto a redundant operator:
\begin{align}
	\label{eq:DeltaLGD}
	\Delta_1(\lwc{GD}{}[][]) &= \frac{3g^2 N_c}{4} \left( 15 - 4 \lt \right) L_G \, .
\end{align}

From the quark two-point function, we obtain the wave-function contributions
\begin{align}
	\label{eq:DeltaZuL}
	\Delta_1\Bigl( Z^{1/2}_{\substack{u,L \\ pr}} \Bigr) &= \frac{1}{2} g^2 C_F \left( 5 + 3 \lt \right) \left( [ L_{uG} M_u ]_{pr} + [ M_u^\dagger L_{uG}^\dagger ]_{pr} \right) \nn
		&\quad - 3 g^4 C_F \delta_{pr} L_G^{(4)} + \frac{1}{4} g^2 C_F (1-2\log(432)) \lwc{uD}{L(4)}[][pr] \, , \\
	\label{eq:DeltaZdL}
	\Delta_1\Bigl( Z^{1/2}_{\substack{d,L \\ pr}} \Bigr) &= \frac{1}{2} g^2 C_F \left( 5 + 3 \lt \right) \left( [ L_{dG} M_d ]_{pr} + [ M_d^\dagger L_{dG}^\dagger ]_{pr} \right) \nn
		&\quad - 3 g^4 C_F \delta_{pr} L_G^{(4)} + \frac{1}{4} g^2 C_F (1-2\log(432)) \lwc{dD}{L(4)}[][pr] \, , \\
	\label{eq:DeltaZuR}
	\Delta_1\Bigl( Z^{1/2}_{\substack{u,R \\ pr}} \Bigr) &= \frac{1}{2} g^2 C_F \left( 5 + 3 \lt \right) \left( [ M_u L_{uG} ]_{pr} + [ L_{uG}^\dagger M_u^\dagger ]_{pr} \right) \nn
		&\quad - 3 g^4 C_F \delta_{pr} L_G^{(4)} + \frac{1}{4} g^2 C_F (1-2\log(432)) \lwc{uD}{R(4)}[][pr] \, , \\
	\label{eq:DeltaZdR}
	\Delta_1\Bigl( Z^{1/2}_{\substack{d,R \\ pr}} \Bigr) &= \frac{1}{2} g^2 C_F \left( 5 + 3 \lt \right) \left( [ M_d L_{dG} ]_{pr} + [ L_{dG}^\dagger M_d^\dagger ]_{pr} \right) \nn
		&\quad - 3 g^4 C_F \delta_{pr} L_G^{(4)} + \frac{1}{4} g^2 C_F (1-2\log(432)) \lwc{dD}{R(4)}[][pr] \, .
\end{align}
Note that there is no lepton wave-function contribution, which could only arise from loops with semileptonic operator insertions, which however are momentum independent. The fermion two-point functions also provide the matching to the mass matrices:
\begin{align}
	\label{eq:DeltaMe}
	\Delta_1([M_e]_{pr}) &= -\frac{N_c}{t} \left( \lwc{eu}{RL\dagger}[S][prvw] [M_u]_{wv} + \lwc{ed}{RL\dagger}[S][prvw] [M_d]_{wv} + \lwc{eu}{RR\dagger}[S][prvw] [M_u^\dagger]_{wv} + \lwc{ed}{RR\dagger}[S][prvw] [M_d^\dagger]_{wv} \right) \nn
		&\quad - \frac{N_c}{3} \left(11 + 6 \lt \right) \begin{aligned}[t]
			& \Big( \lwc{eu}{RL\dagger}[S][prvw] [M_u M_u^\dagger M_u]_{wv} + \lwc{ed}{RL\dagger}[S][prvw] [M_d M_d^\dagger M_d]_{wv} \\
			&+ \lwc{eu}{RR\dagger}[S][prvw] [M_u^\dagger M_u M_u^\dagger]_{wv} + \lwc{ed}{RR\dagger}[S][prvw] [M_d^\dagger M_d M_d^\dagger]_{wv} \Big) \, , \end{aligned} \\
	\label{eq:DeltaMu}
	\Delta_1([M_u]_{pr}) &= -\frac{6g^2C_F}{t} \lwc{uG}{\dagger}[][pr] -\frac{N_c}{t} \left( \lwc{uu}{RR\dagger}[S1][prvw] [M_u^\dagger]_{wv} + \lwc{uu}{RR\dagger}[S1][vwpr] [M_u^\dagger]_{wv} + \lwc{ud}{RR\dagger}[S1][prvw] [M_d^\dagger]_{wv} \right) \nn
		&\quad + \frac{1}{2t} \begin{aligned}[t]
			&\Big( \lwc{uu}{RR\dagger}[S1][pvwr] [M_u^\dagger]_{vw} + \lwc{uu}{RR\dagger}[S1][wrpv] [M_u^\dagger]_{vw} + \lwc{uddu}{RR\dagger}[S1][wrpv] [M_d^\dagger]_{vw} \\
			&+ 4 \lwc{uddu}{LR\dagger}[V1][wrpv] [M_d]_{vw} + 4 \lwc{uu}{LR}[V1][wrpv] [M_u]_{vw} \Big) \end{aligned} \nn
		&\quad - \frac{C_F}{2t} \begin{aligned}[t]
			&\Big( \lwc{uu}{RR\dagger}[S8][pvwr] [M_u^\dagger]_{vw} + \lwc{uu}{RR\dagger}[S8][wrpv] [M_u^\dagger]_{vw} + \lwc{uddu}{RR\dagger}[S8][wrpv] [M_d^\dagger]_{vw} \\
			&+ 4 \lwc{uddu}{LR\dagger}[V8][wrpv] [M_d]_{vw} + 4 \lwc{uu}{LR}[V8][wrpv] [M_u]_{vw} \Big) \end{aligned} \nn
		&\quad - \frac{3}{2} g^2 C_F \left( 9 + 5 \lt \right) \left( [ \lwc{uG}{\dagger}[][] M_u^\dagger M_u ]_{pr} + [  M_u M_u^\dagger\lwc{uG}{\dagger}[][] ]_{pr} \right) \nn
		&\quad - g^2 C_F \left( 5 + 3 \lt \right) [  M_u \lwc{uG}{}[][] M_u ]_{pr} \nn
		&\quad - \frac{N_c}{3} \left( 11 + 6 \lt \right) \begin{aligned}[t]
			&\Big( \lwc{uu}{RR\dagger}[S1][wvpr] [M_u^\dagger M_u M_u^\dagger ]_{vw} + \lwc{uu}{RR\dagger}[S1][prwv] [M_u^\dagger M_u M_u^\dagger ]_{vw} \\
			&+ \lwc{ud}{RR\dagger}[S1][prwv] [M_d^\dagger M_d M_d^\dagger ]_{vw} \Big) \end{aligned} \nn
		&\quad + \frac{1}{6} \left( 11 + 6 \lt \right) \begin{aligned}[t]
			&\Big( \lwc{uu}{RR\dagger}[S1][pvwr] [M_u^\dagger M_u M_u^\dagger ]_{vw} + \lwc{uu}{RR\dagger}[S1][wrpv] [M_u^\dagger M_u M_u^\dagger ]_{vw} \\
			&+ \lwc{uddu}{RR\dagger}[S1][wrpv] [M_d^\dagger M_d M_d^\dagger ]_{vw} + 4 \lwc{uddu}{LR\dagger}[V1][wrpv] [M_d M_d^\dagger M_d ]_{vw} \\
			&+ 4 \lwc{uu}{LR}[V1][wrpv] [M_u M_u^\dagger M_u ]_{vw} \Big) \end{aligned} \nn
		&\quad - \frac{C_F}{6} \left( 11 + 6 \lt \right) \begin{aligned}[t]
			&\Big( \lwc{uu}{RR\dagger}[S8][pvwr] [M_u^\dagger M_u M_u^\dagger ]_{vw} + \lwc{uu}{RR\dagger}[S8][wrpv] [M_u^\dagger M_u M_u^\dagger ]_{vw} \\
			&+ \lwc{uddu}{RR\dagger}[S8][wrpv] [M_d^\dagger M_d M_d^\dagger ]_{vw} + 4 \lwc{uddu}{LR\dagger}[V8][wrpv] [M_d M_d^\dagger M_d ]_{vw} \\
			&+ 4 \lwc{uu}{LR}[V8][wrpv] [M_u M_u^\dagger M_u ]_{vw} \Big) \end{aligned} \nn
		&\quad + 2 g^4 C_F ( 13 + 6 \lt ) [M_u]_{pr} L_G^{(4)} - g^2 C_F \left( 6 + 3 \lt + \log(432) \right) \lwc{Mu}{(3)}[][pr] \nn
		&\quad + \frac{1}{4} g^2 C_F \left( 2 \log(432) - 1 \right) \left( [ \lwc{uD}{R(4)}[][] M_u ]_{pr} + [ M_u \lwc{uD}{L(4)}[][] ]_{pr}  \right)  \, , \\
	\label{eq:DeltaMd}
	\Delta_1([M_d]_{pr}) &= -\frac{6g^2C_F}{t} \lwc{dG}{\dagger}[][pr] -\frac{N_c}{t} \left( \lwc{dd}{RR\dagger}[S1][prvw] [M_d^\dagger]_{wv} + \lwc{dd}{RR\dagger}[S1][vwpr] [M_d^\dagger]_{wv} + \lwc{ud}{RR\dagger}[S1][vwpr] [M_u^\dagger]_{wv} \right) \nn
		&\quad + \frac{1}{2t} \begin{aligned}[t]
			&\Big( \lwc{dd}{RR\dagger}[S1][pvwr] [M_d^\dagger]_{vw} + \lwc{dd}{RR\dagger}[S1][wrpv] [M_d^\dagger]_{vw} + \lwc{uddu}{RR\dagger}[S1][pvwr] [M_u^\dagger]_{vw} \\
			&+ 4 \lwc{uddu}{LR}[V1][wrpv] [M_u]_{vw} + 4 \lwc{dd}{LR}[V1][wrpv] [M_d]_{vw} \Big) \end{aligned} \nn
		&\quad - \frac{C_F}{2t} \begin{aligned}[t]
			&\Big( \lwc{dd}{RR\dagger}[S8][pvwr] [M_d^\dagger]_{vw} + \lwc{dd}{RR\dagger}[S8][wrpv] [M_d^\dagger]_{vw} + \lwc{uddu}{RR\dagger}[S8][pvwr] [M_u^\dagger]_{vw} \\
			&+ 4 \lwc{uddu}{LR}[V8][wrpv] [M_u]_{vw} + 4 \lwc{dd}{LR}[V8][wrpv] [M_d]_{vw} \Big) \end{aligned} \nn
		&\quad - \frac{3}{2} g^2 C_F \left( 9 + 5 \lt \right) \left( [ \lwc{dG}{\dagger}[][] M_d^\dagger M_d ]_{pr} + [  M_d M_d^\dagger\lwc{dG}{\dagger}[][] ]_{pr} \right) \nn
		&\quad - g^2 C_F \left( 5 + 3 \lt \right) [  M_d \lwc{dG}{}[][] M_d ]_{pr} \nn
		&\quad - \frac{N_c}{3} \left( 11 + 6 \lt \right) \begin{aligned}[t]
			&\Big( \lwc{dd}{RR\dagger}[S1][wvpr] [M_d^\dagger M_d M_d^\dagger ]_{vw} + \lwc{dd}{RR\dagger}[S1][prwv] [M_d^\dagger M_d M_d^\dagger ]_{vw} \\
			&+ \lwc{ud}{RR\dagger}[S1][wvpr] [M_u^\dagger M_u M_u^\dagger ]_{vw} \Big) \end{aligned} \nn
		&\quad + \frac{1}{6} \left( 11 + 6 \lt \right) \begin{aligned}[t]
			&\Big( \lwc{dd}{RR\dagger}[S1][pvwr] [M_d^\dagger M_d M_d^\dagger ]_{vw} + \lwc{dd}{RR\dagger}[S1][wrpv] [M_d^\dagger M_d M_d^\dagger ]_{vw} \\
			&+ \lwc{uddu}{RR\dagger}[S1][pvwr] [M_u^\dagger M_u M_u^\dagger ]_{vw} + 4 \lwc{uddu}{LR}[V1][wrpv] [M_u M_u^\dagger M_u ]_{vw} \\
			&+ 4 \lwc{dd}{LR}[V1][wrpv] [M_d M_d^\dagger M_d ]_{vw} \Big) \end{aligned} \nn
		&\quad - \frac{C_F}{6} \left( 11 + 6 \lt \right) \begin{aligned}[t]
			&\Big( \lwc{dd}{RR\dagger}[S8][pvwr] [M_d^\dagger M_d M_d^\dagger ]_{vw} + \lwc{dd}{RR\dagger}[S8][wrpv] [M_d^\dagger M_d M_d^\dagger ]_{vw} \\
			&+ \lwc{uddu}{RR\dagger}[S8][pvwr] [M_u^\dagger M_u M_u^\dagger ]_{vw} + 4 \lwc{uddu}{LR}[V8][wrpv] [M_u M_u^\dagger M_u ]_{vw} \\
			&+ 4 \lwc{dd}{LR}[V8][wrpv] [M_d M_d^\dagger M_d ]_{vw} \Big) \end{aligned} \nn
		&\quad + 2 g^4 C_F ( 13 + 6 \lt ) [M_d]_{pr} L_G^{(4)} - g^2 C_F \left( 6 + 3 \lt + \log(432) \right) \lwc{Md}{(3)}[][pr] \nn
		&\quad + \frac{1}{4} g^2 C_F \left( 2 \log(432) - 1 \right) \left( [ \lwc{dD}{R(4)}[][] M_d ]_{pr} + [ M_d \lwc{dD}{L(4)}[][] ]_{pr}  \right) \, .
\end{align}
In addition, we extract the following matching contributions to redundant operators from the quark two-point functions:
\begin{align}
	\label{eq:DeltaLuD5}
	\Delta_1(\lwc{uD}{(5)}[][pr]) &= -\frac{1}{2} g^2 C_F ( 5 - 12 \lt ) \lwc{uG}{}[][pr] \nn
		&\quad - 2 N_c \Big( \lwc{uu}{RR}[S1][wvpr] [M_u ]_{vw} + \lwc{uu}{RR}[S1][prwv] [M_u ]_{vw} + \lwc{ud}{RR}[S1][prwv] [M_d ]_{vw} \Big) \nn
		&\quad + \begin{aligned}[t]
			&\Big( \lwc{uu}{RR}[S1][pvwr] [M_u ]_{vw} + \lwc{uu}{RR}[S1][wrpv] [M_u ]_{vw} + \lwc{uddu}{RR}[S1][pvwr] [M_d ]_{vw} \\
			&+ 4 \lwc{uddu}{LR}[V1][pvwr] [M_d^\dagger ]_{vw} + 4 \lwc{uu}{LR}[V1][pvwr] [M_u^\dagger ]_{vw} \Big) \end{aligned} \nn
		&\quad - C_F \begin{aligned}[t]
			&\Big( \lwc{uu}{RR}[S8][pvwr] [M_u ]_{vw} + \lwc{uu}{RR}[S8][wrpv] [M_u ]_{vw} + \lwc{uddu}{RR}[S8][pvwr] [M_d ]_{vw} \\
			&+ 4 \lwc{uddu}{LR}[V8][pvwr] [M_d^\dagger ]_{vw} + 4 \lwc{uu}{LR}[V8][pvwr] [M_u^\dagger ]_{vw} \Big)  \, ,\end{aligned} \\
	\label{eq:DeltaLdD5}
	\Delta_1(\lwc{dD}{(5)}[][pr]) &= -\frac{1}{2} g^2 C_F ( 5 - 12 \lt ) \lwc{dG}{}[][pr] \nn
		&\quad - 2 N_c \Big( \lwc{dd}{RR}[S1][wvpr] [M_d ]_{vw} + \lwc{dd}{RR}[S1][prwv] [M_d ]_{vw} + \lwc{ud}{RR}[S1][wvpr] [M_u ]_{vw} \Big) \nn
		&\quad + \begin{aligned}[t]
			&\Big( \lwc{dd}{RR}[S1][pvwr] [M_d ]_{vw} + \lwc{dd}{RR}[S1][wrpv] [M_d ]_{vw} + \lwc{uddu}{RR}[S1][wrpv] [M_u ]_{vw} \\
			&+ 4 \lwc{uddu}{LR\dagger}[V1][pvwr] [M_u^\dagger ]_{vw} + 4 \lwc{dd}{LR}[V1][pvwr] [M_d^\dagger ]_{vw} \Big) \end{aligned} \nn
		&\quad - C_F \begin{aligned}[t]
			&\Big( \lwc{dd}{RR}[S8][pvwr] [M_d ]_{vw} + \lwc{dd}{RR}[S8][wrpv] [M_d ]_{vw} + \lwc{uddu}{RR}[S8][wrpv] [M_u ]_{vw} \\
			&+ 4 \lwc{uddu}{LR\dagger}[V8][pvwr] [M_u^\dagger ]_{vw} + 4 \lwc{dd}{LR}[V8][pvwr] [M_d^\dagger ]_{vw} \Big) \, . \end{aligned}
\end{align}

The fermion--gauge-boson three-point functions lead to the dipole-operator coefficients
\begin{align}
	\label{eq:DeltaLegamma}
	\Delta_1(\lwc{e\gamma}{}[][pr]) &= - 4 N_c \left( 2 + \lt \right) \left( \q_u \lwc{eu}{RR}[T][prwv] [M_u]_{vw} + \q_d \lwc{ed}{RR}[T][prwv] [M_d]_{vw} \right) \, , \\
	\label{eq:DeltaLugamma}
	\Delta_1(\lwc{u\gamma}{}[][pr]) &= g^2 C_F \left( \lt - \log(432) \right) \lwc{u\gamma}{}[][pr] + \frac{1}{2} g^2 C_F \q_u \left( 1 + 8 \lt \right) \lwc{uG}{}[][pr] \nn
		&\quad - N_c \q_u \left( \lwc{uu}{RR}[S1][wvpr] [M_u ]_{vw} + \lwc{uu}{RR}[S1][prwv] [M_u ]_{vw} + \lwc{ud}{RR}[S1][prwv] [M_d ]_{vw} \right) \nn
		&\quad + \frac{1}{2} \left( 3 + \lt \right) \q_u \left( \lwc{uu}{RR}[S1][pvwr] [M_u]_{vw} + \lwc{uu}{RR}[S1][wrpv] [M_u]_{vw} \right) \nn
		&\quad - \frac{1}{2} C_F \left( 3 + \lt \right) \q_u \left( \lwc{uu}{RR}[S8][pvwr] [M_u]_{vw} + \lwc{uu}{RR}[S8][wrpv] [M_u]_{vw} \right) \nn
		&\quad + \frac{1}{2} \left( (2+\lt) \q_d + \q_u \right) \left( \lwc{uddu}{RR}[S1][pvwr] [M_d]_{vw} - C_F \lwc{uddu}{RR}[S8][pvwr] [M_d]_{vw} \right) \nn
		&\quad + 2 \q_u \left( \lwc{uu}{LR}[V1][pvwr] [M_u^\dagger]_{vw} + \lwc{uddu}{LR}[V1][pvwr] [M_d^\dagger]_{vw} \right) \nn
		&\quad - 2 C_F \q_u \left( \lwc{uu}{LR}[V8][pvwr] [M_u^\dagger]_{vw} + \lwc{uddu}{LR}[V8][pvwr] [M_d^\dagger]_{vw} \right) \, , \\
	\label{eq:DeltaLdgamma}
	\Delta_1(\lwc{d\gamma}{}[][pr]) &= g^2 C_F \left( \lt - \log(432) \right) \lwc{d\gamma}{}[][pr] + \frac{1}{2} g^2 C_F \q_d \left( 1 + 8 \lt \right) \lwc{dG}{}[][pr] \nn
		&\quad - N_c \q_d \left( \lwc{dd}{RR}[S1][wvpr] [M_d ]_{vw} + \lwc{dd}{RR}[S1][prwv] [M_d ]_{vw} + \lwc{ud}{RR}[S1][wvpr] [M_u ]_{vw} \right) \nn
		&\quad + \frac{1}{2} \left( 3 + \lt \right) \q_d \left( \lwc{dd}{RR}[S1][pvwr] [M_d]_{vw} + \lwc{dd}{RR}[S1][wrpv] [M_d]_{vw} \right) \nn
		&\quad - \frac{1}{2} C_F \left( 3 + \lt \right) \q_d \left( \lwc{dd}{RR}[S8][pvwr] [M_d]_{vw} + \lwc{dd}{RR}[S8][wrpv] [M_d]_{vw} \right) \nn
		&\quad + \frac{1}{2} \left( (2+\lt) \q_u + \q_d \right) \left( \lwc{uddu}{RR}[S1][wrpv] [M_u]_{vw} - C_F \lwc{uddu}{RR}[S8][wrpv] [M_u]_{vw} \right) \nn
		&\quad + 2 \q_d \left( \lwc{dd}{LR}[V1][pvwr] [M_d^\dagger]_{vw} + \lwc{uddu}{LR\dagger}[V1][pvwr] [M_u^\dagger]_{vw} \right) \nn
		&\quad - 2 C_F \q_d \left( \lwc{dd}{LR}[V8][pvwr] [M_d^\dagger]_{vw} + \lwc{uddu}{LR\dagger}[V8][pvwr] [M_u^\dagger]_{vw} \right) \, , \\
	\label{eq:DeltaLuG}
	\Delta_1(\lwc{uG}{}[][pr]) &= \frac{1}{2} g^2 \left( C_F \left( 1 + 10 \lt - 2 \log(432) \right) - N_c \left( 9 + 4 \lt \right) \right) \lwc{uG}{}[][pr] \nn
		&\quad - N_c \left( \lwc{uu}{RR}[S1][wvpr] [M_u ]_{vw} + \lwc{uu}{RR}[S1][prwv] [M_u ]_{vw} + \lwc{ud}{RR}[S1][prwv] [M_d ]_{vw} \right) \nn
		&\quad + \frac{1}{2} \left( 3 + \lt \right) \left( \lwc{uu}{RR}[S1][pvwr] [M_u]_{vw} + \lwc{uu}{RR}[S1][wrpv] [M_u]_{vw} + \lwc{uddu}{RR}[S1][pvwr] [M_d]_{vw} \right) \nn
		&\quad - \left( \frac{C_F}{2} \left( 3 + \lt \right) - \frac{N_c}{4} \left( 2 + \lt \right) \right) \left( \lwc{uu}{RR}[S8][pvwr] [M_u]_{vw} + \lwc{uu}{RR}[S8][wrpv] [M_u]_{vw} + \lwc{uddu}{RR}[S8][pvwr] [M_d]_{vw} \right) \nn
		&\quad + 2 \left( \lwc{uu}{LR}[V1][pvwr] [M_u^\dagger]_{vw} + \lwc{uddu}{LR}[V1][pvwr] [M_d^\dagger]_{vw} \right) - 2 C_F \left( \lwc{uu}{LR}[V8][pvwr] [M_u^\dagger]_{vw} + \lwc{uddu}{LR}[V8][pvwr] [M_d^\dagger]_{vw} \right) \nn
		&\quad + \frac{N_c}{4} g^4 \left( 11 + 6 \lt \right) [M_u^\dagger]_{pr} \left( L_G - i L_{\widetilde G} \right)  \, , \\
	\label{eq:DeltaLdG}
	\Delta_1(\lwc{dG}{}[][pr]) &= \frac{1}{2} g^2 \left( C_F \left( 1 + 10 \lt - 2 \log(432) \right) - N_c \left( 9 + 4 \lt \right) \right) \lwc{dG}{}[][pr] \nn
		&\quad - N_c \left( \lwc{dd}{RR}[S1][wvpr] [M_d ]_{vw} + \lwc{dd}{RR}[S1][prwv] [M_d ]_{vw} + \lwc{ud}{RR}[S1][wvpr] [M_u ]_{vw} \right) \nn
		&\quad + \frac{1}{2} \left( 3 + \lt \right) \left( \lwc{dd}{RR}[S1][pvwr] [M_d]_{vw} + \lwc{dd}{RR}[S1][wrpv] [M_d]_{vw} + \lwc{uddu}{RR}[S1][wrpv] [M_u]_{vw} \right) \nn
		&\quad - \left( \frac{C_F}{2} \left( 3 + \lt \right) - \frac{N_c}{4} \left( 2 + \lt \right) \right) \left( \lwc{dd}{RR}[S8][pvwr] [M_d]_{vw} + \lwc{dd}{RR}[S8][wrpv] [M_d]_{vw} + \lwc{uddu}{RR}[S8][wrpv] [M_u]_{vw} \right) \nn
		&\quad + 2 \left( \lwc{dd}{LR}[V1][pvwr] [M_d^\dagger]_{vw} + \lwc{uddu}{LR\dagger}[V1][pvwr] [M_u^\dagger]_{vw} \right) - 2 C_F \left( \lwc{dd}{LR}[V8][pvwr] [M_d^\dagger]_{vw} + \lwc{uddu}{LR\dagger}[V8][pvwr] [M_u^\dagger]_{vw} \right) \nn
		&\quad + \frac{N_c}{4} g^4 \left( 11 + 6 \lt \right) [M_d^\dagger]_{pr} \left( L_G - i L_{\widetilde G} \right)  \, ,
\end{align}
as well as further redundant-operator contributions
\begin{align}
	\label{eq:DeltaLnugammaDL}
	\Delta_1(\lwc{\nu\gamma D}{L}[][pr]) &= \frac{N_c}{12} \left( 3 + 8 \lt \right) \left( \q_u \left( \lwc{\nu u}{LL}[V][prvv] + \lwc{\nu u}{LR}[V][prvv] \right) + \q_d \left( \lwc{\nu d}{LL}[V][prvv] + \lwc{\nu d}{LR}[V][prvv] \right) \right) \, , \\
	\label{eq:DeltaLegammaDL}
	\Delta_1(\lwc{e\gamma D}{L}[][pr]) &= \frac{N_c}{12} \left( 3 + 8 \lt \right) \left( \q_u \left( \lwc{eu}{LL}[V][prvv] + \lwc{eu}{LR}[V][prvv] \right) + \q_d \left( \lwc{ed}{LL}[V][prvv] + \lwc{ed}{LR}[V][prvv] \right) \right) \, , \\
	\label{eq:DeltaLegammaDR}
	\Delta_1(\lwc{e\gamma D}{R}[][pr]) &= \frac{N_c}{12} \left( 3 + 8 \lt \right) \left( \q_u \left( \lwc{eu}{RR}[V][prvv] + \lwc{ue}{LR}[V][vvpr] \right) + \q_d \left( \lwc{ed}{RR}[V][prvv] + \lwc{de}{LR}[V][vvpr] \right) \right) \, , \\
	\label{eq:DeltaLugammaDL}
	\Delta_1(\lwc{u\gamma D}{L}[][pr]) &= \frac{1}{12} \left( 3 + 8 \lt \right) \q_u \left( \lwc{uu}{LL}[V][pvvr] + \lwc{uu}{LL}[V][vrpv] \right) \nn
		&\quad + \frac{N_c}{12} \left( 3 + 8 \lt \right) \left(  \q_u \left( \lwc{uu}{LL}[V][prvv] + \lwc{uu}{LL}[V][vvpr] + \lwc{uu}{LR}[V1][prvv] \right) + \q_d \left( \lwc{ud}{LL}[V1][prvv] + \lwc{ud}{LR}[V1][prvv] \right) \right) \, , \\
	\label{eq:DeltaLugammaDR}
	\Delta_1(\lwc{u\gamma D}{R}[][pr]) &= \frac{1}{12} \left( 3 + 8 \lt \right) \q_u \left( \lwc{uu}{RR}[V][pvvr] + \lwc{uu}{RR}[V][vrpv] \right) \nn
		&\quad + \frac{N_c}{12} \left( 3 + 8 \lt \right) \left(  \q_u \left( \lwc{uu}{RR}[V][prvv] + \lwc{uu}{RR}[V][vvpr] + \lwc{uu}{LR}[V1][vvpr] \right) + \q_d \left( \lwc{du}{LR}[V1][vvpr] + \lwc{ud}{RR}[V1][prvv] \right) \right) \, , \\
	\label{eq:DeltaLdgammaDL}
	\Delta_1(\lwc{d\gamma D}{L}[][pr]) &= \frac{1}{12} \left( 3 + 8 \lt \right) \q_d \left( \lwc{dd}{LL}[V][pvvr] + \lwc{dd}{LL}[V][vrpv] \right) \nn
		&\quad + \frac{N_c}{12} \left( 3 + 8 \lt \right) \left(  \q_d \left( \lwc{dd}{LL}[V][prvv] + \lwc{dd}{LL}[V][vvpr] + \lwc{dd}{LR}[V1][prvv] \right) + \q_u \left( \lwc{ud}{LL}[V1][vvpr] + \lwc{du}{LR}[V1][prvv] \right) \right) \, , \\
	\label{eq:DeltaLdgammaDR}
	\Delta_1(\lwc{d\gamma D}{R}[][pr]) &= \frac{1}{12} \left( 3 + 8 \lt \right) \q_d \left( \lwc{dd}{RR}[V][pvvr] + \lwc{dd}{RR}[V][vrpv] \right) \nn
		&\quad + \frac{N_c}{12} \left( 3 + 8 \lt \right) \left(  \q_d \left( \lwc{dd}{RR}[V][prvv] + \lwc{dd}{RR}[V][vvpr] + \lwc{dd}{LR}[V1][vvpr] \right) + \q_u \left( \lwc{ud}{LR}[V1][vvpr] + \lwc{ud}{RR}[V1][vvpr] \right) \right) \, , \\
	\label{eq:DeltaLDuGL}
	\Delta_1(\lwc{DuG}{L}[][pr]) &= \frac{N_c}{4} g^4 \left( L_G + i L_{\widetilde G} \right) \delta_{pr} \, , \\
	\label{eq:DeltaLDuGR}
	\Delta_1(\lwc{DuG}{R}[][pr]) &= \frac{N_c}{4} g^4 \left( L_G - i L_{\widetilde G} \right) \delta_{pr} \, , \\
	\label{eq:DeltaLuDGL}
	\Delta_1(\lwc{uDG}{L}[][pr]) &= - \frac{N_c}{4} g^4 \left( L_G - i L_{\widetilde G} \right) \delta_{pr} \, , \\
	\label{eq:DeltaLuDGR}
	\Delta_1(\lwc{uDG}{R}[][pr]) &= - \frac{N_c}{4} g^4 \left( L_G + i L_{\widetilde G} \right) \delta_{pr} \, , \\
	\label{eq:DeltaLuGDL}
	\Delta_1(\lwc{uGD}{L}[][pr]) &= \frac{1}{12} \left( 3 + 4 \lt \right) \left( \lwc{uu}{LR}[V8][prvv] + \lwc{ud}{LL}[V8][prvv] + \lwc{ud}{LR}[V8][prvv] - 2 \lwc{uu}{LL}[V][pvvr] - 2 \lwc{uu}{LL}[V][vrpv] \right) \nn
		&\quad + N_c g^4 \left( 4 + 3 \lt \right) L_G \, \delta_{pr} \, , \\
	\label{eq:DeltaLuGDR}
	\Delta_1(\lwc{uGD}{R}[][pr]) &= \frac{1}{12} \left( 3 + 4 \lt \right) \left( \lwc{uu}{LR}[V8][vvpr] + \lwc{ud}{RR}[V8][prvv] + \lwc{du}{LR}[V8][vvpr] - 2 \lwc{uu}{RR}[V][pvvr] - 2 \lwc{uu}{RR}[V][vrpv] \right) \nn
		&\quad + N_c g^4 \left( 4 + 3 \lt \right) L_G \, \delta_{pr} \, , \\
	\label{eq:DeltaLDdGL}
	\Delta_1(\lwc{DdG}{L}[][pr]) &= \frac{N_c}{4} g^4 \left( L_G + i L_{\widetilde G} \right) \delta_{pr} \, , \\
	\label{eq:DeltaLDdGR}
	\Delta_1(\lwc{DdG}{R}[][pr]) &= \frac{N_c}{4} g^4 \left( L_G - i L_{\widetilde G} \right) \delta_{pr} \, , \\
	\label{eq:DeltaLdDGL}
	\Delta_1(\lwc{dDG}{L}[][pr]) &= - \frac{N_c}{4} g^4 \left( L_G - i L_{\widetilde G} \right) \delta_{pr} \, , \\
	\label{eq:DeltaLdDGR}
	\Delta_1(\lwc{dDG}{R}[][pr]) &= - \frac{N_c}{4} g^4 \left( L_G + i L_{\widetilde G} \right) \delta_{pr} \, , \\
	\label{eq:DeltaLdGDL}
	\Delta_1(\lwc{dGD}{L}[][pr]) &= \frac{1}{12} \left( 3 + 4 \lt \right) \left( \lwc{dd}{LR}[V8][prvv] + \lwc{ud}{LL}[V8][vvpr] + \lwc{du}{LR}[V8][prvv] - 2 \lwc{dd}{LL}[V][pvvr] - 2 \lwc{dd}{LL}[V][vrpv] \right) \nn
		&\quad + N_c g^4 \left( 4 + 3 \lt \right) L_G \, \delta_{pr} \, , \\
	\label{eq:DeltaLdGDR}
	\Delta_1(\lwc{dGD}{R}[][pr]) &= \frac{1}{12} \left( 3 + 4 \lt \right) \left( \lwc{dd}{LR}[V8][vvpr] + \lwc{ud}{RR}[V8][vvpr] + \lwc{ud}{LR}[V8][vvpr] - 2 \lwc{dd}{RR}[V][pvvr] - 2 \lwc{dd}{RR}[V][vrpv] \right) \nn
		&\quad + N_c g^4 \left( 4 + 3 \lt \right) L_G \, \delta_{pr} \, .
\end{align}

From the three-gluon Green's function, we obtain the matching of the three-gluon operators:
\begin{align}
	\label{eq:DeltaLG}
	\Delta_1(\lwc{G}{}[][]) &= 6 N_c g^2 \lt \lwc{G}{}[][] \, , \\
	\label{eq:DeltaLGt}
	\Delta_1(\lwc{\widetilde G}{}[][]) &= 6 N_c g^2 \lt \lwc{\widetilde G}{}[][] \, .
\end{align}

Finally, the four-fermion contributions can be split into the coefficients of semileptonic operators,
\begin{align}
	\label{eq:DeltaLVLLnuu}
	\Delta_1(\lwc{\nu u}{LL}[V][prst]) &= -\frac{C_F}{2} g^2 \left( 3 + 2 \log(432) \right) \lwc{\nu u}{LL}[V][prst] \, , \\
	\label{eq:DeltaLVLRnuu}
	\Delta_1(\lwc{\nu u}{LR}[V][prst]) &= -\frac{C_F}{2} g^2 \left( 3 + 2 \log(432) \right) \lwc{\nu u}{LR}[V][prst] \, , \\
	\label{eq:DeltaLVLLnud}
	\Delta_1(\lwc{\nu d}{LL}[V][prst]) &= -\frac{C_F}{2} g^2 \left( 3 + 2 \log(432) \right) \lwc{\nu d}{LL}[V][prst] \, , \\
	\label{eq:DeltaLVLRnud}
	\Delta_1(\lwc{\nu d}{LR}[V][prst]) &= -\frac{C_F}{2} g^2 \left( 3 + 2 \log(432) \right) \lwc{\nu d}{LR}[V][prst] \, , \\
	\label{eq:DeltaLVLLnuedu}
	\Delta_1(\lwc{\nu edu}{LL}[V][prst]) &= -\frac{C_F}{2} g^2 \left( 3 + 2 \log(432) \right) \lwc{\nu edu}{LL}[V][prst] \, , \\
	\label{eq:DeltaLVLRnuedu}
	\Delta_1(\lwc{\nu edu}{LR}[V][prst]) &= -\frac{C_F}{2} g^2 \left( 3 + 2 \log(432) \right) \lwc{\nu edu}{LR}[V][prst] \, , \\
	\label{eq:DeltaLSRRnuedu}
	\Delta_1(\lwc{\nu edu}{RR}[S][prst]) &= -C_F g^2 \left( 6 + 3 \lt + \log(432) \right) \lwc{\nu edu}{RR}[S][prst] \, , \\
	\label{eq:DeltaLTRRnuedu}
	\Delta_1(\lwc{\nu edu}{RR}[T][prst]) &= C_F g^2 \left( \lt - \log(432) \right) \lwc{\nu edu}{RR}[T][prst] \, , \\
	\label{eq:DeltaLSRLnuedu}
	\Delta_1(\lwc{\nu edu}{RL}[S][prst]) &= -C_F g^2 \left( 6 + 3 \lt + \log(432) \right) \lwc{\nu edu}{RL}[S][prst] \, , \\
	\label{eq:DeltaLVLLeu}
	\Delta_1(\lwc{eu}{LL}[V][prst]) &= -\frac{C_F}{2} g^2 \left( 3 + 2 \log(432) \right) \lwc{eu}{LL}[V][prst] \, , \\
	\label{eq:DeltaLVLReu}
	\Delta_1(\lwc{eu}{RR}[V][prst]) &= -\frac{C_F}{2} g^2 \left( 3 + 2 \log(432) \right) \lwc{eu}{RR}[V][prst] \, , \\
	\label{eq:DeltaLVLReu}
	\Delta_1(\lwc{eu}{LR}[V][prst]) &= -\frac{C_F}{2} g^2 \left( 3 + 2 \log(432) \right) \lwc{eu}{LR}[V][prst] \, , \\
	\label{eq:DeltaLVLRue}
	\Delta_1(\lwc{ue}{LR}[V][prst]) &= -\frac{C_F}{2} g^2 \left( 3 + 2 \log(432) \right) \lwc{ue}{LR}[V][prst] \, , \\
	\label{eq:DeltaLSRReu}
	\Delta_1(\lwc{eu}{RR}[S][prst]) &= -C_F g^2 \left( 6 + 3 \lt + \log(432) \right) \lwc{eu}{RR}[S][prst] \, , \\
	\label{eq:DeltaLTRReu}
	\Delta_1(\lwc{eu}{RR}[T][prst]) &= C_F g^2 \left( \lt - \log(432) \right) \lwc{eu}{RR}[T][prst] \, , \\
	\label{eq:DeltaLSRLeu}
	\Delta_1(\lwc{eu}{RL}[S][prst]) &= -C_F g^2 \left( 6 + 3 \lt + \log(432) \right) \lwc{eu}{RL}[S][prst] \, , \\
	\label{eq:DeltaLVLLed}
	\Delta_1(\lwc{ed}{LL}[V][prst]) &= -\frac{C_F}{2} g^2 \left( 3 + 2 \log(432) \right) \lwc{ed}{LL}[V][prst] \, , \\
	\label{eq:DeltaLVLRed}
	\Delta_1(\lwc{ed}{RR}[V][prst]) &= -\frac{C_F}{2} g^2 \left( 3 + 2 \log(432) \right) \lwc{ed}{RR}[V][prst] \, , \\
	\label{eq:DeltaLVLRed}
	\Delta_1(\lwc{ed}{LR}[V][prst]) &= -\frac{C_F}{2} g^2 \left( 3 + 2 \log(432) \right) \lwc{ed}{LR}[V][prst] \, , \\
	\label{eq:DeltaLVLRde}
	\Delta_1(\lwc{de}{LR}[V][prst]) &= -\frac{C_F}{2} g^2 \left( 3 + 2 \log(432) \right) \lwc{de}{LR}[V][prst] \, , \\
	\label{eq:DeltaLSRRed}
	\Delta_1(\lwc{ed}{RR}[S][prst]) &= -C_F g^2 \left( 6 + 3 \lt + \log(432) \right) \lwc{ed}{RR}[S][prst] \, , \\
	\label{eq:DeltaLTRRed}
	\Delta_1(\lwc{ed}{RR}[T][prst]) &= C_F g^2 \left( \lt - \log(432) \right) \lwc{ed}{RR}[T][prst] \, , \\
	\label{eq:DeltaLSRLed}
	\Delta_1(\lwc{ed}{RL}[S][prst]) &= -C_F g^2 \left( 6 + 3 \lt + \log(432) \right) \lwc{ed}{RL}[S][prst] \, .
\end{align}
as well as four-quark-operator coefficients,
\begin{align}
	% uuuu
	\label{eq:DeltaLS1RRuu}
	\Delta_1(\lwc{uu}{RR}[S1][prst]) &= - 2 g^2 C_F \left( 6 + 3\lt + \log(432) \right) \lwc{uu}{RR}[S1][prst] + \frac{4C_F}{N_c} g^2 \left( 3 + 2\lt \right) \lwc{uu}{RR}[S1][ptsr] \nn
		&\quad - \frac{2C_F}{N_c} g^2 \left( 2 + \lt \right) \lwc{uu}{RR}[S8][prst] - \frac{C_F (N_c^2-2)}{N_c^2} g^2 \left( 3 + 2\lt \right) \lwc{uu}{RR}[S8][ptsr] \, , \\
	\label{eq:DeltaLS8RRuu}
	\Delta_1(\lwc{uu}{RR}[S8][prst]) &= - 4 g^2 \left( 2 + \lt \right) \lwc{uu}{RR}[S1][prst] + \frac{4}{N_c} g^2 \left( 3 + 2\lt \right) \lwc{uu}{RR}[S1][ptsr] \nn
		&\quad + g^2 \left( 2 C_F \left( 2 + \lt - \log(432) \right) - N_c \right) \lwc{uu}{RR}[S8][prst] \nn
		&\quad + \left( \frac{2C_F}{N_c} + \frac{3}{N_c^2} \right) g^2 \left( 3 + 2\lt \right) \lwc{uu}{RR}[S8][ptsr] \, , \\
	\label{eq:DeltaLVLLuu}
	\Delta_1(\lwc{uu}{LL}[V][prst]) &= - g^2 \left( C_F \left( 3 + 2\log(432) \right) + \frac{1}{2N_c} \left( 11 + 6\lt \right) \right) \lwc{uu}{LL}[V][prst] + \frac{1}{2} g^2 \left( 11 + 6\lt \right) \lwc{uu}{LL}[V][ptsr] \, , \\
	\label{eq:DeltaLVRRuu}
	\Delta_1(\lwc{uu}{RR}[V][prst]) &= - g^2 \left( C_F \left( 3 + 2\log(432) \right) + \frac{1}{2N_c} \left( 11 + 6\lt \right) \right) \lwc{uu}{RR}[V][prst] + \frac{1}{2} g^2 \left( 11 + 6\lt \right) \lwc{uu}{RR}[V][ptsr] \, , \\
	\label{eq:DeltaLV1LRuu}
	\Delta_1(\lwc{uu}{LR}[V1][prst]) &= - C_F g^2 \left( 3 + 2\log(432) \right) \lwc{uu}{LR}[V1][prst] + \frac{C_F}{2N_c} g^2 \left( 7 + 6\lt \right) \lwc{uu}{LR}[V8][prst] \, , \\
	\label{eq:DeltaLV8LRuu}
	\Delta_1(\lwc{uu}{LR}[V8][prst]) &= g^2 \left( 7 + 6\lt \right) \lwc{uu}{LR}[V1][prst] + \frac{g^2}{2} \left( N_c \left( 7 + 6\lt \right) - 2 C_F\left( 17 + 12\lt + 2 \log(432) \right) \right) \lwc{uu}{LR}[V8][prst] \, , \\
	% dddd
	\label{eq:DeltaLS1RRdd}
	\Delta_1(\lwc{dd}{RR}[S1][prst]) &= - 2 g^2 C_F \left( 6 + 3\lt + \log(432) \right) \lwc{dd}{RR}[S1][prst] + \frac{4C_F}{N_c} g^2 \left( 3 + 2\lt \right) \lwc{dd}{RR}[S1][ptsr] \nn
		&\quad - \frac{2C_F}{N_c} g^2 \left( 2 + \lt \right) \lwc{dd}{RR}[S8][prst] - \frac{C_F (N_c^2-2)}{N_c^2} g^2 \left( 3 + 2\lt \right) \lwc{dd}{RR}[S8][ptsr] \, , \\
	\label{eq:DeltaLS8RRdd}
	\Delta_1(\lwc{dd}{RR}[S8][prst]) &= - 4 g^2 \left( 2 + \lt \right) \lwc{dd}{RR}[S1][prst] + \frac{4}{N_c} g^2 \left( 3 + 2\lt \right) \lwc{dd}{RR}[S1][ptsr] \nn
		&\quad + g^2 \left( 2 C_F \left( 2 + \lt - \log(432) \right) - N_c \right) \lwc{dd}{RR}[S8][prst] \nn
		&\quad + \left( \frac{2C_F}{N_c} + \frac{3}{N_c^2} \right) g^2 \left( 3 + 2\lt \right) \lwc{dd}{RR}[S8][ptsr] \, , \\
	\label{eq:DeltaLVLLdd}
	\Delta_1(\lwc{dd}{LL}[V][prst]) &= - g^2 \left( C_F \left( 3 + 2\log(432) \right) + \frac{1}{2N_c} \left( 11 + 6\lt \right) \right) \lwc{dd}{LL}[V][prst] + \frac{1}{2} g^2 \left( 11 + 6\lt \right) \lwc{dd}{LL}[V][ptsr] \, , \\
	\label{eq:DeltaLVRRdd}
	\Delta_1(\lwc{dd}{RR}[V][prst]) &= - g^2 \left( C_F \left( 3 + 2\log(432) \right) + \frac{1}{2N_c} \left( 11 + 6\lt \right) \right) \lwc{dd}{RR}[V][prst] + \frac{1}{2} g^2 \left( 11 + 6\lt \right) \lwc{dd}{RR}[V][ptsr] \, , \\
	\label{eq:DeltaLV1LRdd}
	\Delta_1(\lwc{dd}{LR}[V1][prst]) &= - C_F g^2 \left( 3 + 2\log(432) \right) \lwc{dd}{LR}[V1][prst] + \frac{C_F}{2N_c} g^2 \left( 7 + 6\lt \right) \lwc{dd}{LR}[V8][prst] \, , \\
	\label{eq:DeltaLV8LRdd}
	\Delta_1(\lwc{dd}{LR}[V8][prst]) &= g^2 \left( 7 + 6\lt \right) \lwc{dd}{LR}[V1][prst] + \frac{g^2}{2} \left( N_c \left( 7 + 6\lt \right) - 2 C_F\left( 17 + 12\lt + 2 \log(432) \right) \right) \lwc{dd}{LR}[V8][prst] \, , \\
	% uudd
	\label{eq:DeltaLS1RRud}
	\Delta_1(\lwc{ud}{RR}[S1][prst]) &= - 2 g^2 C_F \left( 6 + 3\lt + \log(432) \right) \lwc{ud}{RR}[S1][prst] + \frac{4C_F}{N_c} g^2 \left( 3 + 2\lt \right) \lwc{uddu}{RR}[S1][ptsr] \nn
		&\quad - \frac{2C_F}{N_c} g^2 \left( 2 + \lt \right) \lwc{ud}{RR}[S8][prst] - \frac{C_F (N_c^2-2)}{N_c^2} g^2 \left( 3 + 2\lt \right) \lwc{uddu}{RR}[S8][ptsr] \, , \\
	\label{eq:DeltaLS1RRuddu}
	\Delta_1(\lwc{uddu}{RR}[S1][prst]) &= - 2 g^2 C_F \left( 6 + 3\lt + \log(432) \right) \lwc{uddu}{RR}[S1][prst] + \frac{4C_F}{N_c} g^2 \left( 3 + 2\lt \right) \lwc{ud}{RR}[S1][ptsr] \nn
		&\quad - \frac{2C_F}{N_c} g^2 \left( 2 + \lt \right) \lwc{uddu}{RR}[S8][prst] - \frac{C_F (N_c^2-2)}{N_c^2} g^2 \left( 3 + 2\lt \right) \lwc{ud}{RR}[S8][ptsr] \, , \\
	\label{eq:DeltaLS8RRud}
	\Delta_1(\lwc{ud}{RR}[S8][prst]) &= - 4 g^2 \left( 2 + \lt \right) \lwc{ud}{RR}[S1][prst] + \frac{4}{N_c} g^2 \left( 3 + 2\lt \right) \lwc{uddu}{RR}[S1][ptsr] \nn
		&\quad + g^2 \left( 2C_F \left( 2 + \lt - \log(432) \right) - N_c \right) \lwc{ud}{RR}[S8][prst] \nn
		&\quad + \left( \frac{2C_F}{N_c} + \frac{3}{N_c^2} \right) g^2 \left( 3 + 2\lt \right) \lwc{uddu}{RR}[S8][ptsr] \, , \\
	\label{eq:DeltaLS8RRuddu}
	\Delta_1(\lwc{uddu}{RR}[S8][prst]) &= - 4 g^2 \left( 2 + \lt \right) \lwc{uddu}{RR}[S1][prst] + \frac{4}{N_c} g^2 \left( 3 + 2\lt \right) \lwc{ud}{RR}[S1][ptsr] \nn
		&\quad + g^2 \left( 2C_F \left( 2 + \lt - \log(432) \right) - N_c \right) \lwc{uddu}{RR}[S8][prst] \nn
		&\quad + \left( \frac{2C_F}{N_c} + \frac{3}{N_c^2} \right) g^2 \left( 3 + 2\lt \right) \lwc{ud}{RR}[S8][ptsr] \, , \\
	\label{eq:DeltaLV1LLud}
	\Delta_1(\lwc{ud}{LL}[V1][prst]) &= - C_F g^2 \left( 3 + 2\log(432) \right) \lwc{ud}{LL}[V1][prst] - \frac{C_F}{2N_c} g^2 \left( 11 + 6\lt \right) \lwc{ud}{LL}[V8][prst] \, , \\
	\label{eq:DeltaLV1RRud}
	\Delta_1(\lwc{ud}{RR}[V1][prst]) &= - C_F g^2 \left( 3 + 2\log(432) \right) \lwc{ud}{RR}[V1][prst] - \frac{C_F}{2N_c} g^2 \left( 11 + 6\lt \right) \lwc{ud}{RR}[V8][prst] \, , \\
	\label{eq:DeltaLV1LRud}
	\Delta_1(\lwc{ud}{LR}[V1][prst]) &= - C_F g^2 \left( 3 + 2\log(432) \right) \lwc{ud}{LR}[V1][prst] + \frac{C_F}{2N_c} g^2 \left( 7 + 6\lt \right) \lwc{ud}{LR}[V8][prst] \, , \\
	\label{eq:DeltaLV1LRdu}
	\Delta_1(\lwc{du}{LR}[V1][prst]) &= - C_F g^2 \left( 3 + 2\log(432) \right) \lwc{du}{LR}[V1][prst] + \frac{C_F}{2N_c} g^2 \left( 7 + 6\lt \right) \lwc{du}{LR}[V8][prst] \, , \\
	\label{eq:DeltaLV1LRuddu}
	\Delta_1(\lwc{uddu}{LR}[V1][prst]) &= - C_F g^2 \left( 3 + 2\log(432) \right) \lwc{uddu}{LR}[V1][prst] + \frac{C_F}{2N_c} g^2 \left( 7 + 6\lt \right) \lwc{uddu}{LR}[V8][prst] \, , \\
	\label{eq:DeltaLV8LLud}
	\Delta_1(\lwc{ud}{LL}[V8][prst]) &= -g^2 \left( 11 + 6\lt \right) \lwc{ud}{LL}[V1][prst] - g^2 \left( 2N_c \left( 5 + 3\lt \right) - C_F\left( 19 + 12\lt - 2 \log(432) \right) \right) \lwc{ud}{LL}[V8][prst] \, , \\
	\label{eq:DeltaLV8RRud}
	\Delta_1(\lwc{ud}{RR}[V8][prst]) &= -g^2 \left( 11 + 6\lt \right) \lwc{ud}{RR}[V1][prst] - g^2 \left( 2N_c \left( 5 + 3\lt \right) - C_F\left( 19 + 12\lt - 2 \log(432) \right) \right) \lwc{ud}{RR}[V8][prst] \, , \\
	\label{eq:DeltaLV8LRud}
	\Delta_1(\lwc{ud}{LR}[V8][prst]) &= g^2 \left( 7 + 6\lt \right) \lwc{ud}{LR}[V1][prst] + \frac{g^2}{2} \left( N_c \left( 7 + 6\lt \right) - 2 C_F\left( 17 + 12\lt + 2 \log(432) \right) \right) \lwc{ud}{LR}[V8][prst] \, , \\
	\label{eq:DeltaLV8LRdu}
	\Delta_1(\lwc{du}{LR}[V8][prst]) &= g^2 \left( 7 + 6\lt \right) \lwc{du}{LR}[V1][prst] + \frac{g^2}{2} \left( N_c \left( 7 + 6\lt \right) - 2 C_F\left( 17 + 12\lt + 2 \log(432) \right) \right) \lwc{du}{LR}[V8][prst] \, , \\
	\label{eq:DeltaLV8LRuddu}
	\Delta_1(\lwc{uddu}{LR}[V8][prst]) &= g^2 \left( 7 + 6\lt \right) \lwc{uddu}{LR}[V1][prst] + \frac{g^2}{2} \left( N_c \left( 7 + 6\lt \right) - 2 C_F\left( 17 + 12\lt + 2 \log(432) \right) \right) \lwc{uddu}{LR}[V8][prst] \, .
\end{align}
Further operator coefficients not listed explicitly do not receive a one-loop matching contribution. (As stated before, here we disregard lepton- and baryon-number-violating operators as well as QED corrections.)

\subsection{After field redefinitions}
\label{sec:OnShellResults}

Field redefinitions are used to remove redundant operators from the LEFT basis. In the basis of Ref.~\cite{Naterop:2023dek}, the redundant operators do not vanish by the equations of motion, but they are related to the physical operators. Therefore, the field redefinitions do not only remove the redundant operators, but they induce a shift in the coefficients of the physical operators.

The cosmological constant, the gauge couplings and theta terms, as well as the three-gluon operator coefficients are unaffected by the field redefinitions. The mass matrices, dipole-operator coefficients, and four-fermion-operator coefficients however are shifted. If we neglect operators that do not receive a matching contribution at one loop and disregard $\O(\alpha_\mathrm{QED})$ corrections, the relevant shifts can be expressed as follows. The quark masses are shifted as
\begin{equation}
	\Delta_1(M_q) \mapsto \Delta_1(M_q) + \frac{1}{4}\left(M_q M_q^\dagger \Delta_1(L_{q D}^{(5)\dagger}) + 2 M_q \Delta_1(L_{q D}^{(5)}) M_q + \Delta_1(L_{q D}^{(5)\dagger}) M_q^\dagger M_q \right)
\end{equation}
with $q \in \{u,d\}$, whereas the quark chromo-dipole-operator coefficients receive the shift
\begin{equation}
	\Delta_1(\lwc{qG}{}[][]) \mapsto \Delta_1(\lwc{qG}{}[][]) + M_q^\dagger \Delta_1(\lwc{DqG}{R}[][]) - \Delta_1(\lwc{qDG}{L}[][]) M_q^\dagger \, .
\end{equation}
Finally, the four-quark-operator coefficients are shifted by the field redefinitions according to
\begin{align}
	\Delta_1(\lwc{qq}{LL}[V][prst]) &\mapsto \Delta_1(\lwc{qq}{LL}[V][prst]) - \frac{g^2}{4} \left( \delta_{pt} \Delta_1(\lwc{qGD}{L}[][sr]) + \delta_{sr} \Delta_1(\lwc{qGD}{L}[][pt]) \right) \nn
		&\qquad + \frac{g^2}{4N_c} \left( \delta_{pr} \Delta_1(\lwc{qGD}{L}[][st]) + \delta_{st} \Delta_1(\lwc{qGD}{L}[][pr]) \right) + \frac{g^4}{2} \left( \frac{\delta_{pr}\delta_{st}}{N_c} - \delta_{pt} \delta_{sr} \right) \Delta_1(\lwc{GD}{}[][]) \, , \nn
	\Delta_1(\lwc{qq}{RR}[V][prst]) &\mapsto \Delta_1(\lwc{qq}{RR}[V][prst]) - \frac{g^2}{4} \left( \delta_{pt} \Delta_1(\lwc{qGD}{R}[][sr]) + \delta_{sr} \Delta_1(\lwc{qGD}{R}[][pt]) \right) \nn
		&\qquad + \frac{g^2}{4N_c} \left( \delta_{pr} \Delta_1(\lwc{qGD}{R}[][st]) + \delta_{st} \Delta_1(\lwc{qGD}{R}[][pr]) \right) + \frac{g^4}{2} \left( \frac{\delta_{pr}\delta_{st}}{N_c} - \delta_{pt} \delta_{sr} \right) \Delta_1(\lwc{GD}{}[][]) \, , \nn
	\Delta_1(\lwc{ud}{LL}[V8][prst]) &\mapsto \Delta_1(\lwc{ud}{LL}[V8][prst]) + g^2 \delta_{pr} \Delta_1(\lwc{dGD}{L}[][st]) + g^2 \delta_{st} \Delta_1(\lwc{uGD}{L}[][pr]) + 2 g^4 \delta_{pr} \delta_{st} \Delta_1(\lwc{GD}{}[][]) \, , \nn
	\Delta_1(\lwc{ud}{RR}[V8][prst]) &\mapsto \Delta_1(\lwc{ud}{RR}[V8][prst]) + g^2 \delta_{pr} \Delta_1(\lwc{dGD}{R}[][st]) + g^2 \delta_{st} \Delta_1(\lwc{uGD}{R}[][pr]) + 2 g^4 \delta_{pr} \delta_{st} \Delta_1(\lwc{GD}{}[][]) \, , \nn
	\Delta_1(\lwc{q \tilde q}{LR}[V8][prst]) &\mapsto \Delta_1(\lwc{q \tilde q}{LR}[V8][prst]) + g^2 \delta_{pr} \Delta_1(\lwc{\tilde q GD}{R}[][st]) + g^2 \delta_{st} \Delta_1(\lwc{qGD}{L}[][pr]) + 2 g^4 \delta_{pr} \delta_{st} \Delta_1(\lwc{GD}{}[][]) \, ,
\end{align}
where $q, \tilde q \in \{ u, d \}$.

With these prescriptions, the effects of the field redefinitions can be easily obtained from the off-shell results in App.~\ref{sec:OffShellResults}. For convenience, we provide the results after field redefinitions for those parameters that are shifted.

All coefficients of redundant operators are shifted to zero. The results for the quark masses after field redefinitions are given by
\begin{align}
	\label{eq:DeltaMuOnShell}
	\Delta_1([M_u]_{pr}) &= -\frac{6g^2C_F}{t} \lwc{uG}{\dagger}[][pr] -\frac{N_c}{t} \left( \lwc{uu}{RR\dagger}[S1][prvw] [M_u^\dagger]_{wv} + \lwc{uu}{RR\dagger}[S1][vwpr] [M_u^\dagger]_{wv} + \lwc{ud}{RR\dagger}[S1][prvw] [M_d^\dagger]_{wv} \right) \nn
		&\quad + \frac{1}{2t} \begin{aligned}[t]
			&\Big( \lwc{uu}{RR\dagger}[S1][pvwr] [M_u^\dagger]_{vw} + \lwc{uu}{RR\dagger}[S1][wrpv] [M_u^\dagger]_{vw} + \lwc{uddu}{RR\dagger}[S1][wrpv] [M_d^\dagger]_{vw} \\
			&+ 4 \lwc{uddu}{LR\dagger}[V1][wrpv] [M_d]_{vw} + 4 \lwc{uu}{LR}[V1][wrpv] [M_u]_{vw} \Big) \end{aligned} \nn
		&\quad - \frac{C_F}{2t} \begin{aligned}[t]
			&\Big( \lwc{uu}{RR\dagger}[S8][pvwr] [M_u^\dagger]_{vw} + \lwc{uu}{RR\dagger}[S8][wrpv] [M_u^\dagger]_{vw} + \lwc{uddu}{RR\dagger}[S8][wrpv] [M_d^\dagger]_{vw} \\
			&+ 4 \lwc{uddu}{LR\dagger}[V8][wrpv] [M_d]_{vw} + 4 \lwc{uu}{LR}[V8][wrpv] [M_u]_{vw} \Big) \end{aligned} \nn
		&\quad - \frac{1}{8} g^2 C_F \left( 113 + 48 \lt \right) \left( [ \lwc{uG}{\dagger}[][] M_u^\dagger M_u ]_{pr} + [  M_u M_u^\dagger\lwc{uG}{\dagger}[][] ]_{pr} \right) - \frac{25 g^2 C_F}{4} [  M_u \lwc{uG}{}[][] M_u ]_{pr} \nn
		&\quad - \frac{N_c}{3} \left( 11 + 6 \lt \right) \begin{aligned}[t]
			&\Big( \lwc{uu}{RR\dagger}[S1][wvpr] [M_u^\dagger M_u M_u^\dagger ]_{vw} + \lwc{uu}{RR\dagger}[S1][prwv] [M_u^\dagger M_u M_u^\dagger ]_{vw} \\
			&+ \lwc{ud}{RR\dagger}[S1][prwv] [M_d^\dagger M_d M_d^\dagger ]_{vw} \Big) \end{aligned} \nn
		&\quad + \frac{1}{6} \left( 11 + 6 \lt \right) \begin{aligned}[t]
			&\Big( \lwc{uu}{RR\dagger}[S1][pvwr] [M_u^\dagger M_u M_u^\dagger ]_{vw} + \lwc{uu}{RR\dagger}[S1][wrpv] [M_u^\dagger M_u M_u^\dagger ]_{vw} \\
			&+ \lwc{uddu}{RR\dagger}[S1][wrpv] [M_d^\dagger M_d M_d^\dagger ]_{vw} + 4 \lwc{uddu}{LR\dagger}[V1][wrpv] [M_d M_d^\dagger M_d ]_{vw} \\
			&+ 4 \lwc{uu}{LR}[V1][wrpv] [M_u M_u^\dagger M_u ]_{vw} \Big) \end{aligned} \nn
		&\quad - \frac{C_F}{6} \left( 11 + 6 \lt \right) \begin{aligned}[t]
			&\Big( \lwc{uu}{RR\dagger}[S8][pvwr] [M_u^\dagger M_u M_u^\dagger ]_{vw} + \lwc{uu}{RR\dagger}[S8][wrpv] [M_u^\dagger M_u M_u^\dagger ]_{vw} \\
			&+ \lwc{uddu}{RR\dagger}[S8][wrpv] [M_d^\dagger M_d M_d^\dagger ]_{vw} + 4 \lwc{uddu}{LR\dagger}[V8][wrpv] [M_d M_d^\dagger M_d ]_{vw} \\
			&+ 4 \lwc{uu}{LR}[V8][wrpv] [M_u M_u^\dagger M_u ]_{vw} \Big) \end{aligned} \nn
		%%%%
		&\quad - \frac{1}{4} \left( [M_u M_u^\dagger]_{pu} \delta_{xr} + \delta_{pu} [M_u^\dagger M_u]_{xr} \right) \nn
		&\quad \qquad \times \Bigg( 2N_c \Big( \lwc{uu}{RR\dagger}[S1][wvux] [M_u^\dagger ]_{vw} + \lwc{uu}{RR^\dagger}[S1][uxwv] [M_u^\dagger ]_{vw} + \lwc{ud}{RR}[S1][uxwv] [M_d ]_{vw} \Big) \nn
		&\quad \qquad \qquad - \begin{aligned}[t]
			& \Big( \lwc{uu}{RR^\dagger}[S1][uvwx] [M_u^\dagger ]_{vw} + \lwc{uu}{RR^\dagger}[S1][wxuv] [M_u^\dagger ]_{vw} + \lwc{uddu}{RR^\dagger}[S1][uvwx] [M_d^\dagger ]_{vw} \\
			&+ 4 \lwc{uddu}{LR^\dagger}[V1][uvwx] [M_d ]_{vw} + 4 \lwc{uu}{LR}[V1][uvwx] [M_u ]_{vw} \Big) \end{aligned} \nn
		&\quad \qquad \qquad + C_F \begin{aligned}[t]
			&\Big( \lwc{uu}{RR^\dagger}[S8][uvwx] [M_u^\dagger ]_{vw} + \lwc{uu}{RR^\dagger}[S8][wxuv] [M_u^\dagger ]_{vw} + \lwc{uddu}{RR^\dagger}[S8][uvwx] [M_d^\dagger ]_{vw} \\
			&+ 4 \lwc{uddu}{LR^\dagger}[V8][uvwx] [M_d ]_{vw} + 4 \lwc{uu}{LR}[V8][uvwx] [M_u ]_{vw} \Big) \Bigg) \end{aligned} \nn
		%%%%
		&\quad - \frac{1}{2} [M_u]_{pu} [M_u]_{xr} \nn
		&\quad\qquad \times \Bigg( 2 N_c\Big( \lwc{uu}{RR}[S1][wvux] [M_u ]_{vw} + \lwc{uu}{RR}[S1][uxwv] [M_u ]_{vw} + \lwc{ud}{RR}[S1][uxwv] [M_d ]_{vw} \Big) \nn
		&\quad\qquad\qquad - \begin{aligned}[t]
			&\Big( \lwc{uu}{RR}[S1][uvwx] [M_u ]_{vw} + \lwc{uu}{RR}[S1][wxuv] [M_u ]_{vw} + \lwc{uddu}{RR}[S1][uvwx] [M_d ]_{vw} \\
			&+ 4 \lwc{uddu}{LR}[V1][uvwx] [M_d^\dagger ]_{vw} + 4 \lwc{uu}{LR}[V1][uvwx] [M_u^\dagger ]_{vw} \Big) \end{aligned} \nn
		&\quad\qquad\qquad + C_F \begin{aligned}[t]
			&\Big( \lwc{uu}{RR}[S8][uvwx] [M_u ]_{vw} + \lwc{uu}{RR}[S8][wxuv] [M_u ]_{vw} + \lwc{uddu}{RR}[S8][uvwx] [M_d ]_{vw} \\
			&+ 4 \lwc{uddu}{LR}[V8][uvwx] [M_d^\dagger ]_{vw} + 4 \lwc{uu}{LR}[V8][uvwx] [M_u^\dagger ]_{vw} \Big) \Bigg) \end{aligned} \nn
		%%%%
		&\quad + 2 g^4 C_F ( 13 + 6 \lt ) [M_u]_{pr} L_G^{(4)} - g^2 C_F \left( 6 + 3 \lt + \log(432) \right) \lwc{Mu}{(3)}[][pr] \nn
		&\quad + \frac{1}{4} g^2 C_F \left( 2 \log(432) - 1 \right) \left( [ \lwc{uD}{R(4)}[][] M_u ]_{pr} + [ M_u \lwc{uD}{L(4)}[][] ]_{pr}  \right)  \, , \\
	\label{eq:DeltaMdOnShell}
	\Delta_1([M_d]_{pr}) &= -\frac{6g^2C_F}{t} \lwc{dG}{\dagger}[][pr] -\frac{N_c}{t} \left( \lwc{dd}{RR\dagger}[S1][prvw] [M_d^\dagger]_{wv} + \lwc{dd}{RR\dagger}[S1][vwpr] [M_d^\dagger]_{wv} + \lwc{ud}{RR\dagger}[S1][vwpr] [M_u^\dagger]_{wv} \right) \nn
		&\quad + \frac{1}{2t} \begin{aligned}[t]
			&\Big( \lwc{dd}{RR\dagger}[S1][pvwr] [M_d^\dagger]_{vw} + \lwc{dd}{RR\dagger}[S1][wrpv] [M_d^\dagger]_{vw} + \lwc{uddu}{RR\dagger}[S1][pvwr] [M_u^\dagger]_{vw} \\
			&+ 4 \lwc{uddu}{LR}[V1][wrpv] [M_u]_{vw} + 4 \lwc{dd}{LR}[V1][wrpv] [M_d]_{vw} \Big) \end{aligned} \nn
		&\quad - \frac{C_F}{2t} \begin{aligned}[t]
			&\Big( \lwc{dd}{RR\dagger}[S8][pvwr] [M_d^\dagger]_{vw} + \lwc{dd}{RR\dagger}[S8][wrpv] [M_d^\dagger]_{vw} + \lwc{uddu}{RR\dagger}[S8][pvwr] [M_u^\dagger]_{vw} \\
			&+ 4 \lwc{uddu}{LR}[V8][wrpv] [M_u]_{vw} + 4 \lwc{dd}{LR}[V8][wrpv] [M_d]_{vw} \Big) \end{aligned} \nn
		&\quad - \frac{1}{8} g^2 C_F \left( 113 + 48 \lt \right) \left( [ \lwc{dG}{\dagger}[][] M_d^\dagger M_d ]_{pr} + [  M_d M_d^\dagger\lwc{dG}{\dagger}[][] ]_{pr} \right) - \frac{25 g^2 C_F}{4} [  M_d \lwc{dG}{}[][] M_d ]_{pr} \nn
		&\quad - \frac{N_c}{3} \left( 11 + 6 \lt \right) \begin{aligned}[t]
			&\Big( \lwc{dd}{RR\dagger}[S1][wvpr] [M_d^\dagger M_d M_d^\dagger ]_{vw} + \lwc{dd}{RR\dagger}[S1][prwv] [M_d^\dagger M_d M_d^\dagger ]_{vw} \\
			&+ \lwc{ud}{RR\dagger}[S1][wvpr] [M_u^\dagger M_u M_u^\dagger ]_{vw} \Big) \end{aligned} \nn
		&\quad + \frac{1}{6} \left( 11 + 6 \lt \right) \begin{aligned}[t]
			&\Big( \lwc{dd}{RR\dagger}[S1][pvwr] [M_d^\dagger M_d M_d^\dagger ]_{vw} + \lwc{dd}{RR\dagger}[S1][wrpv] [M_d^\dagger M_d M_d^\dagger ]_{vw} \\
			&+ \lwc{uddu}{RR\dagger}[S1][pvwr] [M_u^\dagger M_u M_u^\dagger ]_{vw} + 4 \lwc{uddu}{LR}[V1][wrpv] [M_u M_u^\dagger M_u ]_{vw} \\
			&+ 4 \lwc{dd}{LR}[V1][wrpv] [M_d M_d^\dagger M_d ]_{vw} \Big) \end{aligned} \nn
		&\quad - \frac{C_F}{6} \left( 11 + 6 \lt \right) \begin{aligned}[t]
			&\Big( \lwc{dd}{RR\dagger}[S8][pvwr] [M_d^\dagger M_d M_d^\dagger ]_{vw} + \lwc{dd}{RR\dagger}[S8][wrpv] [M_d^\dagger M_d M_d^\dagger ]_{vw} \\
			&+ \lwc{uddu}{RR\dagger}[S8][pvwr] [M_u^\dagger M_u M_u^\dagger ]_{vw} + 4 \lwc{uddu}{LR}[V8][wrpv] [M_u M_u^\dagger M_u ]_{vw} \\
			&+ 4 \lwc{dd}{LR}[V8][wrpv] [M_d M_d^\dagger M_d ]_{vw} \Big) \end{aligned} \nn
		%%%%
		&\quad - \frac{1}{4} \left( [M_d M_d^\dagger]_{pu} \delta_{xr} + \delta_{pu} [M_d^\dagger M_d]_{xr} \right) \nn
			&\quad\qquad \times \Bigg( 2 N_c \Big( \lwc{dd}{RR\dagger}[S1][wvux] [M_d^\dagger ]_{vw} + \lwc{dd}{RR\dagger}[S1][uxwv] [M_d^\dagger ]_{vw} + \lwc{ud}{RR\dagger}[S1][wvux] [M_u^\dagger ]_{vw} \Big) \nn
		&\quad\qquad\qquad - \begin{aligned}[t]
			&\Big( \lwc{dd}{RR\dagger}[S1][uvwx] [M_d^\dagger ]_{vw} + \lwc{dd}{RR\dagger}[S1][wxuv] [M_d^\dagger ]_{vw} + \lwc{uddu}{RR\dagger}[S1][wxuv] [M_u^\dagger ]_{vw} \\
			&+ 4 \lwc{uddu}{LR}[V1][uvwx] [M_u ]_{vw} + 4 \lwc{dd}{LR}[V1][uvwx] [M_d ]_{vw} \Big) \end{aligned} \nn
		&\quad\qquad\qquad + C_F \begin{aligned}[t]
			&\Big( \lwc{dd}{RR\dagger}[S8][uvwx] [M_d^\dagger ]_{vw} + \lwc{dd}{RR\dagger}[S8][wxuv] [M_d^\dagger ]_{vw} + \lwc{uddu}{RR\dagger}[S8][wxuv] [M_u^\dagger ]_{vw} \\
			&+ 4 \lwc{uddu}{LR}[V8][uvwx] [M_u ]_{vw} + 4 \lwc{dd}{LR}[V8][uvwx] [M_d ]_{vw} \Big) \Bigg) \end{aligned} \nn
		%%%%
		&\quad - \frac{1}{2} [M_d]_{pu} [M_d]_{xr} \nn
			&\quad\qquad \times \Bigg(  2 N_c \Big( \lwc{dd}{RR}[S1][wvux] [M_d ]_{vw} + \lwc{dd}{RR}[S1][uxwv] [M_d ]_{vw} + \lwc{ud}{RR}[S1][wvux] [M_u ]_{vw} \Big) \nn
			&\quad\qquad\qquad - \begin{aligned}[t]
				&\Big( \lwc{dd}{RR}[S1][uvwx] [M_d ]_{vw} + \lwc{dd}{RR}[S1][wxuv] [M_d ]_{vw} + \lwc{uddu}{RR}[S1][wxuv] [M_u ]_{vw} \\
				&+ 4 \lwc{uddu}{LR\dagger}[V1][uvwx] [M_u^\dagger ]_{vw} + 4 \lwc{dd}{LR}[V1][uvwx] [M_d^\dagger ]_{vw} \Big) \end{aligned} \nn
			&\quad\qquad\qquad + C_F \begin{aligned}[t]
				&\Big( \lwc{dd}{RR}[S8][uvwx] [M_d ]_{vw} + \lwc{dd}{RR}[S8][wxuv] [M_d ]_{vw} + \lwc{uddu}{RR}[S8][wxuv] [M_u ]_{vw} \\
				&+ 4 \lwc{uddu}{LR\dagger}[V8][uvwx] [M_u^\dagger ]_{vw} + 4 \lwc{dd}{LR}[V8][uvwx] [M_d^\dagger ]_{vw} \Big) \Bigg) \end{aligned} \bigg ) \nn
		&\quad + 2 g^4 C_F ( 13 + 6 \lt ) [M_d]_{pr} L_G^{(4)} - g^2 C_F \left( 6 + 3 \lt + \log(432) \right) \lwc{Md}{(3)}[][pr] \nn
		&\quad + \frac{1}{4} g^2 C_F \left( 2 \log(432) - 1 \right) \left( [ \lwc{dD}{R(4)}[][] M_d ]_{pr} + [ M_d \lwc{dD}{L(4)}[][] ]_{pr}  \right)  \, . 
\end{align}
The quark chromo-dipole-operator coefficients are shifted to
\begin{align}
	\label{eq:DeltaLuGOnShell}
	\Delta_1(\lwc{uG}{}[][pr]) &= \frac{1}{2} g^2 \left( C_F \left( 1 + 10 \lt - 2 \log(432) \right) - N_c \left( 9 + 4 \lt \right) \right) \lwc{uG}{}[][pr] \nn
		&\quad - N_c \left( \lwc{uu}{RR}[S1][wvpr] [M_u ]_{vw} + \lwc{uu}{RR}[S1][prwv] [M_u ]_{vw} + \lwc{ud}{RR}[S1][prwv] [M_d ]_{vw} \right) \nn
		&\quad + \frac{1}{2} \left( 3 + \lt \right) \left( \lwc{uu}{RR}[S1][pvwr] [M_u]_{vw} + \lwc{uu}{RR}[S1][wrpv] [M_u]_{vw} + \lwc{uddu}{RR}[S1][pvwr] [M_d]_{vw} \right) \nn
		&\quad - \left( \frac{C_F}{2} \left( 3 + \lt \right) - \frac{N_c}{4} \left( 2 + \lt \right) \right) \left( \lwc{uu}{RR}[S8][pvwr] [M_u]_{vw} + \lwc{uu}{RR}[S8][wrpv] [M_u]_{vw} + \lwc{uddu}{RR}[S8][pvwr] [M_d]_{vw} \right) \nn
		&\quad + 2 \left( \lwc{uu}{LR}[V1][pvwr] [M_u^\dagger]_{vw} + \lwc{uddu}{LR}[V1][pvwr] [M_d^\dagger]_{vw} \right) - 2 C_F \left( \lwc{uu}{LR}[V8][pvwr] [M_u^\dagger]_{vw} + \lwc{uddu}{LR}[V8][pvwr] [M_d^\dagger]_{vw} \right) \nn
		&\quad + \frac{N_c}{4} g^4 \left( 13 + 6 \lt \right) [M_u^\dagger]_{pr} \left( L_G - i L_{\widetilde G} \right)  \, , \\
	\label{eq:DeltaLdGOnShell}
	\Delta_1(\lwc{dG}{}[][pr]) &= \frac{1}{2} g^2 \left( C_F \left( 1 + 10 \lt - 2 \log(432) \right) - N_c \left( 9 + 4 \lt \right) \right) \lwc{dG}{}[][pr] \nn
		&\quad - N_c \left( \lwc{dd}{RR}[S1][wvpr] [M_d ]_{vw} + \lwc{dd}{RR}[S1][prwv] [M_d ]_{vw} + \lwc{ud}{RR}[S1][wvpr] [M_u ]_{vw} \right) \nn
		&\quad + \frac{1}{2} \left( 3 + \lt \right) \left( \lwc{dd}{RR}[S1][pvwr] [M_d]_{vw} + \lwc{dd}{RR}[S1][wrpv] [M_d]_{vw} + \lwc{uddu}{RR}[S1][wrpv] [M_u]_{vw} \right) \nn
		&\quad - \left( \frac{C_F}{2} \left( 3 + \lt \right) - \frac{N_c}{4} \left( 2 + \lt \right) \right) \left( \lwc{dd}{RR}[S8][pvwr] [M_d]_{vw} + \lwc{dd}{RR}[S8][wrpv] [M_d]_{vw} + \lwc{uddu}{RR}[S8][wrpv] [M_u]_{vw} \right) \nn
		&\quad + 2 \left( \lwc{dd}{LR}[V1][pvwr] [M_d^\dagger]_{vw} + \lwc{uddu}{LR\dagger}[V1][pvwr] [M_u^\dagger]_{vw} \right) - 2 C_F \left( \lwc{dd}{LR}[V8][pvwr] [M_d^\dagger]_{vw} + \lwc{uddu}{LR\dagger}[V8][pvwr] [M_u^\dagger]_{vw} \right) \nn
		&\quad + \frac{N_c}{4} g^4 \left( 13 + 6 \lt \right) [M_d^\dagger]_{pr} \left( L_G - i L_{\widetilde G} \right)  \, .
\end{align}
Finally, the following four-quark operators are changed after field redefinitions:
\begin{align}
	% uuuu
	\label{eq:DeltaLVLLuuOnShell}
	\Delta_1(\lwc{uu}{LL}[V][prst]) &= - g^2 \left( C_F \left( 3 + 2\log(432) \right) + \frac{1}{2N_c} \left( 11 + 6\lt \right) \right) \lwc{uu}{LL}[V][prst] + \frac{1}{2} g^2 \left( 11 + 6\lt \right) \lwc{uu}{LL}[V][ptsr] \nn
		&\quad - \frac{g^2}{48} \left( 3 + 4 \lt \right) \delta_{pt} \left( \lwc{uu}{LR}[V8][srvv] + \lwc{ud}{LL}[V8][srvv] + \lwc{ud}{LR}[V8][srvv] - 2 \lwc{uu}{LL}[V][svvr] - 2 \lwc{uu}{LL}[V][vrsv] \right) \nn
		&\quad - \frac{g^2}{48} \left( 3 + 4 \lt \right) \delta_{sr} \left( \lwc{uu}{LR}[V8][ptvv] + \lwc{ud}{LL}[V8][ptvv] + \lwc{ud}{LR}[V8][ptvv] - 2 \lwc{uu}{LL}[V][pvvt] - 2 \lwc{uu}{LL}[V][vtpv] \right) \nn
		&\quad + \frac{g^2}{48 N_c} \left( 3 + 4 \lt \right) \delta_{pr} \left( \lwc{uu}{LR}[V8][stvv] + \lwc{ud}{LL}[V8][stvv] + \lwc{ud}{LR}[V8][stvv] - 2 \lwc{uu}{LL}[V][svvt] - 2 \lwc{uu}{LL}[V][vtsv] \right) \nn
		&\quad + \frac{g^2}{48N_c} \left( 3 + 4 \lt \right) \delta_{st} \left( \lwc{uu}{LR}[V8][prvv] + \lwc{ud}{LL}[V8][prvv] + \lwc{ud}{LR}[V8][prvv] - 2 \lwc{uu}{LL}[V][pvvr] - 2 \lwc{uu}{LL}[V][vrpv] \right) \nn
		&\quad + \left( \delta_{pr}\delta_{st} - N_c \delta_{pt} \delta_{sr} \right)  \frac{61 g^6}{8} L_G \, , \\
	\label{eq:DeltaLVRRuuOnShell}
	\Delta_1(\lwc{uu}{RR}[V][prst]) &= - g^2 \left( C_F \left( 3 + 2\log(432) \right) + \frac{1}{2N_c} \left( 11 + 6\lt \right) \right) \lwc{uu}{RR}[V][prst] + \frac{1}{2} g^2 \left( 11 + 6\lt \right) \lwc{uu}{RR}[V][ptsr] \nn
		&\quad - \frac{g^2}{48} \left( 3 + 4 \lt \right) \delta_{pt} \left( \lwc{uu}{LR}[V8][vvsr] + \lwc{ud}{RR}[V8][srvv] + \lwc{du}{LR}[V8][vvsr] - 2 \lwc{uu}{RR}[V][svvr] - 2 \lwc{uu}{RR}[V][vrsv] \right) \nn
		&\quad - \frac{g^2}{48} \left( 3 + 4 \lt \right) \delta_{sr} \left( \lwc{uu}{LR}[V8][vvpt] + \lwc{ud}{RR}[V8][ptvv] + \lwc{du}{LR}[V8][vvpt] - 2 \lwc{uu}{RR}[V][pvvt] - 2 \lwc{uu}{RR}[V][vtpv] \right) \nn
		&\quad + \frac{g^2}{48 N_c} \left( 3 + 4 \lt \right) \delta_{pr} \left( \lwc{uu}{LR}[V8][vvst] + \lwc{ud}{RR}[V8][stvv] + \lwc{du}{LR}[V8][vvst] - 2 \lwc{uu}{RR}[V][svvt] - 2 \lwc{uu}{RR}[V][vtsv] \right) \nn
		&\quad + \frac{g^2}{48N_c} \left( 3 + 4 \lt \right) \delta_{st} \left( \lwc{uu}{LR}[V8][vvpr] + \lwc{ud}{RR}[V8][prvv] + \lwc{du}{LR}[V8][vvpr] - 2 \lwc{uu}{RR}[V][pvvr] - 2 \lwc{uu}{RR}[V][vrpv] \right) \nn
		&\quad + \left( \delta_{pr}\delta_{st} - N_c \delta_{pt} \delta_{sr} \right)  \frac{61 g^6}{8} L_G \, , \\
	\label{eq:DeltaLV8LRuuOnShell}
	\Delta_1(\lwc{uu}{LR}[V8][prst]) &= g^2 \left( 7 + 6\lt \right) \lwc{uu}{LR}[V1][prst] + \frac{g^2}{2} \left( N_c \left( 7 + 6\lt \right) - 2 C_F\left( 17 + 12\lt + 2 \log(432) \right) \right) \lwc{uu}{LR}[V8][prst] \nn
		&\quad + \frac{g^2}{12} \left( 3 + 4 \lt \right) \delta_{pr} \left( \lwc{uu}{LR}[V8][vvst] + \lwc{ud}{RR}[V8][stvv] + \lwc{du}{LR}[V8][vvst] - 2 \lwc{uu}{RR}[V][svvt] - 2 \lwc{uu}{RR}[V][vtsv] \right) \nn
		&\quad + \frac{g^2 }{12} \left( 3 + 4 \lt \right) \delta_{st} \left( \lwc{uu}{LR}[V8][prvv] + \lwc{ud}{LL}[V8][prvv] + \lwc{ud}{LR}[V8][prvv] - 2 \lwc{uu}{LL}[V][pvvr] - 2 \lwc{uu}{LL}[V][vrpv] \right) \nn
		&\quad + \frac{61 g^6 N_c}{2} L_G \, \delta_{pr} \delta_{st} \, , \\
	% dddd
	\label{eq:DeltaLVLLddOnShell}
	\Delta_1(\lwc{dd}{LL}[V][prst]) &= - g^2 \left( C_F \left( 3 + 2\log(432) \right) + \frac{1}{2N_c} \left( 11 + 6\lt \right) \right) \lwc{dd}{LL}[V][prst] + \frac{1}{2} g^2 \left( 11 + 6\lt \right) \lwc{dd}{LL}[V][ptsr] \nn
		&\quad - \frac{g^2}{48} \left( 3 + 4 \lt \right) \delta_{pt} \left( \lwc{dd}{LR}[V8][srvv] + \lwc{ud}{LL}[V8][vvsr] + \lwc{du}{LR}[V8][srvv] - 2 \lwc{dd}{LL}[V][svvr] - 2 \lwc{dd}{LL}[V][vrsv] \right) \nn
		&\quad - \frac{g^2}{48} \left( 3 + 4 \lt \right) \delta_{sr} \left( \lwc{dd}{LR}[V8][ptvv] + \lwc{ud}{LL}[V8][vvpt] + \lwc{du}{LR}[V8][ptvv] - 2 \lwc{dd}{LL}[V][pvvt] - 2 \lwc{dd}{LL}[V][vtpv] \right) \nn
		&\quad + \frac{g^2}{48 N_c} \left( 3 + 4 \lt \right) \delta_{pr} \left( \lwc{dd}{LR}[V8][stvv] + \lwc{ud}{LL}[V8][vvst] + \lwc{du}{LR}[V8][stvv] - 2 \lwc{dd}{LL}[V][svvt] - 2 \lwc{dd}{LL}[V][vtsv] \right) \nn
		&\quad + \frac{g^2}{48N_c} \left( 3 + 4 \lt \right) \delta_{st} \left( \lwc{dd}{LR}[V8][prvv] + \lwc{ud}{LL}[V8][vvpr] + \lwc{du}{LR}[V8][prvv] - 2 \lwc{dd}{LL}[V][pvvr] - 2 \lwc{dd}{LL}[V][vrpv] \right) \nn
		&\quad + \left( \delta_{pr}\delta_{st} - N_c \delta_{pt} \delta_{sr} \right)  \frac{61 g^6}{8} L_G \, , \\
	\label{eq:DeltaLVRRddOnShell}
	\Delta_1(\lwc{dd}{RR}[V][prst]) &= - g^2 \left( C_F \left( 3 + 2\log(432) \right) + \frac{1}{2N_c} \left( 11 + 6\lt \right) \right) \lwc{dd}{RR}[V][prst] + \frac{1}{2} g^2 \left( 11 + 6\lt \right) \lwc{dd}{RR}[V][ptsr] \nn
		&\quad - \frac{g^2}{48} \left( 3 + 4 \lt \right) \delta_{pt} \left( \lwc{dd}{LR}[V8][vvsr] + \lwc{ud}{RR}[V8][vvsr] + \lwc{ud}{LR}[V8][vvsr] - 2 \lwc{dd}{RR}[V][svvr] - 2 \lwc{dd}{RR}[V][vrsv] \right) \nn
		&\quad - \frac{g^2}{48} \left( 3 + 4 \lt \right) \delta_{sr} \left( \lwc{dd}{LR}[V8][vvpt] + \lwc{ud}{RR}[V8][vvpt] + \lwc{ud}{LR}[V8][vvpt] - 2 \lwc{dd}{RR}[V][pvvt] - 2 \lwc{dd}{RR}[V][vtpv] \right) \nn
		&\quad + \frac{g^2}{48 N_c} \left( 3 + 4 \lt \right) \delta_{pr} \left( \lwc{dd}{LR}[V8][vvst] + \lwc{ud}{RR}[V8][vvst] + \lwc{ud}{LR}[V8][vvst] - 2 \lwc{dd}{RR}[V][svvt] - 2 \lwc{dd}{RR}[V][vtsv] \right) \nn
		&\quad + \frac{g^2}{48N_c} \left( 3 + 4 \lt \right) \delta_{st} \left( \lwc{dd}{LR}[V8][vvpr] + \lwc{ud}{RR}[V8][vvpr] + \lwc{ud}{LR}[V8][vvpr] - 2 \lwc{dd}{RR}[V][pvvr] - 2 \lwc{dd}{RR}[V][vrpv] \right) \nn
		&\quad + \left( \delta_{pr}\delta_{st} - N_c \delta_{pt} \delta_{sr} \right)  \frac{61 g^6}{8} L_G \, , \\
	\label{eq:DeltaLV8LRddOnShell}
	\Delta_1(\lwc{dd}{LR}[V8][prst]) &= g^2 \left( 7 + 6\lt \right) \lwc{dd}{LR}[V1][prst] + \frac{g^2}{2} \left( N_c \left( 7 + 6\lt \right) - 2 C_F\left( 17 + 12\lt + 2 \log(432) \right) \right) \lwc{dd}{LR}[V8][prst] \nn
		&\quad + \frac{g^2}{12} \left( 3 + 4 \lt \right) \delta_{pr} \left( \lwc{dd}{LR}[V8][vvst] + \lwc{ud}{RR}[V8][vvst] + \lwc{ud}{LR}[V8][vvst] - 2 \lwc{dd}{RR}[V][svvt] - 2 \lwc{dd}{RR}[V][vtsv] \right) \nn
		&\quad + \frac{g^2 }{12} \left( 3 + 4 \lt \right) \delta_{st} \left( \lwc{dd}{LR}[V8][prvv] + \lwc{ud}{LL}[V8][vvpr] + \lwc{du}{LR}[V8][prvv] - 2 \lwc{dd}{LL}[V][pvvr] - 2 \lwc{dd}{LL}[V][vrpv] \right) \nn
		&\quad + \frac{61 g^6 N_c}{2} L_G \, \delta_{pr} \delta_{st} \, , \\
	% uudd
	\label{eq:DeltaLV8LLudOnShell}
	\Delta_1(\lwc{ud}{LL}[V8][prst]) &= -g^2 \left( 11 + 6\lt \right) \lwc{ud}{LL}[V1][prst] - g^2 \left( 2N_c \left( 5 + 3\lt \right) - C_F\left( 19 + 12\lt - 2 \log(432) \right) \right) \lwc{ud}{LL}[V8][prst] \nn
		&\quad + \frac{g^2}{12} \left( 3 + 4 \lt \right) \delta_{pr} \left( \lwc{dd}{LR}[V8][stvv] + \lwc{ud}{LL}[V8][vvst] + \lwc{du}{LR}[V8][stvv] - 2 \lwc{dd}{LL}[V][svvt] - 2 \lwc{dd}{LL}[V][vtsv] \right) \nn
		&\quad + \frac{g^2 }{12} \left( 3 + 4 \lt \right) \delta_{st} \left( \lwc{uu}{LR}[V8][prvv] + \lwc{ud}{LL}[V8][prvv] + \lwc{ud}{LR}[V8][prvv] - 2 \lwc{uu}{LL}[V][pvvr] - 2 \lwc{uu}{LL}[V][vrpv] \right) \nn
		&\quad + \frac{61 g^6 N_c}{2} L_G \, \delta_{pr} \delta_{st} \, , \\
	\label{eq:DeltaLV8RRudOnShell}
	\Delta_1(\lwc{ud}{RR}[V8][prst]) &= -g^2 \left( 11 + 6\lt \right) \lwc{ud}{RR}[V1][prst] - g^2 \left( 2N_c \left( 5 + 3\lt \right) - C_F\left( 19 + 12\lt - 2 \log(432) \right) \right) \lwc{ud}{RR}[V8][prst] \nn
		&\quad + \frac{g^2}{12} \left( 3 + 4 \lt \right) \delta_{pr} \left( \lwc{dd}{LR}[V8][vvst] + \lwc{ud}{RR}[V8][vvst] + \lwc{ud}{LR}[V8][vvst] - 2 \lwc{dd}{RR}[V][svvt] - 2 \lwc{dd}{RR}[V][vtsv] \right) \nn
		&\quad + \frac{g^2 }{12} \left( 3 + 4 \lt \right) \delta_{st} \left( \lwc{uu}{LR}[V8][vvpr] + \lwc{ud}{RR}[V8][prvv] + \lwc{du}{LR}[V8][vvpr] - 2 \lwc{uu}{RR}[V][pvvr] - 2 \lwc{uu}{RR}[V][vrpv] \right) \nn
		&\quad + \frac{61 g^6 N_c}{2} L_G \, \delta_{pr} \delta_{st} \, , \\
	\label{eq:DeltaLV8LRudOnShell}
	\Delta_1(\lwc{ud}{LR}[V8][prst]) &= g^2 \left( 7 + 6\lt \right) \lwc{ud}{LR}[V1][prst] + \frac{g^2}{2} \left( N_c \left( 7 + 6\lt \right) - 2 C_F\left( 17 + 12\lt + 2 \log(432) \right) \right) \lwc{ud}{LR}[V8][prst] \nn
		&\quad + \frac{g^2}{12} \left( 3 + 4 \lt \right) \delta_{pr} \left( \lwc{dd}{LR}[V8][vvst] + \lwc{ud}{RR}[V8][vvst] + \lwc{ud}{LR}[V8][vvst] - 2 \lwc{dd}{RR}[V][svvt] - 2 \lwc{dd}{RR}[V][vtsv] \right) \nn
		&\quad + \frac{g^2 }{12} \left( 3 + 4 \lt \right) \delta_{st} \left( \lwc{uu}{LR}[V8][prvv] + \lwc{ud}{LL}[V8][prvv] + \lwc{ud}{LR}[V8][prvv] - 2 \lwc{uu}{LL}[V][pvvr] - 2 \lwc{uu}{LL}[V][vrpv] \right) \nn
		&\quad + \frac{61 g^6 N_c}{2} L_G \, \delta_{pr} \delta_{st} \, , \\
	\label{eq:DeltaLV8LRduOnShell}
	\Delta_1(\lwc{du}{LR}[V8][prst]) &= g^2 \left( 7 + 6\lt \right) \lwc{du}{LR}[V1][prst] + \frac{g^2}{2} \left( N_c \left( 7 + 6\lt \right) - 2 C_F\left( 17 + 12\lt + 2 \log(432) \right) \right) \lwc{du}{LR}[V8][prst] \nn
		&\quad + \frac{g^2}{12} \left( 3 + 4 \lt \right) \delta_{pr} \left( \lwc{uu}{LR}[V8][vvst] + \lwc{ud}{RR}[V8][stvv] + \lwc{du}{LR}[V8][vvst] - 2 \lwc{uu}{RR}[V][svvt] - 2 \lwc{uu}{RR}[V][vtsv] \right) \nn
		&\quad + \frac{g^2 }{12} \left( 3 + 4 \lt \right) \delta_{st} \left( \lwc{dd}{LR}[V8][prvv] + \lwc{ud}{LL}[V8][vvpr] + \lwc{du}{LR}[V8][prvv] - 2 \lwc{dd}{LL}[V][pvvr] - 2 \lwc{dd}{LL}[V][vrpv] \right) \nn
		&\quad + \frac{61 g^6 N_c}{2} L_G \, \delta_{pr} \delta_{st} \, .
\end{align}
All other non-redundant parameters remain unchanged compared to App.~\ref{sec:OffShellResults}.

	\end{myfmf}

	\clearpage

	\phantomsection
	\addcontentsline{toc}{section}{\numberline{}References}
	\bibliographystyle{utphysmod}
	\bibliography{Literature}
	
\end{document}